\documentclass[11pt]{article}
\pdfoutput = 1

\usepackage[utf8]{inputenc}
\usepackage{color,graphicx}
\usepackage{amsmath}
\usepackage{verbatim}
\usepackage{amssymb}
\usepackage{physics}
\usepackage{amsfonts}
\usepackage{cite}
\usepackage{array}
\usepackage{setspace}
\usepackage{float}
\usepackage{url}
\usepackage{slashed}
\usepackage{tikz}
\usepackage{graphicx}
\usepackage{fancybox}
\usepackage{tensor}
\usepackage{tikz}
\usepackage{cite}
\usepackage{caption}
\usepackage{subcaption}
\usepackage{mathrsfs}
\allowdisplaybreaks
\newcommand{\ubone}[2]{\underbrace{#1}_{\scriptscriptstyle #2}}
\newcommand{\ubtwo}[3]{\underbrace{#1}_{\substack{\scriptscriptstyle #2\\[-1pt]\scriptscriptstyle #3}}}
\newcommand{\ubbarone}[2]{\underbrace{#1}_{\scriptscriptstyle \overline{#2}}}
\newcommand{\ubbartwo}[3]{\underbrace{#1}_{\substack{\scriptscriptstyle \overline{#2}\\[-1pt]\scriptscriptstyle \overline{#3}}}}

\usepackage[margin = 2.2cm]{geometry}
\usepackage[ragged]{footmisc}
\usepackage{hyperref}
\hypersetup{
    colorlinks,
    citecolor=blue,
    filecolor=black,
    linkcolor=blue,
    urlcolor=blue,
    linktocpage=true
}

\def\nn{\nonumber}

\def\r{\rho}

\newcommand{\be}{\begin{equation}}
\newcommand{\ee}{\end{equation}}
\newcommand{\bea}{\begin{align}}
\newcommand{\eea}{\end{align}}
\newcommand{\bi}{\begin{itemize}}
\newcommand{\ei}{\end{itemize}}

\def\nn\nonumber
\def\ve{\varepsilon}
\def\nn{\nonumber}

\numberwithin{equation}{section}

\begin{document}

\hypersetup{pageanchor=false}
\thispagestyle{empty}

 \begin{center}
 \vspace*{.4cm}
      {\Large \bf Constrained particle on a group: from propagators to correlators}
      \\[0.45cm]
      {\large Guanda Lin}
      \\[0.2cm]
      {\it Leinweber Institute for Theoretical Physics and Department of Physics,\\
      University of California, Berkeley, California 94720, U.S.A.}
      \\[0.15cm]
      {\tt geoff\_guanda\_lin@berkeley.edu}

 \end{center}

 \begin{abstract}

We develop a particle-on-a-group formulation of super-JT gravity aimed at computing supersymmetric correlators.
We show that the (super)JT gravity can be described by a particle moving on the isometry group satisfying constraints from boundary conditions of (super)JT gravity.
In this language the $\mathcal N=2$ and $\mathcal N=4$ theories are described by constrained particles on $SU(1,1|1)$ and $PSU(1,1|2)$.
Solving the constraints gives the super-Schwarzian actions.
We also quantize the reduced superparticle, with careful treatment of the fermionic constraints.
We then derive the physical worldline supercharges from the requirement that the transformations preserve the constraints.
These charges allow us to construct supersymmetric interval propagators in invariant variables and to formulate boundary-anchored Wilson-line operators for both superconformal primary and descendant insertions.
Finally, we use these ingredients to build an algorithm for correlators.
We obtain the $\mathcal N=2$ and $\mathcal N=4$ three-point composition kernels and zero-energy scalar three-point functions.
For four-point functions, the same method reproduces the standard bosonic JT OTOC, gives an explicit zero-energy OTOC in $\mathcal N=2$ and $\mathcal N=4$ SJT, which is potentially useful for studying Berry curvature and BPS chaos. 

 \end{abstract}

\pagebreak
\hypersetup{pageanchor=true}
\setcounter{page}{1}
\tableofcontents
\newpage

\setcounter{section}{0}

\section{Introduction}
\label{sec:introduction}

Recent years have witnessed significant progress in understanding the universal low-energy dynamics of near-extremal and near-BPS black holes.
For a large class of black holes, the near-horizon region develops a long nearly-$\mathrm{AdS}_2$ throat.
After dimensional reduction, the universal gravitational dynamics near the edge of this throat is described by JT gravity.
In the simplest case the only boundary mode is the reparametrization of the throat boundary, whose dynamics is the Schwarzian theory, equivalently the disk-level description of Jackiw--Teitelboim gravity \cite{Jackiw:1984je,Teitelboim:1983ux,Almheiri:2014cka,Jensen:2016pah,Maldacena:2016upp,Engelsoy:2016xyb}.

For supersymmetric black holes the corresponding low-energy theory is super-JT gravity.
The dimensional reduction now keeps the reparametrization mode as well as the boundary supersymmetries and $R$-symmetry modes fixed by the near-horizon superisometry.
Different amounts of supersymmetry for black holes give $\mathcal N\leq 4$ super-JT.
These structures control the low-energy dynamics of near-BPS black holes, including the BPS ground-state sector and the continuum of near-BPS excitations \cite{Heydeman:2020hhw,Boruch:2022tno,Heydeman:2025vcc}.

A useful heuristic in these discussions is to follow the relevant fields and operators from the asymptotic region to the edge of the near-horizon region, and then map the low-energy problem to JT or super-JT.
This UV-to-throat map lets one study the black hole as a quantum mechanical system: its spectrum and density of states are obtained from the JT or super-JT partition function \cite{Iliesiu:2020qvm,Moitra:2019bub,Kapec:2023ruw,Rakic:2023vhv,Maulik:2024dwq,Kapec:2024zdj,Heydeman:2020hhw,Boruch:2022tno,Heydeman:2025vcc}, while its response to perturbations is probed by reducing higher-dimensional fields to matter insertions dressed by the same near-horizon dynamics \cite{Mertens:2019bvy,Blommaert:2020yeo,Bai:2023hpd,Brown:2024ajk,Emparan:2023ypa,Kolanowski:2024zrq,Emparan:2025sao,Biggs:2025nzs,Maulik:2025hax,Lin:2025wof,Betzios:2025sct}.
In these discussions the properties of the low-temperature black hole are encoded in correlators, mostly two-point functions, receiving quantum corrections from the Schwarzian fluctuations of the near-horizon geometry.

Beyond the spectrum and response, one also wants observables related to chaotic behavior, and they often involve more complicated higher-point correlators.
The standard example is the out-of-time-order correlator (OTOC) \cite{Shenker:2013pqa,Roberts:2014isa,Roberts:2014ifa,Hosur:2015ylk,Maldacena:2015waa,Maldacena:2016hyu,Xu:2024otoc,Stanford:2023npy}, which measures how a simple perturbation is scrambled by black hole dynamics.
In nearly-$\mathrm{AdS}_2$ systems it gives the familiar maximal chaos signal of the Schwarzian/JT mode \cite{Jensen:2016pah,Engelsoy:2016xyb,Maldacena:2016upp,Kitaev:2017awl,Mertens:2017mtv,Blommaert:2018oro,Mertens:2022irh}.
Interestingly, even in a protected BPS sector there can still be a useful notion of chaos \cite{Lin:2022rzw,Lin:2022zxd,Chen:2024oqv}.
This is special because the relevant states are exactly degenerate ground states, so there is neither scrambling-type chaotic time evolution nor level-statisics notion of chaoswithin the protected subspace.
The chaos is instead in the eigenvectors: it describes how the degenerate ground states mix with each other under perturbations.
One specific probe is the non-Abelian Berry curvature, which measures the geometric phase acquired by the degenerate ground states as the parameters of the system are adiabatically varied \cite{Chen:2026vml}.
Gravitional prediction should give statistical properties of these Berry curvatures, in particular, moments of Berry curvatures can be captured by (higher-point) correlators in gravity: for example, the second moment involves time-ordered and out-of-time-ordered four-point correlators.

With this motivation for computing correlators, we next ask how they can be determined in practice.
From the mathematical point of view, these correlators are standard representation-theoretic objects.
For example, the ordinary JT OTOC can be written in terms of the $SL(2,\mathbb R)$ $6j$ symbol.
This means that the problem can be separated into a few pieces: the representations which can appear in an intermediate channel, the dressed three-point functions, and the crossing data which relates different pairings of the operators.
For the supergroups relevant to $\mathcal N=2,4$ super-JT, the analogous channel data, three-point structures and crossing kernels have not been worked out, at least to our knowledge.
One goal of this paper is to derive some physicscally relevant parts of these data.

Among the different formulations of JT gravity, the particle-on-a-group description is especially convenient for this purpose.
It gives a physicist's route to the same representation-theoretic answers.
The reason is that the particle-on-a-group formulation keeps the symmetry explicit: gravity and matter multiplets are organized by their transformation properties, so the correlators have a clear representation-theoretic meaning.
Conversely, the same symmetry also organizes the final answers, since the natural building blocks are group invariants and covariants.
This is the approach we take in this paper.

\paragraph{Summary of results.}
In this paper, we set up and clarify the particle-on-a-group description of (super-)JT gravity, and study applications of this formalism to supersymmetric correlators.
We first review the bosonic prototype in Section~\ref{sec:n0-jt-particle-main} as a warm-up.
Ordinary JT can be written as an $SL(2,\mathbb R)$ BF theory whose boundary dynamics is a constrained particle on $SL(2,\mathbb R)$.
We explain how the JT boundary condition becomes the particle constraint, why the constrained particle describes the same dynamics, and how the constraint should be imposed quantum mechanically.
The same logic will be supersymmetrized in the rest of the paper.
For reference, we provide a self-contained derivation in Appendix~\ref{app:n0-jt-bf-particle}.

We then perform the supersymmetric generalization in Section~\ref{sec:superjt-boundary-particle}, with emphasis on the $\mathcal N=2$ and $\mathcal N=4$ theories.
The starting point is to identify how the supergravity boundary conditions become constraints for the particle on $SU(1,1|1)$ or $PSU(1,1|2)$.
We translate the fields among the supergravity, super-BF, and particle-on-a-group descriptions, identify the constraint equations, and derive the super-Schwarzian action from the constrained particle on a supergroup.
Therefore, at least classically, the particle-on-a-group formalism gives a group-theoretic realization of the same (super-)Schwarzian boundary dynamics.

We then quantize the reduced theory and address the subtleties associated with the constraints.
In the supersymmetric case, the positive fermionic constraints satisfy an algebra that closes onto the bosonic parabolic charge.
Once this bosonic charge is fixed, these fermionic constraints become second class and cannot all be imposed directly on ordinary scalar wavefunctions.
Our particle-on-a-group derivation explains the auxiliary fermion $\chi$ in the LMRS Hilbert-space construction \cite{Lin:2022zxd}: it is the degree of freedom associated with the positive-fermion sector of the constrained particle, and the reduced wavefunctions are schematically $F(x,\rho,a,\theta_-,\bar\theta_-,\chi)$.
The same discussion on constraints also fixes the particle measure used when wavefunctions are glued.

In Section~\ref{sec:worldline-supercharges-propagator}, we clarify the distinction and relation between two symmetries: the (super)symmetry of the target space, which is the group, and the symmetry of the worldline.
We often refer to them as bulk and boundary symmetries.
The bulk symmetry is the $SU(1,1|1)$ or $PSU(1,1|2)$ isometry of the super-$\mathrm{AdS}_2$ spacetime, while the boundary symmetry is the $\mathcal N=2$ or $\mathcal N=4$ super-Poincare symmetry of the boundary superline.
The first result is a first-principles derivation of the relation between bulk and boundary supercharges.
Although closely related, a bare bulk supercharge does not preserve the super-JT boundary condition in general; rather, it must be supplemented by a compensating large diffeomorphism.
In the particle-on-a-group language, a bare $SU(1,1|1)$ or $PSU(1,1|2)$ translation would cause the trajectory to fail to satisfy the constraints.
We find an equivalent constraint-preserving supercharge, which is precisely the $\mathcal N=2$ or $\mathcal N=4$ super-Poincare supercharge \cite{Lin:2022zxd,Fan:2021wsb,Belaey:2024dde,Belaey:2023jtr,Lin:2025wof}.
Using these charges, we construct supersymmetric propagators organized by bulk-supergroup invariants and explain the reduction to Liouville quantum mechanics \cite{Lin:2022zxd,Lin:2025wof}.

Moreover, the same distinction between bulk and boundary supersymmetry controls Wilson-line operators.
A bulk field is a representation of the bulk isometry group, but its boundary-anchored two-point function has a more complicated transformation law than a bare $SU(1,1|1)$ or $PSU(1,1|2)$ transformation, because of the compensating diffeomorphisms required by the constraint.
We show that for superconformal primaries, the bulk matter two-point function becomes a simple open Wilson line.
For superdescendants, however, the construction is more subtle and must use the physical super-Poincare supercharge.
This extends the familiar bosonic BF Wilson-line/Schwarzian bilocal correspondence \cite{Blommaert:2018oro}.
For practical purposes, we also express the Wilson-line operators in terms of bulk-supergroup invariants.
As an example, we discuss the marginal operator used for the D1--D5--P moduli deformation in Section~\ref{sec:discussion-future}, whose bilocal is obtained by acting with $\mathcal N=4$ supercharges on the short-multiplet primary bilocal.

In Section~\ref{sec:correlators-length-kernel}, we turn these ingredients into an algorithm for computing correlation functions.
Starting from the reduced propagators and Wilson-line operators, we glue boundary intervals with the reduced particle measure and quotient by the diagonal global symmetry.
The resulting integral is written in invariant length variables.
The basic building block is the three-length composition kernel $I_3$, obtained by composing two interval propagators at a common endpoint.
For ordinary JT this is the familiar bosonic $I_3$ kernel \cite{Yang:2018gdb,Jafferis:2022wez,Jafferis:2022uhu}.
In the supersymmetric theories the same construction gives $\mathcal N=2$ and $\mathcal N=4$ kernels with a finite component structure reflecting the fermions and the $R$-symmetry data of the propagator.
As a first application, we compute zero-energy scalar three-point functions in the $\mathcal N=2$~\cite{Penington:2024sum} and $\mathcal N=4$ theories.
A technical point in this calculation is that the kernels are involved and the length-space integrands are complicated.
We develop an integration-by-parts method in Appendix~\ref{app:IBP-four-point-kernels} to simplify them.

We finally apply the same length-space algorithm to four-point functions.
The time-ordered channel is obtained from a product of two-point functions, while the crossed channel gives the OTOC.
As a check on the method, Appendix~\ref{app:N0-I3-gamma-Wilson} applies the same length-integral technology to the bosonic theory and reproduces the standard $\mathcal N=0$ JT OTOC answer \cite{Maldacena:2016upp,Stanford:2017thb,Mertens:2017mtv,Turiaci:2017zwd,Lam:2018pvp,Blommaert:2018oro}.
In the $\mathcal N=2$ zero-energy case the length integrals can be reduced by integration by parts to an explicit finite answer, and we also analyze the corresponding large-$\Delta$ approximation.
In the $\mathcal N=4$ BPS case the exact integral structure is more involved, but the same method gives a controlled large-$\Delta$ saddle approximation.

We end the paper with Section~\ref{sec:discussion-future}, where we discuss concrete protocols for two future calculations: the representation-theoretic data needed for a general closed-form super-JT OTOC, and the application to Berry curvature for D1--D5--P black hole microstates.

\setcounter{section}{1}
\section{Warm-up: Bosonic-JT from constrained particle on a group}
\label{sec:n0-jt-particle-main}

In this section we review ordinary JT gravity in the form used as the bosonic
template for the rest of the paper.  The detailed derivations are
collected in Appendix~\ref{app:n0-jt-bf-particle}; here we only keep the
ingredients needed in the main text.  
We start from the metric dilaton-gravity
description, pass to the Schwarzian boundary mode, rewrite the same boundary
problem in the $SL(2,\mathbb R)$ BF language, and finally obtain the
constrained particle on $SL(2,\mathbb R)$.  The metric and BF descriptions
give the same boundary dynamics, but they emphasize different
structures.  The metric derivation makes the boundary curve geometrical, while
the BF/group-particle description makes the $SL(2,\mathbb R)$ representation
theory and boundary Wilson lines manifest.  This also fixes the notation for
later use.

The starting point is the metric formulation of Euclidean JT gravity.  The
theory contains a two-dimensional metric $G_{\mu\nu}$ and a dilaton
$\Phi$ \cite{Jackiw:1984je,Teitelboim:1983ux}, with action
\begin{equation}
  I_{\rm JT}
  =
  -\frac12\int_M d^2X\,\sqrt G\,\Phi(R+2)
  -
  \int_{\partial M}ds\,\Phi_\partial(K-1).
  \label{eq:main-bosonic-jt-action}
\end{equation}
The dilaton equation sets the curvature to $R=-2$, and hence the bulk metric
is locally $\mathrm{AdS}_2$.  The nontrivial mode is therefore the regulated
boundary curve.  This is the standard nearly-$\mathrm{AdS}_2$ derivation of
the Schwarzian boundary action
\cite{Maldacena:2016upp,Engelsoy:2016xyb}.

The nearly-$\mathrm{AdS}_2$ boundary condition fixes the leading proper length
of the boundary together with the leading dilaton profile,
\begin{equation}
  h_{uu}\big|_{\partial M}
  =
  \left(\frac{q}{\epsilon}\right)^2,
  \qquad
  \Phi\big|_{\partial M}
  =
  \frac{\phi_r}{\epsilon},
  \qquad
  \epsilon\to0 .
  \label{eq:main-bosonic-jt-bc}
\end{equation}
Many Schwarzian conventions set the boundary einbein to one and absorb the
physical length into the range of $u$.  We instead keep the constant $q$
explicit; we will see why later.  The boundary condition fixes the
asymptotic scale, but it still allows the boundary curve to fluctuate through
the map $u\mapsto x(u)$.  We call this reparametrization $x(u)$; common
alternatives are $t(u)$ and $f(u)$.

Solving the constant-curvature constraint and evaluating the boundary
extrinsic-curvature term gives the Schwarzian action
\begin{equation}
  I_{\rm Sch}
  =
  -\phi_r
  \int du\,\{x,u\},
  \qquad
  \{x,u\}
  =
  \frac{x'''}{x'}
  -
  \frac32\left(\frac{x''}{x'}\right)^2 .
  \label{eq:main-bosonic-schwarzian-action}
\end{equation}
Thus the metric derivation describes JT by the boundary reparametrization
$x(u)$, modulo the global $SL(2,\mathbb R)$ redundancy of
$\mathrm{AdS}_2$.

One convenient way to encode the Schwarzian in the metric and bridge to the
first-order description is to use Fefferman--Graham gauge.  Writing
$ds^2=dr^2+\left(qe^r+e^{-r}\mathcal L(u)+\cdots\right)^2du^2$ and fixing
$r$ to be a large cutoff value, one can solve the boundary condition
\eqref{eq:main-bosonic-jt-bc}.  In the AdS/CFT language, the growing piece is
the source and is held fixed, while the decaying piece is the
normalizable data, which we call the vev data in this paper.  The
extrinsic-curvature condition then determines
$\mathcal L(u)=-(2q)^{-1}\{x,u\}$.  We now translate this source/vev
organization to the first-order $SL(2,\mathbb R)$ BF description.

Now we turn to the first-order formulation.  This is simply the passage to
first-order variables, ``taking the square root of the metric,'' as with the
vielbein in ordinary gravity; see Appendix~\ref{app:n0-jt-bf-particle} for
details.  A convenient
way to package the data is to introduce an $SL(2,\mathbb R)$ gauge connection
$\mathcal A$ and an adjoint scalar $\mathcal X$, whose components include
the dilaton.  This is the BF description
\cite{Mertens:2018fds,Blommaert:2018oro,Iliesiu:2019xuh,Mertens:2022irh}.
In this language the metric variables $(G_{\mu\nu},\Phi)$ are replaced by
$(\mathcal A,\mathcal X)$, and the topological nature of the bulk theory
becomes manifest.  The bulk action is
\begin{equation}
  I_{\rm BF,bulk}
  =
  -\int_M
  \left\langle
    \mathcal X,\mathcal F[\mathcal A]
  \right\rangle,
  \qquad
  \mathcal F[\mathcal A]=d\mathcal A+\mathcal A\wedge\mathcal A .
  \label{eq:main-bosonic-bf-bulk}
\end{equation}
Here $\langle\,,\,\rangle$ denotes the invariant bilinear form on
$\mathfrak{sl}(2,\mathbb R)$, with generator conventions fixed in
Appendix~\ref{app:supergroup-conventions}.  The equations of motion impose
$\mathcal F[\mathcal A]=0$, i.e. $\mathcal A$ is a flat connection, and make
$\mathcal X$ covariantly constant.  Together these equations reproduce the
metric JT equations through the first-order formalism.

The boundary condition must also be translated into BF variables.  In the BF
description, the Dirichlet boundary condition in the metric formalism becomes a
mixed boundary condition \cite{Blommaert:2018oro,Cardenas:2018krd}.  We
implement it by adding the boundary term
\begin{equation}
  I_{\rm BF}
  =
  -\int_M
  \left\langle
    \mathcal X,\mathcal F[\mathcal A]
  \right\rangle
  +
  \int_{\partial M}du
  \left[
    \left\langle
      \mathcal X,\mathcal A_u
    \right\rangle
    -
    \frac{1}{2\phi_r}
    \left\langle
      \mathcal X,\mathcal X
    \right\rangle
  \right].
  \label{eq:main-bosonic-bf-action}
\end{equation}
Varying the full action gives
\begin{equation}
  \mathcal X\big|_{\partial M}
  =
  \phi_r\,\mathcal A_u\big|_{\partial M}.
  \label{eq:main-bosonic-mixed-bc}
\end{equation}

The source/vev split described above is encoded in the BF connection: the
Fefferman--Graham gauge above translates to the following radial-stripped
boundary connection,
\begin{equation}
  \mathcal A_u
  =
  q\,\mathsf L_-
  +
  \mathcal L(u)\,\mathsf L_+ .
  \label{eq:main-bosonic-source-vev}
\end{equation}
The coefficient $\mathcal L(u)$ is the same vev mode that appeared in
the Fefferman--Graham expansion, now written as the $\mathsf L_+$ component
of the radial-stripped connection.

We now extract the boundary particle from the flat BF connection.  Since
$\mathcal F[\mathcal A]=0$, we can write
$\mathcal A=\mathcal G^{-1}d\mathcal G$, with
$\mathcal G(r,u)\in SL(2,\mathbb R)$.  Restricting to the boundary gives
$\mathcal A_u=\mathcal G^{-1}\dot{\mathcal G}$.  This
$\mathcal G^{-1}\dot{\mathcal G}$ can be interpreted as a group current, which
we denote by $\mathscr J(u)$.  The BF boundary action then gives the
first-order action for a particle on the group,
\begin{equation}
  I_{\rm pG}[\mathcal G,\mathcal P]
  =
  \phi_r
  \int du
  \left[
    \left\langle
      \mathcal P,
      \mathcal G^{-1}\dot{\mathcal G}
    \right\rangle
    -
    \frac12
    \left\langle
      \mathcal P,\mathcal P
    \right\rangle
  \right].
  \label{eq:main-bosonic-first-order-pG}
\end{equation}
Here $\mathcal P=\mathcal X/\phi_r$ is the normalized particle momentum.  If
we integrate out $\mathcal P$, we obtain the second-order form
\begin{equation}
  I_{\rm pG}[\mathcal G]
  =
  \frac{\phi_r}{2}
  \int du\,
  \left\langle
    \mathcal G^{-1}\dot{\mathcal G},
    \mathcal G^{-1}\dot{\mathcal G}
  \right\rangle .
  \label{eq:main-bosonic-second-order-pG}
\end{equation}
This is still the unconstrained particle on $SL(2,\mathbb R)$. 

Ordinary JT is obtained only after imposing the condition inherited from the fixed boundary
source.  We highlight the following simple calculation.  Using the Gauss
parameterization
$\mathcal G(u)=e^{x(u)\mathsf L_-}e^{\rho(u)\mathsf L_0}e^{\gamma(u)\mathsf L_+}$,
the corresponding current is
\begin{equation}
  \mathcal G^{-1}\dot{\mathcal G}
  =
  e^\rho\dot x\,\mathsf L_-
  +
  \left(
    \dot\rho-2\gamma e^\rho\dot x
  \right)\mathsf L_0
  +
  \left(
    \dot\gamma+\gamma\dot\rho-\gamma^2e^\rho\dot x
  \right)\mathsf L_+ .
  \label{eq:main-bosonic-gauss-current}
\end{equation}
Imposing
\begin{equation}
  \mathcal G^{-1}\dot{\mathcal G}
  =
  q\,\mathsf L_-
  +
  \mathcal L(u)\,\mathsf L_+
  \label{eq:main-bosonic-current-constraint}
\end{equation}
gives
\begin{equation}
  e^\rho=\frac{q}{\dot x},
  \qquad
  \gamma=\frac{\dot\rho}{2q}
  =
  -\frac{\ddot x}{2q\dot x},
  \qquad
  \mathcal L(u)
  =
  -\frac{1}{2q}\{x,u\}.
  \label{eq:main-bosonic-schwarzian-solution}
\end{equation}
We mention that closely related derivations of this BF and group-particle
route to the Schwarzian can be found in \cite{Alkalaev:2022qfc,Cardenas:2018krd}.
In our normalization
$\langle q\mathsf L_-+\mathcal L\mathsf L_+,q\mathsf L_-+\mathcal L\mathsf L_+\rangle=4q\mathcal L$.
Hence the quadratic particle action reduces to
\begin{equation}
  I_{\rm pG}
  =
  -\phi_r
  \int du\,\{x,u\}.
  \label{eq:main-bosonic-schwarzian-from-particle}
\end{equation}
This reproduces the Schwarzian action from the constrained group particle.  This
also gives the explicit dictionary between the boundary curve and the group particle:
the Schwarzian reparametrization $x(u)$ is the lower-triangular coordinate of
$\mathcal G(u)$, while the Schwarzian derivative is encoded in the
$\mathsf L_+$ component $\mathcal L(u)$ of the reduced current.

Having established the particle-on-a-group formalism at the action level, we
now turn to its quantization.  The problem is a quantum-mechanical system with a
constraint.  Here we only summarize the result needed later; the more detailed
Hamiltonian and path-integral derivation is in
Appendix~\ref{app:n0-jt-bf-particle}.  The source condition is imposed as a
right parabolic charge constraint,
$\langle\mathcal P,\mathsf L_+\rangle=q$.  The appearance of
$\mathsf L_+$ follows from the invariant pairing: it is the momentum
conjugate to the $\mathsf L_-$ source in
\eqref{eq:main-bosonic-source-vev}.  A first-class form of the constrained
particle is obtained by introducing a Lagrange multiplier, $\lambda(u)$,
\begin{equation}
  I[\mathcal G,\mathcal P,\lambda]
  =
  \phi_r
  \int du
  \left[
    \left\langle
      \mathcal P,
      \mathcal G^{-1}\dot{\mathcal G}
      -
      \lambda\mathsf L_+
    \right\rangle
    -
    \frac12
    \left\langle
      \mathcal P,\mathcal P
    \right\rangle
    +
    q\lambda
  \right].
  \label{eq:main-bosonic-lagrange-reduction}
\end{equation}
This action has a local right $N_+$ redundancy,
\begin{equation}
  \mathcal G(u)\to \mathcal G(u)e^{\alpha(u)\mathsf L_+},
  \qquad
  \lambda(u)\to \lambda(u)+\dot\alpha(u),
  \qquad
  N_+=\exp(\mathbb R\mathsf L_+),
  \label{eq:main-bosonic-right-redundancy}
\end{equation}
with the corresponding adjoint action on $\mathcal P$.  Thus the constrained
particle involves both a fixed parabolic charge and a quotient by the gauge
orbit generated by this constraint.  

The effect of this gauge redundancy is clearest in Hamiltonian language.
Physical wavefunctions are equivariant under the right $N_+$ action:
$\Psi_q(x,\rho,\gamma)=e^{-q\gamma}F(x,\rho)$, or infinitesimally
$L_+^R\Psi_q=-q\Psi_q$.  Acting on these wavefunctions,
the $SL(2,\mathbb R)$ Casimir becomes the reduced Hamiltonian, while the Haar
measure descends to the quotient measure,
\begin{equation}
  \begin{aligned}
    \widehat H_{{\rm JT},q}
    &=
    \frac{1}{2\phi_r}
    \left(
      \partial_\rho^2+\partial_\rho
      +
      q e^{-\rho}\partial_x
    \right),
    \qquad
    d\mu_{\rm phys}
    =
    \frac{e^\rho\,d\rho\,dx\,d\gamma}{d\gamma}
    =
    e^\rho\,dx\,d\rho .
  \end{aligned}
  \label{eq:main-bosonic-reduction-summary}
\end{equation}
The eigenfunctions of this reduced Hamiltonian give the Schwarzian propagator;
in the length variable used below they become the familiar Bessel profiles
$\Psi_s(L)\propto K_{2is}(2qL)$, up to normalization.  The measure in
\eqref{eq:main-bosonic-reduction-summary} is the one used whenever interval
wavefunctions are glued.

Let us end with one more reason for using the particle-on-a-group approach: it organizes operator insertions well by symmetry.
A matter field transforms in a representation $\mathcal R$ of the isometry group, and on a flat connection its Wilson line reduces to the endpoint matrix element $\mathcal U_\Delta(i,j)=\mathcal R_\Delta(\mathcal G_i^{-1}\mathcal G_j)$.
Choosing scalar-primary endpoint polarizations and using the source constraint $e^\rho\dot x=q$ at each endpoint gives the usual Schwarzian bilocal as
\begin{equation}
  \begin{aligned}
  \mathcal O_\Delta(i,j)
  &=
  q^{2\Delta}
  {}_\Delta\langle {\rm primary}|
  \mathcal R_\Delta(\mathcal G_i^{-1}\mathcal G_j)
  |{\rm primary}\rangle_\Delta                                      \\
  &=
  \left[
    \frac{q e^{-\frac12(\rho_i+\rho_j)}}{x_j-x_i}
  \right]^{2\Delta}
  =
  \left[
    \frac{\dot x_i\dot x_j}{(x_i-x_j)^2}
  \right]^\Delta .
  \end{aligned}
  \label{eq:main-n0-schwarzian-bilocal}
\end{equation}
Thus, already in bosonic JT, the particle-on-a-group language gives a compact way to organize correlators by reduced wavefunctions, the physical measure, and Wilson-line bilocals.
Appendix~\ref{app:N0-I3-gamma-Wilson} uses precisely these ingredients to reproduce the standard OTOC kernel in length space.

\section{Super-JT from the particle on a group}
\label{sec:superjt-boundary-particle}
Section~\ref{sec:n0-jt-particle-main} described bosonic JT as a constrained particle on $SL(2,\mathbb R)$.  
We now apply the same logic for super-JT.
We begin in
Section~\ref{subsec:boundary-data-super-fg} by setting up notation and
explaining the bridge from dilaton supergravity to first-order BF theory,
which is the natural bridge to the particle on a group.  The
standard reference for this (super)gravity-to-BF map in the
$\mathcal N=4$ case is \cite{Heydeman:2020hhw}.\footnote{In
\cite{Heydeman:2020hhw} the
boundary super-reparametrization is parametrized by $f(\tau)$,
$g(\tau)\in SU(2)$, and fermions $\eta^p(\tau),\bar\eta_p(\tau)$.  These
become our particle variables $x(u)$, $U(u)\in SU(2)_{\mathfrak R}$, and
$\theta_-^p(u),\bar\theta_{-,p}(u)$.  Their component super-Schwarzian
multiplet $(S_b,S_f^p,\bar S_{f,p},S_T^i)$, or equivalently the BF boundary
fields $(\tilde{L},\psi^p,\bar\psi_p,B_\tau^i)$, is represented here by the reduced
current components $(\mathcal L,\mathcal S^p,\bar{\mathcal S}_p,\mathcal R^i)$.}
We follow the same source/vev organization, but rewrite it in a form adapted to
the particle on a group.
Section~\ref{subsec:n2-super-schwarzian-particle}
then gives the main classical check of the particle on a supergroup formalism.  After
imposing the constraints from super-JT boundary conditions, we show that the particle action reduces to the
$\mathcal N=2$ or $\mathcal N=4$ super-Schwarzian action, meaning that the
constrained particle is a group-theoretic
manifestation of super-Schwarzian dynamics.
Finally, Section~\ref{subsec:susy-hamiltonian-reduction} addresses the
quantization of this constrained system.  The fixed parabolic charge condition
is standard, while the new fermionic constraints require a more careful treatment
because they become second class after the parabolic charge is fixed.  This is
the natural origin of the auxiliary fermions in the LMRS construction
\cite{Lin:2022zxd}, and it also fixes the reduced Hilbert space and the
path-integral or wavefunction gluing measure used in later sections.

\subsection{Boundary data and radial weights}
\label{subsec:boundary-data-super-fg}

As in Section~\ref{sec:n0-jt-particle-main}, we treat super-JT gravity in the usual
AdS/CFT way: the gravitational theory is a two-dimensional dilaton gravity with
a prescribed boundary condition, and that boundary condition is most transparent
after choosing a holographic radial gauge near infinity.  We then explain the
passage to first-order variables in a little more detail, introducing the
frame fields and packaging the supergravity fields into a gauge theory
connection.  This gives a direct route from (super)BF theory to the particle on
a group.

\paragraph{Coordinates and fields.}
We use the superspace coordinates $\mathbf Z^M=(X^\mu=(r,u);\Theta^{p\alpha},\bar\Theta_p{}^\alpha)$ and $z^m=(u,\vartheta^p,\bar\vartheta_p)$.
Here $\mathbf Z^M$ are bulk supercoordinates and $z^m$ are the boundary superline coordinates.
In the main text we will only use component fields; the flat superline basis and the corresponding derivatives are recorded in Appendix~\ref{app:superspace-notation}.\footnote{In particular, Appendix~\ref{app:superspace-notation} gives $\Delta u$, $d\vartheta^p$, $d\bar\vartheta_p$, and the derivatives $D_p,\bar D^p$.}
The index $p$ has one value for $\mathcal N=2$ and two values for $\mathcal N=4$.
We write the R-symmetry group as $\mathfrak R=U(1)_{\mathfrak R}$ for $\mathcal N=2$, and $\mathfrak R=SU(2)_{\mathfrak R}$ for $\mathcal N=4$.
The index $I$ denotes an $\mathfrak R$-algebra index: $I=\mathfrak R$ for $U(1)_{\mathfrak R}$, and $I=i=1,2,3$ for $SU(2)_{\mathfrak R}$.
The component supergravity fields are
\begin{equation}
  E_\mu{}^a,\qquad
  \omega_\mu,\qquad
  A_\mu{}^I,\qquad
  \Psi_\mu^p,\qquad
  \bar\Psi_{\mu,p}.
  \label{eq:component-supergravity-fields}
\end{equation}
They are the frame, spin connection, R-symmetry gauge field, and gravitini.

\paragraph{Asymptotic modes and boundary conditions in radial gauge.}
As in the bosonic discussion, radial weights distinguish fixed sources from vev
data.  We now apply this separation to the frame, gravitini, and
$\mathfrak R$-symmetry gauge field.  The growing modes are held fixed at the
boundary, while the decaying modes are the vev data\footnote{We use ``vev'' as
shorthand for vacuum expectation value, or more generally the holographic
one-point-function data conjugate to a source.} integrated over in the
gravitational path integral.

It is convenient to make this separation in radial gauge.
We impose radial gauge on the component fields,
\begin{equation}
  E^{\hat r}=dr,
  \qquad
  E_u{}^{\hat r}=0,
  \qquad
  \Psi_r^p=\bar\Psi_{r,p}=0,
  \qquad
  A_r{}^I=0 .
  \label{eq:sugra-radial-gauge-components}
\end{equation}
The boundary data are carried by the transverse components
\begin{equation}
  E_u{}^{\hat u},\qquad
  \Psi_u^p,\qquad
  \bar\Psi_{u,p},\qquad
  A_u{}^I .
\end{equation}

We now solve the radial equations mode by mode.
For the frame field radial gauge leaves
\begin{equation}
  E^{\hat u}=E_u{}^{\hat u}(r,u)\,du,
  \qquad
  ds^2=dr^2+\left(E_u{}^{\hat u}(r,u)\right)^2du^2 .
  \label{eq:fg-frame-and-metric}
\end{equation}
The asymptotic solution is
\begin{equation}
  E_u{}^{\hat u}(r,u)
  =
  e^r e_u^{(0)}(u)
  +e^{-r}\mathcal L(u)
  +\cdots .
  \label{eq:transverse-frame-eigenmodes}
\end{equation}
The leading coefficient $e_u^{(0)}$ is the metric source, while $\mathcal L$ is the stress-tensor vev.
The spin connection follows from the torsion constraint,
\begin{equation}
  \omega^{\hat r\hat u}
  =
  \partial_rE_u{}^{\hat u}(r,u)\,du
  =
  \left(e^r e_u^{(0)}-e^{-r}\mathcal L+\cdots\right)du .
  \label{eq:transverse-spin-connection}
\end{equation}

The gravitino equation is the fermionic curvature equation.
In radial gauge the leading radial part and its two eigensolutions are
\begin{equation}
\begin{aligned}
  0
  &=
  \left(\partial_r+\frac12\gamma_{\hat r}\right)\Psi_u^p+\cdots,
  \\
  \frac12(1-\gamma_{\hat r})\Psi_u^p
  &=
  e^{r/2}v_u^{(0)p}
  +\cdots,
  \qquad
  \frac12(1+\gamma_{\hat r})\Psi_u^p
  &=
  e^{-r/2}\mathcal S^p
  +\cdots,
  \\
  0
  &=
  \left(\partial_r+\frac12\gamma_{\hat r}\right)\bar\Psi_{u,p}+\cdots,
  \\
  \frac12(1-\gamma_{\hat r})\bar\Psi_{u,p}
  &=
  e^{r/2}\bar v_{u,p}^{(0)}
  +\cdots,
  \qquad
  \frac12(1+\gamma_{\hat r})\bar\Psi_{u,p}
  &=
  e^{-r/2}\bar{\mathcal S}_p
  +\cdots .
\end{aligned}
  \label{eq:transverse-gravitino-eigenmodes}
\end{equation}
The $e^{r/2}$ coefficients $v_u^{(0)p},\bar v_{u,p}^{(0)}$ are the gravitino sources.
The $e^{-r/2}$ coefficients $\mathcal S^p,\bar{\mathcal S}_p$ are the supercurrent vevs.

The R-symmetry gauge field equation is the $\mathfrak R$-curvature equation.
Its radial component is
\begin{equation}
  F_{\mathfrak R,ru}{}^{I}
  =
  \partial_r A_u{}^I
  -\partial_u A_r{}^I
  +f^I{}_{JK}A_r{}^J A_u{}^K
  =
  0 .
  \label{eq:r-symmetry-radial-eom}
\end{equation}
Using $A_r{}^I=0$, this becomes
\begin{equation}
  \partial_r A_u{}^I=0 .
  \label{eq:r-symmetry-radial-eom-gauge-fixed}
\end{equation}
Thus the R-symmetry gauge field has zero radial weight.
Since it has zero weight, the radial equation does not by itself split $A_u{}^I$ into a growing source mode and a decaying vev mode.
The boundary condition depends on the ensemble: one may fix the boundary holonomy, or one may fix the conjugate $\mathfrak R$-flux/charge.
In the rest of the paper we use the fixed $\mathfrak R$-flux, equivalently fixed $\mathfrak R$-charge, ensemble.
The $SU(2)_{\mathfrak R}$ group element $U(u)$ is therefore retained as a particle variable in the $\mathcal N=4$ theory, and its conjugate current $\mathcal R^i(u)$ appears as part of the super-Schwarzian vev multiplet.

The Dirichlet boundary condition fixes the source coefficients and lets the vev coefficients fluctuate.
For the zero-weight R-symmetry field this statement is understood with the fixed-flux ensemble just specified.
The bosonic metric source and vanishing gravitino source condition is
\begin{equation}
  e_u^{(0)}=\mathfrak q,
  \qquad
  v_u^{(0)p}=\bar v_{u,p}^{(0)}=0 .
  \label{eq:bosonic-source-ensemble}
\end{equation}
This condition does not constrain the vev data $\mathcal L,\mathcal S^p,\bar{\mathcal S}_p$, which are allowed to fluctuate.
In the supersymmetric discussion we denote this fixed boundary source by
$\mathfrak q$, following the LMRS notation \cite{Lin:2022zxd}.

\paragraph{First-order form as a bridge.}
We now package the same boundary data into the super-BF variables.  This is the
supergravity analogue of the first-order map reviewed in
Section~\ref{sec:n0-jt-particle-main}: the frame, spin connection,
$\mathfrak R$-symmetry gauge field, and gravitini become components of a
single superconnection $\mathcal A$, while the dilaton multiplet becomes the
adjoint scalar $\mathcal X$.  We use $\mathsf L_\pm,\mathsf L_0$ for the
$SL(2)$ generators, $\mathsf R_I$ for the $\mathfrak R$-symmetry
generators, and $\mathsf F_{\pm,p},\bar{\mathsf F}_{\pm}^{p}$ for the
fermionic generators.  With the radial weights used below, the
$\mathsf F_{-,p},\bar{\mathsf F}_{-}^{p}$ coefficients are growing source
modes, while the $\mathsf F_{+,p},\bar{\mathsf F}_{+}^{p}$ coefficients are
decaying vev modes; see Appendix~\ref{app:supergroup-conventions} for the
algebra and pairing.
We write this connection schematically as
\begin{equation}
  \mathcal A
  =
  \mathcal A^-\,\mathsf L_-
  +\mathcal A^0\,\mathsf L_0
  +\mathcal A^+\,\mathsf L_+
  +\mathsf A^I\,\mathsf R_I
  +\Psi_-^p\,\mathsf F_{-,p}
  +\bar\Psi_{-,p}\,\bar{\mathsf F}_{-}^{p}
  +\Psi_+^p\,\mathsf F_{+,p}
  +\bar\Psi_{+,p}\,\bar{\mathsf F}_{+}^{p}.
  \label{eq:first-order-connection-package}
\end{equation}
The first-order equations of motion are the BF equations
\begin{equation}
  \mathcal F[\mathcal A]
  \equiv
  d\mathcal A+\mathcal A\wedge\mathcal A
  =
  0,
  \qquad
  D_{\mathcal A}\mathcal X
  \equiv
  d\mathcal X+[\mathcal A,\mathcal X]
  =
  0 .
  \label{eq:first-order-bf-equations}
\end{equation}
The flatness equation packages the component equations for the frame, spin
connection, $\mathfrak R$-symmetry gauge field, and gravitini; the second
equation is the dilaton-multiplet equation.
The first-order radial gauge is
\begin{equation}
  \mathcal A_r=\mathsf L_0 .
  \label{eq:first-order-radial-gauge}
\end{equation}
The radial part of the flatness equation is
\begin{equation}
  \partial_r\mathcal A_u+[\mathsf L_0,\mathcal A_u]=0 .
  \label{eq:first-order-radial-equation}
\end{equation}
If $[\mathsf L_0,\mathsf T_w]=w\,\mathsf T_w$, the coefficient of $\mathsf T_w$ is $e^{-wr}$ times a boundary field.
Therefore
\begin{align}
  \mathcal A(r,u)
  &=
  \mathsf L_0\,dr \nonumber
  \\
  & +du\Big[
    e^r e_u^{(0)}\,\mathsf L_-
    +e^{r/2}v_u^{(0)p}\,\mathsf F_{-,p}
    +e^{r/2}\bar v_{u,p}^{(0)}\,\bar{\mathsf F}_{-}^{p}
    +\mathsf A_u{}^I\,\mathsf R_I
    +e^{-r/2}\mathcal S^p\,\mathsf F_{+,p}
    +e^{-r/2}\bar{\mathcal S}_p\,\bar{\mathsf F}_{+}^{p}
    +e^{-r}\mathcal L\,\mathsf L_+
  \Big] .
  \label{eq:full-radial-boundary-connection}
\end{align}
The minus-components are the sources and the plus-components are the vev data.
The zero-weight coefficient $\mathsf A_u{}^I$ is treated according to the chosen $\mathfrak R$-ensemble; in a fixed-flux ensemble its conjugate charge is the R-current $\mathcal R^I$.
For $\mathcal N=4$, the vev fields $(\mathcal L,\mathcal S^p,\bar{\mathcal S}_p,\mathcal R^i)$ form the component super-Schwarzian multiplet, and the $\mathcal N=2$ case is the $U(1)_{\mathfrak R}$ truncation \cite{Heydeman:2020hhw}.

The adjoint scalar $\mathcal X$ is the first-order dilaton multiplet.  Besides
the ordinary JT dilaton and auxiliary torsion fields, it contains the dilatini
and the $\mathfrak R$-symmetry multipliers.  Its radial equation gives
\begin{equation}
  \mathcal X(r,u)
  =
  e^r X_-\,\mathsf L_-
  +e^{r/2}\zeta_-^p\,\mathsf F_{-,p}
  +e^{r/2}\bar\zeta_{-,p}\,\bar{\mathsf F}_{-}^{p}
  +X_{\mathcal R}{}^I\,\mathsf R_I
  +e^{-r/2}\zeta_+^p\,\mathsf F_{+,p}
  +e^{-r/2}\bar\zeta_{+,p}\,\bar{\mathsf F}_{+}^{p}
  +e^{-r}X_+\,\mathsf L_+ .
  \label{eq:full-radial-dilaton-multiplet}
\end{equation}
The minus-components are fixed by the renormalized dilaton data, while the
plus-components are conjugate vev data.  We only use the radial part of
$D_{\mathcal A}\mathcal X=0$ here; the transverse equation will reappear
below.

The BF action with the JT boundary term is
\begin{equation}
  I[\mathcal A,\mathcal X]
  =
  -\int_M
  \left\langle
    \mathcal X,\mathcal F[\mathcal A]
  \right\rangle
  +
  \int_{\partial M}du\,
  \left[
    \left\langle
      \mathcal X,\mathcal A_u
    \right\rangle
    -
    \frac{1}{2\phi_r}
    \left\langle
      \mathcal X,\mathcal X
    \right\rangle
	  \right].
	  \label{eq:super-bf-boundary-action}
	\end{equation}
Here $\langle\,,\,\rangle$ is the invariant supertrace pairing, with the gravity normalization fixed in Appendix~\ref{app:supergroup-conventions}.
The boundary term imposes the JT mixed boundary condition
\begin{equation}
  \left.\mathcal X\right|_{\partial M}
  =
  \phi_r\left.\mathcal A_u\right|_{\partial M},
  \qquad
  \left.\delta\mathcal X\right|_{\partial M}
  =
  \phi_r\left.\delta\mathcal A_u\right|_{\partial M},
\end{equation}
as reviewed in Section~\ref{sec:n0-jt-particle-main}, with details in
Appendix~\ref{subsec:n0-bf-particle}.
For related references on this BF/boundary-particle viewpoint in the bosonic and $\mathcal N=4$ cases, see also \cite{Blommaert:2018oro,Valach:2019jzv,Heydeman:2020hhw}.

For the bosonic metric source and vanishing gravitino sources, the boundary
value of the connection is
\begin{equation}
  \left.\mathcal A_u\right|_{\partial M}
  =
  e^r\mathfrak q\,\mathsf L_-
  +0\times\mathsf F_{-,p}
  +0\times\bar{\mathsf F}_{-}^{p}
  +0\times\mathsf L_0
  +\mathsf A_u{}^I\,\mathsf R_I
  +e^{-r/2}\mathcal S^p\,\mathsf F_{+,p}
  +e^{-r/2}\bar{\mathcal S}_p\,\bar{\mathsf F}_{+}^{p}
  +e^{-r}\mathcal L\,\mathsf L_+ .
  \label{eq:bosonic-source-full-radial-connection}
\end{equation}
The displayed zero components, together with the fixed $\mathsf L_-$ source,
will be important later: after stripping the radial weights, they become the
constraint equations in \eqref{eq:n2-super-source-conditions}.  The
plus-components $\mathcal L,\mathcal S^p,\bar{\mathcal S}_p$ are still
dynamical vev data.\footnote{If one later imposes the remaining transverse
equation of motion, namely the $u$-component of
$D_{\mathcal A}\mathcal X=0$, this gives $\partial_u\mathcal L=0$,
$\partial_u\mathcal S^p=0$, and
$\partial_u\bar{\mathcal S}_p=0$.  Equivalently, these constant vev data solve
the Schwarzian equations of motion.}

Flatness, $\mathcal F[\mathcal A]=0$, lets us write the connection locally as
$\mathcal A=\mathcal G^{-1}d\mathcal G$, where $\mathcal G(r,u)$ is a
group-valued field on AdS$_2$.
After stripping the radial weights, the boundary condition on
$\mathcal A_u$ becomes a condition on the boundary current
$\mathcal{G}^{-1}(u)\dot{\mathcal{G}}(u)$, with
$\mathcal{G}(u)=\left.\mathcal G(r,u)\right|_{\partial M}$.
The detailed boundary reduction is reviewed in Section~\ref{sec:n0-jt-particle-main}
and Appendix~\ref{subsec:n0-bf-particle}.

\subsection{\texorpdfstring{Super-Schwarzian from the particle on a group}{Super-Schwarzian from the particle on a group}}
\label{subsec:n2-super-schwarzian-particle}

We now perform the supersymmetric analogue of the current calculation in
Section~\ref{sec:n0-jt-particle-main}.  The goal is to show that imposing the
super-JT source constraints on the group particle produces the
super-Schwarzian multiplet.  For the explicit calculation we use the
$\mathcal N=2$ theory with $U(1)_{\mathfrak R}$ generator
$\mathsf J_{\mathfrak R}$, and we comment on the $\mathcal N=4$ case at the
end.  The particle-on-a-group action follows from the BF boundary action
\eqref{eq:super-bf-boundary-action} by replacing the boundary gauge connection
$\mathcal A_u$ with $\mathcal G^{-1}(u)\dot{\mathcal G}(u)$, with normalized
particle momentum $\mathcal P=\mathcal X/\phi_r$.  In this language the same
$\mathfrak q$ is the fixed parabolic charge of the group particle, identified
with the boundary source above.  The first-order particle action is
\begin{equation}
  I[\mathcal G,\mathcal P]
  =
  \phi_r\int du\,
  \left[
    \left\langle
      \mathcal P,\mathcal G^{-1}\dot{\mathcal G}
    \right\rangle
    -
    \frac{1}{2}
    \left\langle
      \mathcal P,\mathcal P
    \right\rangle
  \right].
  \label{eq:n2-first-order-superparticle-action}
\end{equation}
One can integrate out $\mathcal P$ to get
\begin{equation}
  I[\mathcal G]
  =
  \frac{\phi_r}{2}
  \int du\,
  \left\langle
    \mathcal G^{-1}\dot{\mathcal G},
    \mathcal G^{-1}\dot{\mathcal G}
  \right\rangle .
  \label{eq:n2-second-order-superparticle-action}
\end{equation}

\paragraph{Variables and constraints.}
We use the ordered parametrization
\begin{equation}
  \mathcal G
  =
  e^{x\mathsf L_-}
  e^{\theta_-\mathsf F_-+\bar\theta_-\bar{\mathsf F}_-}
  e^{\rho\mathsf L_0}
  e^{a\mathsf J_{\mathfrak R}}
  e^{\theta_+\mathsf F_++\bar\theta_+\bar{\mathsf F}_+}
  e^{\gamma\mathsf L_+}.
  \label{eq:n2-super-gauss-param}
\end{equation}
The variables $x,\theta_-,\bar\theta_-,a$ are the surviving boundary variables after the source conditions are imposed.
The variables $\rho,\theta_+,\bar\theta_+,\gamma$ are solved for by the minus-component constraints.
The $U(1)_{\mathfrak R}$ variable $a(u)$ is kept because the zero-weight R-symmetry boundary condition is an ensemble choice.
We use $a$ as a Lie-algebra coordinate; the unitary convention $e^{iaJ}$ is recovered by $a\to ia$.

Define
\begin{equation}
  \mathscr J=\mathcal G^{-1}\dot{\mathcal G}.
  \label{eq:n2-boundary-current-definition}
\end{equation}
and we denote by $\mathscr J_{\mathsf T}$ its coefficient along the generator $\mathsf T$.
The component which replaces the bosonic $\dot x$ is
\begin{equation}
  \Pi
  =
  \dot x+\bar\theta_-\dot\theta_-+\theta_-\dot{\bar\theta}_-.
  \label{eq:n2-pi-definition}
\end{equation}
This is the component form of the invariant superline one-form.
A direct multiplication gives the components of $\mathscr J$ needed for the source constraint:
\begin{align}
  \mathscr J_{\mathsf L_-}
  &=
  e^\rho\Pi,
  \nonumber\\
  \mathscr J_{\mathsf F_-}
  &=
  e^{a+\rho/2}\dot\theta_-
  +e^\rho\Pi\,\theta_+
  +O(\theta_-^3),
  \nonumber\\
  \mathscr J_{\bar{\mathsf F}_-}
  &=
  e^{-a+\rho/2}\dot{\bar\theta}_-
  -e^\rho\Pi\,\bar\theta_+
  +O(\theta_-^3),
  \nonumber\\
  \mathscr J_{\mathsf L_0}
  &=
  \dot\rho
  -2\gamma e^\rho\Pi
  +O(\theta_-^4).
  \label{eq:n2-minus-current-components}
\end{align}
The absence of a quadratic correction in $\mathscr J_{\mathsf L_0}$ gives a non-trivial check: the two terms which would come from solving $\mathscr J_{\mathsf F_-}=0$ and $\mathscr J_{\bar{\mathsf F}_-}=0$ cancel.

We now impose the source conditions obtained by stripping the radial weights from \eqref{eq:bosonic-source-full-radial-connection},
\begin{equation}
  \mathscr J_{\mathsf L_-}=\mathfrak q,
  \qquad
  \mathscr J_{\mathsf F_-}=0,
  \qquad
  \mathscr J_{\bar{\mathsf F}_-}=0,
  \qquad
  \mathscr J_{\mathsf L_0}=0 .
  \label{eq:n2-super-source-conditions}
\end{equation}
These equations fix the metric source, turn off the gravitino sources, and choose the simple $L_0$ representative.
They do not set the plus-component vev data to zero.
Solving them gives
\begin{equation}
  e^\rho=\frac{\mathfrak q}{\Pi},
  \qquad
  \theta_+=-\frac{e^{a+\rho/2}}{\mathfrak q}\dot\theta_-+O(\theta_-^3),
  \qquad
  \bar\theta_+=\frac{e^{-a+\rho/2}}{\mathfrak q}\dot{\bar\theta}_-+O(\theta_-^3),
  \qquad
  \gamma=\frac{\dot\rho}{2\mathfrak q}+O(\theta_-^4).
  \label{eq:n2-auxiliary-solution}
\end{equation}
This is the supersymmetric version of the bosonic solution
\eqref{eq:main-bosonic-schwarzian-solution}, with $\dot x$ replaced by
$\Pi$.

\paragraph{Reduced current and super-Schwarzian.}
After the auxiliary variables are eliminated, the reduced current has the same form as in Appendix~\ref{app:superspace-notation},
\begin{equation}
  \mathscr J_{\rm red}^{(2)}
  =
  \mathfrak q\,\mathsf L_-
  +\mathcal R\,\mathsf J_{\mathfrak R}
  +\mathcal S\,\mathsf F_+
  +\bar{\mathcal S}\,\bar{\mathsf F}_+
  +\mathcal L\,\mathsf L_+ .
  \label{eq:n2-reduced-current}
\end{equation}
To quadratic order in the boundary fermions,
\begin{equation}
  \mathcal R
  =
  \dot a
  -
  \frac{\dot\theta_-\dot{\bar\theta}_-}{\Pi}
  +O(\theta_-^4),
  \label{eq:n2-r-current-component}
\end{equation}
and the plus fermionic components begin as
\begin{align}
  \mathcal S
  &=
  -\frac{e^{a+\rho/2}}{\mathfrak q}
  \left(
    \ddot\theta_- -\frac{\dot\Pi}{\Pi}\dot\theta_-
  \right)
  +O(\theta_-^3),
  \nonumber\\
  \bar{\mathcal S}
  &=
  \frac{e^{-a+\rho/2}}{\mathfrak q}
  \left(
    \ddot{\bar\theta}_- -\frac{\dot\Pi}{\Pi}\dot{\bar\theta}_-
  \right)
  +O(\theta_-^3).
  \label{eq:n2-supercurrent-components}
\end{align}
The stress-tensor component is
\begin{equation}
  \mathcal L
  =
  -\frac{1}{2\mathfrak q}
  \left[
    \frac{\ddot\Pi}{\Pi}
    -
    \frac{3}{2}
    \left(\frac{\dot\Pi}{\Pi}\right)^2
  \right]
  +
  \frac{1}{\mathfrak q\Pi}
  \left(
    \ddot\theta_-\dot{\bar\theta}_-
    -
    \dot\theta_-\ddot{\bar\theta}_-
  \right)
  +O(\theta_-^4).
  \label{eq:n2-stress-component}
\end{equation}
These vev fields are the plus-components of the same reduced current after the source components have been fixed.

Substituting \eqref{eq:n2-reduced-current} into the quadratic action gives
\begin{equation}
  \frac{1}{2}\left\langle \mathscr J_{\rm red}^{(2)},\mathscr J_{\rm red}^{(2)}\right\rangle
  =
  2\mathfrak q\,\mathcal L
  -2\mathcal R^2
  +2\mathcal S\bar{\mathcal S}\text{ terms}.
  \label{eq:n2-reduced-current-norm-schematic}
\end{equation}
The resulting particle action has the form
\begin{equation}
  I_{\rm red}
  =
  2\phi_r \mathfrak q\int du\,\mathcal L_{\mathcal N=2},
  \label{eq:n2-super-schwarzian-action-form}
\end{equation}
up to total derivatives and the normalization convention for the $U(1)_{\mathfrak R}$ generator.
Equations \eqref{eq:n2-reduced-current-norm-schematic} and \eqref{eq:n2-super-schwarzian-action-form} are therefore the component form of the $\mathcal N=2$ super-Schwarzian action written in superspace in Appendix~\ref{app:n2-superspace-super-schwarzian}; see also \cite{Fu:2016vas,Heydeman:2020hhw}.

\paragraph{\texorpdfstring{$\mathcal N=4$}{N=4} case.}
For $\mathcal N=4$, we perform the same calculation in $PSU(1,1|2)$.
The single minus pair $\theta_-,\bar\theta_-$ becomes
$\theta_-^p,\bar\theta_{-,p}$, and the $U(1)_{\mathfrak R}$ factor
$e^{a\mathsf J_{\mathfrak R}}$ is replaced by $U(u)\in SU(2)_{\mathfrak R}$.
The invariant superline one-form becomes
\begin{equation}
  \Pi
  =
  \dot x
  +\bar\theta_{-,p}\dot\theta_-^p
  +\theta_-^p\dot{\bar\theta}_{-,p}.
  \label{eq:n4-pi-definition-main-text}
\end{equation}
Solving the $\mathcal N=4$ source constraints gives
\begin{equation}
  e^\rho=\frac{\mathfrak q}{\Pi},
  \qquad
  \theta_+^p=-\frac{e^{\rho/2}}{\mathfrak q}(U^{-1})^p{}_{q}\dot\theta_-^q+O(\theta^3),
  \qquad
  \bar\theta_{+,p}=\frac{e^{\rho/2}}{\mathfrak q}\dot{\bar\theta}_{-,q}U^q{}_{p}+O(\theta^3),
  \qquad
  \gamma=\frac{\dot\rho}{2\mathfrak q}+O(\theta^4).
  \label{eq:n4-auxiliary-solution-main-text}
\end{equation}
The reduced current is most cleanly written in terms of the group-current
coefficients
$(\mathcal R^i,\mathcal S^p,\bar{\mathcal S}_p,\mathcal L)$.
\begin{equation}
  \mathscr J_{\rm red}^{(4)}
  =
  \mathfrak q\,\mathsf L_-
  +\mathcal R^i\,\mathsf R_i
  +\mathcal S^p\,\mathsf F_{+,p}
  +\bar{\mathcal S}_p\,\bar{\mathsf F}_{+}^{p}
  +\mathcal L\,\mathsf L_+ .
  \label{eq:n4-reduced-current-main-text}
\end{equation}
The analog of \eqref{eq:n2-r-current-component}--\eqref{eq:n2-stress-component}
is
\begin{align}
  \mathcal R^i
  &=
  \frac{i}{2}\operatorname{Tr}\!\left(\sigma^iU^{-1}\dot U\right)
  -
  \frac{i}{2\Pi}
  (\sigma^i)^q{}_{p}
  (U^{-1})^p{}_{r}\dot\theta_-^r\,
  \dot{\bar\theta}_{-,s}U^s{}_{q}
  +O(\theta^4),
  \nonumber\\
  \mathcal S^p
  &=
  -\frac{e^{\rho/2}}{\mathfrak q}
  \left[
    (U^{-1}\ddot\theta_-)^p
    -
    \frac{\dot\Pi}{\Pi}(U^{-1}\dot\theta_-)^p
  \right]
  +O(\theta^3),
  \nonumber\\
  \bar{\mathcal S}_p
  &=
  \frac{e^{\rho/2}}{\mathfrak q}
  \left[
    (\ddot{\bar\theta}_-U)_p
    -
    \frac{\dot\Pi}{\Pi}(\dot{\bar\theta}_-U)_p
  \right]
  +O(\theta^3),
  \nonumber\\
  \mathcal L
  &=
  -\frac{1}{2\mathfrak q}
  \left[
    \frac{\ddot\Pi}{\Pi}
    -
    \frac32\left(\frac{\dot\Pi}{\Pi}\right)^2
  \right]
  +
  \frac{1}{\mathfrak q\Pi}
  \left[
    (U^{-1}\ddot\theta_-)^p(\dot{\bar\theta}_-U)_p
    -
    (U^{-1}\dot\theta_-)^p(\ddot{\bar\theta}_-U)_p
  \right]
  +O(\theta^4).
  \label{eq:n4-raw-current-components-main}
\end{align}
These are the direct coefficients of $\mathcal G^{-1}\dot{\mathcal G}$.
The reduced action is
\begin{equation}
  \frac{1}{2}
  \left\langle
    \mathscr J_{\rm red}^{(4)},
    \mathscr J_{\rm red}^{(4)}
  \right\rangle
  =
  2\mathfrak q\,\mathcal L
  -\mathcal R^i\mathcal R^i
  +\mathcal S^p\bar{\mathcal S}_p\text{ terms}.
  \label{eq:n4-reduced-action-main-text}
\end{equation}
Using the pairings in \eqref{eq:n4-pairings}, this gives the reduced action
$I_{\rm red}^{(4)}=2\phi_r\mathfrak q\int du\,\mathcal L_{\mathcal N=4}$,
where $\mathcal L_{\mathcal N=4}$ is the component Lagrangian.
Thus the particle-on-a-group calculation reproduces the $\mathcal N=4$
super-Schwarzian action \cite{Matsuda:1989kp,Aoyama:2018lfc,Heydeman:2020hhw}.

\subsection{Hamiltonian reduction and the fermionic polarization}
\label{subsec:susy-hamiltonian-reduction}

As explained in Section~\ref{sec:n0-jt-particle-main}, 
to quantize the constrained particle on a group, we not only need to impose the constraint equation but also need to quotient out a (right) subgroup $N_+$.
In super-JT we impose the same bosonic charge together with the positive fermionic charges.
The new point is that, after $L_+^R=-\mathfrak q$ is fixed, the positive fermionic
constraints satisfy some non-trivial algebra..
We will show how to deal with this and how the LMRS auxiliary fermions $\chi^p$ appears.

\paragraph{How should the fermionic constraints be imposed?}
The super-FG boundary condition fixes the minus-component sources in \eqref{eq:n2-super-source-conditions}.
Using the invariant pairing, this becomes a fixed right-charge condition for the generators $\mathsf L_+$, $\mathsf F_+$, and $\bar{\mathsf F}_+$:
\begin{equation}
  \mathfrak n_+^{\rm susy}
  =
  \operatorname{span}
  \left\{
    \mathsf L_+,\,
    \mathsf F_{+,p},\,
    \bar{\mathsf F}_{+}^{p}
  \right\},
  \qquad p=1,2 ,
  \label{eq:positive-super-parabolic}
\end{equation}
with one value of $p$ in the $\mathcal N=2$ truncation.
We impose
\begin{equation}
  \mu_{\mathsf L_+}^R=-\mathfrak q,
  \qquad
  \mu_{\mathsf F_{+,p}}^R=0,
  \qquad
  \mu_{\bar{\mathsf F}_{+}^{p}}^R=0,
  \label{eq:fixed-positive-generator-value}
\end{equation}
or, quantum mechanically, $L_+^R\Psi=-\mathfrak q\Psi$.
The bosonic equation fixes the dependence of the wavefunction along the right $N_+$ coordinate.
The fermionic equations require more care, because after the $\mathsf L_+$ charge is fixed the $\mathsf F_+$ and $\bar{\mathsf F}_+$ constraints no longer close into a quantity that vanishes on the constrained surface.
Indeed,
\begin{equation}
  \left\{
    \mathsf F_{+,p}^R,
    \bar{\mathsf F}_{+}^{R,q}
  \right\}=
  \delta_p{}^q\,\mathsf L_+^R .
  \label{eq:positive-fermion-right-algebra}
\end{equation}
If we tried to impose both fermionic constraints directly as scalar annihilation conditions,
\begin{equation}
  \mathsf F_{+,p}^R\Psi=0,
  \qquad
  \bar{\mathsf F}_{+}^{R,q}\Psi=0,
\end{equation}
then their anticommutator would give
\begin{equation}
  0=
  \left\{
    \mathsf F_{+,p}^R,
    \bar{\mathsf F}_{+}^{R,q}
  \right\}\Psi =
  -\mathfrak q\,\delta_p{}^q\Psi .
\end{equation}
Thus the obstruction is precise: the bracket of two constraints which are both set to zero becomes the nonzero number $-\mathfrak q$.

This is the standard distinction between first-class and second-class constraints.
For a first-class set, the brackets close into the constraints themselves, and therefore vanish after the constraints are imposed.
Here the $\mathsf F_+,\bar{\mathsf F}_+$ pair is second class once the $\mathsf L_+$ charge is fixed.
The fixed-value equations are still legitimate equations on phase space, but they should not be imposed simultaneously as annihilation equations on an ordinary scalar wavefunction.
One can either solve the second-class system with the Dirac bracket, or enlarge the phase space so that an equivalent first-class system is obtained before quantization.

We first illustrate the two equivalent treatments in a minimal model.
Consider one pair of independent Grassmann variables $\theta,\bar\theta$, with conjugate momenta $\pi_\theta,\pi_{\bar\theta}$, and impose the two constraints 
\begin{equation}
  C=\pi_\theta+\mathfrak q\bar\theta,
  \qquad
  \bar C=\pi_{\bar\theta}+\mathfrak q\theta .
\end{equation}
With the standard brackets one obtains $\{C,\bar C\}\sim 2\mathfrak q$.
For $\mathfrak q\neq0$, $C$ and $\bar C$ are therefore a second-class pair.

The direct treatment is to use the Dirac bracket.
For any second-class constraint set $\phi_A=0$, define the Dirac bracket by 
\begin{equation}
  \{f,g\}_D
  =
  \{f,g\}
  -
  \{f,\phi_A\}
  (\Delta^{-1})^{AB}
  \{\phi_B,g\},
  \label{eq:dirac-bracket-definition}
\end{equation}
if the constraint-bracket matrix $\Delta_{AB}=\{\phi_A,\phi_B\}$
is invertible on the constraint surface.
It is constructed so that, for all functions $f$,
\begin{equation}
  \{f,\phi_A\}_D=0 .
\end{equation}
In the present example, this gives $
  \{\theta,\bar\theta\}_D= {2\mathfrak q}^{-1}$ and $
  \{\theta,\theta\}_D=\{\bar\theta,\bar\theta\}_D=0$.
Thus $\theta$ and $\bar\theta$ become conjugate fermionic variables on the reduced phase space.
The original coordinate-momentum bracket therefore becomes a bracket between $\theta$ and $\bar\theta$ themselves.\footnote{Equivalently, the Dirac bracket is the inverse of the reduced fermionic symplectic form $\Omega_{\rm red}\sim2\mathfrak q\,d\bar\theta\,d\theta$. This form is nondegenerate on the remaining $(\theta,\bar\theta)$ directions, so those directions are not null directions of the reduced phase space. The reduced Berezin measure is the one associated with this fermionic symplectic form, up to the same convention-dependent signs and factors.}

For the supergroup quotient it is more convenient to use an equivalent first-class construction.
Instead of replacing the Poisson bracket by the Dirac bracket, we enlarge the phase space by an auxiliary fermionic pair $(\chi,\pi_\chi)$, with $\{\chi,\pi_\chi\}=1$.
The old second-class equations $C=\bar C=0$ are then replaced by the shifted constraints
\begin{equation}
  \widehat C=C+ 2 \mathfrak q\chi,
  \qquad
  \widehat{\bar C}=\bar C-\pi_\chi .
\end{equation}
The coefficient of $\chi$ is fixed by the same number that appears in the original bracket $\{C,\bar C\}$.
With this choice, the bracket of $\mathfrak q\chi$ with $\pi_\chi$ cancels the nonzero term in $\{C,\bar C\}$, and $\widehat C,\widehat{\bar C}$ become first class.
Upon quantization, $\chi$ acts by multiplication and its conjugate acts by differentiation with respect to $\chi$.
This construction is equivalent to the Dirac-bracket reduction, but keeps the reduction in a form adapted to imposing constraints by a quotient.

We now apply the same construction to the supergroup particle.
The role of $C,\bar C$ is played by the two fixed fermionic right charges $\mu_{\mathsf F_{+,p}}^R$ and $\mu_{\bar{\mathsf F}_{+}^{p}}^R$.
Their bracket is nonzero because $\mu_{\mathsf L_+}^R$ has already been fixed to $-\mathfrak q$.
For each $p$, we therefore introduce a new Grassmann variable $\chi^p$ and its conjugate momentum $\pi_{\chi,p}$, with $\{\chi^p,\pi_{\chi,q}\}=\delta^p{}_q$.
The variable $\chi^p$ is therefore introduced only to supply the missing bracket: the terms $\mathfrak q\chi^p$ and $\pi_{\chi,p}$ cancel the obstruction in \eqref{eq:positive-fermion-right-algebra}.
This gives a first-class formulation of the fixed fermionic right charges, which can then be imposed before quantization in the same way as the bosonic right charge.

\paragraph{Constrained action and reduced wavefunctions.}
We now implement the first-class constraints in the worldline action.
Let $\lambda$ denote the even Lagrange multiplier for the $\mathsf L_+$ condition, and let $\epsilon^p,\bar\epsilon_p$ denote the odd Lagrange multipliers for the two fermionic conditions.
With the conventions used below, the shifted constraints are
\begin{equation}
  \mu_{\mathsf L_+}^R+\mathfrak q=0,
  \qquad
  \mu_{\mathsf F_{+,p}}^R+\mathfrak q\,\chi^p=0,
  \qquad
  \mu_{\bar{\mathsf F}_{+}^{p}}^R-\pi_{\chi,p}=0 .
  \label{eq:total-super-constraints}
\end{equation}
The auxiliary first-order term is
\begin{equation}
  I_\chi
  =
  \int du\,\pi_{\chi,p}\dot\chi^p ,
\end{equation}
and the constrained first-order particle action is
\begin{equation}
  I_{\rm red}
  =
  \phi_r\int du
  \left[
    \left\langle
      \mathcal P,\mathcal G^{-1}\dot{\mathcal G}
    \right\rangle
    -
    \frac{1}{2}
    \left\langle
      \mathcal P,\mathcal P
    \right\rangle
    -
    \lambda\left(\mu_{\mathsf L_+}^R+\mathfrak q\right)
    -
    \epsilon^p\left(\mu_{\mathsf F_{+,p}}^R+\mathfrak q\,\chi^p\right)
    -
    \bar\epsilon_p\left(\mu_{\bar{\mathsf F}_{+}^{p}}^R-\pi_{\chi,p}\right)
  \right]
  +
  I_\chi .
  \label{eq:susy-reduced-first-order-action}
\end{equation}
The constraints follow directly from this action.

It is useful to spell out the local form of these equations.
We isolate the $\mathsf F_+,\bar{\mathsf F}_+,\mathsf L_+$ coordinates by writing their block as $e^{\Theta_+}e^{y\mathsf L_+}$.
The right charges are represented by momenta conjugate to $y,\theta_+^p,\bar\theta_{+,p}$, and after using the bosonic equation $p_y=-\mathfrak q$ in the fermionic charges one obtains
\begin{equation}
  \mu_{\mathsf L_+}^R=p_y,
  \qquad
  \mu_{\mathsf F_{+,p}}^R
  =
  \pi_{\theta_+^p}-\frac{\mathfrak q}{2}\bar\theta_{+,p},
  \qquad
  \mu_{\bar{\mathsf F}_{+}^{p}}^R
  =
  \pi_{\bar\theta_{+,p}}-\frac{\mathfrak q}{2}\theta_+^p .
  \label{eq:positive-sector-local-right-charges}
\end{equation}
Substituting \eqref{eq:positive-sector-local-right-charges} into \eqref{eq:total-super-constraints} gives
\begin{equation}
  p_y=-\mathfrak q,
  \qquad
  \pi_{\theta_+^p}-\frac{\mathfrak q}{2}\bar\theta_{+,p}=-\mathfrak q\chi^p,
  \qquad
  \pi_{\bar\theta_{+,p}}-\frac{\mathfrak q}{2}\theta_+^p=\pi_{\chi,p}.
  \label{eq:positive-sector-coordinate-constraints}
\end{equation}
The last two equations are the local first-class version of the second-class fermionic pair discussed above.\footnote{Keeping a single value of $p$ gives the $\mathcal N=2$ truncation.
In \cite{Lin:2022zxd}, the operators $G_+^R,\bar G_+^R$ are the analogues of our constrained right charges $\mu_{\mathsf F_+}^R,\mu_{\bar{\mathsf F}_+}^R$.
The related covariant derivatives $D_+=\partial_{\theta_+}+\bar\theta_+\partial_\gamma$ and $\bar D_+=\partial_{\bar\theta_+}+\theta_+\partial_\gamma$ enter the remaining supergroup generators.
Compared with $G_+^R,\bar G_+^R$, they have the opposite sign in the $\theta_+\partial_\gamma$ or $\bar\theta_+\partial_\gamma$ term, with the $e^{\pm ia}$ phases stripped off.
On the sector with the $G_+^R,\bar G_+^R$ constraints and $L_+^R=-\mathfrak q$, the construction of \cite{Lin:2022zxd} replaces the constrained $(\theta_+,\bar\theta_+)$ dependence by one fermion, $D_+\to2e^{ia}\chi\,\partial_\gamma$ and $\bar D_+\to e^{-ia}\partial_\chi$, which is the quantized version of our shifted constraints: $\mathfrak q\chi^p$ acts by multiplication and $\pi_{\chi,p}$ by $\partial_{\chi^p}$.
Here $y,\chi$ correspond to the $\gamma,\chi$ variables of \cite{Lin:2022zxd} up to normalization, while our $(\theta_+^p,\bar\theta_{+,p})$ variables differ from the $(\theta_+,\bar\theta_+)$ variables of \cite{Lin:2022zxd} by the $e^{\pm ia}$ rotation, so the coordinate equations look slightly different.}
Thus, after introducing $\chi^p$, one can safely eliminate $\theta_+^p$ and $\bar\theta_{+,p}$; the constrained dependence on $\theta_+^p$ and $\bar\theta_{+,p}$ is represented by the single variable $\chi^p$.
The first equation fixes the $y$-dependence, so the reduced wavefunction can be written as
\begin{equation}
  \Psi
  =
  e^{-\mathfrak q y}
  F(x,\rho,U,\theta_-^p,\bar\theta_{-,p},\chi^p),
  \qquad
  U\in SU(2)_{\mathfrak R}.
  \label{eq:n4-reduced-wavefunction}
\end{equation}
\paragraph{Reduced measure.}
We finally describe the reduced measure.
At the path-integral level, the momentum integrations over $\mathcal P$ and $\pi_{\chi,p}$ impose the usual conjugate-momentum relations.
After these variables are integrated out, the remaining path integral is over the group variable $\mathcal G$, the multipliers $\lambda,\epsilon^p,\bar\epsilon_p$, and the auxiliary fermions $\chi^p$.
Here $\lambda$ imposes the $\mathsf L_+$ constraint, while $\epsilon^p$ and $\bar\epsilon_p$ impose the two fermionic constraints in \eqref{eq:total-super-constraints}.
Since the total constraints are first class, they generate local right shifts by $\mathbf{N}_+^{\rm susy}=\exp\mathfrak n_+^{\rm susy}$.
The path integral must therefore divide by the formal volume of this redundant gauge direction.
Schematically,
\begin{equation}
  Z_{\rm quot}
  =
  \int
  \frac{
    D\mathcal G\,
    D\lambda\,D\epsilon\,D\bar\epsilon\,
    D\chi
  }{
    \text{Vol}(\mathbf{N}_+^{\rm susy})
  }\,
  e^{-I_{\rm red}}.
  \label{eq:susy-quotient-path-integral}
\end{equation}
This is the same gauge-fixing and measure argument reviewed in
Section~\ref{sec:n0-jt-particle-main} for the bosonic particle, with the
fermionic right shifts included.
In current variables, the quotient is what permits the simple source representative.
For the bosonic particle this is the choice $s(u)=0$ in
$\mathscr J=\mathcal G^{-1}\dot{\mathcal G}
=\mathfrak q\mathsf L_-+s(u)\mathsf L_0+\mathcal L(u)\mathsf L_+$; in the
supersymmetric case it is the corresponding source condition
\eqref{eq:n2-super-source-conditions}.

In the Hamiltonian description, the same fixed-source condition is imposed by the right constraints \eqref{eq:total-super-constraints}.
Those constraints give equivariance under right multiplication by the $\mathsf L_+,\mathsf F_+,\bar{\mathsf F}_+$ subgroup.
The quotient just described is the corresponding statement for the physical inner product.
The wavefunction gluing measure is therefore read off from the same gauge-fixed path-integral measure.\footnote{For comparison, in the $\mathcal N=2$ truncation the convention of \cite{Lin:2022zxd} uses $d\mu_{\rm phys}^{\mathcal N=2}=(2\pi\mathfrak q)^{-1}d\rho\,dx\,d\theta_-\,d\bar\theta_-\,da\,d\chi$.  With $\Psi_0^{\mathcal N=2}(L)=\frac{2\mathfrak q L}{\sqrt{\pi}}K_{1/2}(2\mathfrak q L)$ and $L=e^{-\ell/2}$, their $j=0$ norm is $\int d\ell\,2(\Psi_0^{\mathcal N=2}(L))^2=1$, where the factor of $2$ counts the two fermionic structures.  We also reproduce this measure with our method.}
\begin{equation}
  d\mu_{\rm phys}
  \sim
  d\mu_{\rm BH}^{\rm red}\,
  d\chi^1d\chi^2
  =
  e^{-\rho}
  d\rho\,dx\,d\mu_{SU(2)}
  d\theta_-^1d\theta_-^2
  d\bar\theta_{-,1}d\bar\theta_{-,2}
  d\chi^1d\chi^2 ,
  \label{eq:susy-gluing-measure-schematic}
\end{equation}
so inserting complete reduced states between two interval propagators uses this quotient measure, not the unreduced Berezin--Haar measure on the full supergroup.
The auxiliary $\chi^p$ variables remain in the physical measure because they are the fermionic variables left by the first-class implementation of the constraints.

\section{Worldline supercharges, supersymmetric propagator and Wilson-line operators}
\label{sec:worldline-supercharges-propagator}

In this section we turn from the construction of the constrained particle to
the objects that it computes: supercharges, propagators, and boundary-anchored
Wilson-line operators.  The central point is that the boundary theory does not
realize the full bulk supergroup in a naive way.  In the bulk, matter fields
and Wilson lines transform under the full superisometry group, such as
$PSU(1,1|2)$ for the $\mathcal N=4$ theory, while the boundary particle
realizes the $\mathcal N=4$ super-Poincare symmetry preserved by the super-JT
boundary conditions.  Section~\ref{subsec:physical-worldline-supercharges}
derives this relation: the boundary supercharges arise from bulk supergroup
generators only after supplementing them by the compensating transformations
required to stay in the fixed-source representative.  This gives the
particle-on-a-group derivation of the worldline supercharges used in
\cite{Lin:2022zxd,Fan:2021wsb}.  Section~\ref{subsec:supersymmetric-propagator}
then constructs the corresponding supersymmetric propagator.  We organize the
two-endpoint dependence in terms of invariants of the diagonal bulk supergroup
action, and then impose the boundary supercharge conditions; the resulting
invariant functions are the building blocks that will later be glued into
length kernels.  Finally,
Section~\ref{subsec:bulk-matter-boundary-operators} applies the same logic
to matter insertions.  A boundary-anchored two-point function is represented by
a bulk Wilson line, but the boundary endpoints are not arbitrary bulk points:
they are restricted to the same fixed-source embedding used in the Hamiltonian
reduction.  As a result, primary components give the familiar invariant
bilocals, while supercharge descendants must be obtained using the compensated
boundary supercharges rather than bare bulk generators.  This distinction is
essential for the moduli-operator bilocal used in the Berry-curvature
discussion of Section~\ref{sec:discussion-future}.

\subsection{The physical worldline supercharges}
\label{subsec:physical-worldline-supercharges}

\paragraph{From boundary supertranslations to reduced worldline supercharges.}
We begin by identifying the charge to be constructed.
The physical worldline supercharge is the Noether charge for a rigid supertranslation of the boundary superline with coordinates $z=(u,\vartheta^p,\bar\vartheta_p)$ and flat one-form $\Delta u=du+i\,d\vartheta^p\,\bar\vartheta_p+i\,d\bar\vartheta_p\,\vartheta^p$.
A constant supertranslation acts schematically as
\begin{equation}
  \delta_\epsilon\vartheta^p=\epsilon^p,\qquad
  \delta_\epsilon\bar\vartheta_p=0,\qquad
  \delta_\epsilon u
  =
  -\epsilon^p\bar\vartheta_p ,
  \label{eq:boundary-supertranslation-variation}
\end{equation}
with the barred transformation obtained by conjugation.

We next translate this statement into the image variables used by the particle action.
The coordinate $x$ is the reparametrized boundary time $u$.
We restrict to the minus fermions and regard $\theta_-^p,\bar\theta_{-,p}$ as the reparametrized versions of $\vartheta^p,\bar\vartheta_p$, because the reduction above has eliminated the independent positive variables $\theta_+^p,\bar\theta_{+,p}$.\footnote{More precisely, the group coordinates are target-space variables, while $(u,\vartheta^p,\bar\vartheta_p)$ are worldline variables.  Eliminating the independent positive variables is not the same as setting the target $\theta_+^p,\bar\theta_{+,p}$ to zero; already in \eqref{eq:n2-auxiliary-solution}, they are functions of the minus variables.  Solving this embedding directly is expected to reproduce the supercharge expressions of \cite{Heydeman:2020hhw}; here we instead improve the leading minus charge so that it preserves the positive-data constraints.}
Thus the bare charge is supplied by the minus part of the boundary representative, $\mathcal G_-=e^{x\mathsf L_-}e^{\theta_-^p\mathsf F_{-,p}+\bar\theta_{-,p}\bar{\mathsf F}_{-}^{p}}$.
The needed minus supertranslation algebra is
\begin{equation}
  \{\mathsf F_{-,p},\bar{\mathsf F}_{-}^{q}\}
  \sim
  \delta_p{}^q\mathsf L_-,
  \qquad
  [\mathsf L_-,\mathsf F_{-,p}]
  =
  [\mathsf L_-,\bar{\mathsf F}_{-}^{p}]
  =
  0 .
  \label{eq:minus-supertranslation-algebra-main}
\end{equation}
Acting from the left by $\epsilon^p\mathsf F_{-,p}$ then gives
\begin{equation}
  \delta_\epsilon\theta_-^p=\epsilon^p,\qquad
  \delta_\epsilon\bar\theta_{-,p}=0,\qquad
  \delta_\epsilon x
  =
  -\epsilon^p\bar\theta_{-,p}.
  \label{eq:minus-supertranslation-variation-main}
\end{equation}
Equation~\eqref{eq:minus-supertranslation-variation-main} preserves
$\Pi=\dot x+\bar\theta_{-,p}\dot\theta_-^p+\theta_-^p\dot{\bar\theta}_{-,p}$,
the component form of $\Delta u$, and is therefore the target-space version of
\eqref{eq:boundary-supertranslation-variation}.
Thus, before any positive data are fixed as in \eqref{eq:n2-super-source-conditions}, the algebraic bare charges are $q_{{\rm bare},p}=\mathsf F_{-,p}$ and $\bar q_{\rm bare}^{p}=\bar{\mathsf F}_{-}^{p}$.

The $q_{\rm bare}$ charges have nontrivial commutators with the positive generators appearing in the constraints, and therefore cannot be the physical supercharges of the superline:
\begin{equation}
  \{\mathsf L_+-\mathfrak q,\mathsf F_{-,p}\}
  \sim
  \mathsf F_{+,p},
  \qquad
  \{\bar{\mathsf F}_{+}^{q},\mathsf F_{-,p}\}
  \sim
  \delta_p{}^q\,\mathsf L_0
  +(\mathfrak R\text{-generators}) .
  \label{eq:minus-supertranslation-obstruction-main}
\end{equation}
The first bracket vanishes on the constrained surface, while the second one does not.
Equivalently, the bare minus supertranslation moves the particle trajectory out of the chosen fixed-source slice in phase space.
In the gravitational description, a bare supertranslation generated by $\mathsf F_{-,p}$ would break FG gauge.

To fix this, we perform the procedure of ``fixing constraints with constraints", which is possible if the constraints satisfy non-trivial algebra. 
If $\phi_A$ denotes the positive equations $(\mathsf L_+-\mathfrak q,\mathsf F_+,\bar{\mathsf F}_+)$, a physical worldline supercharge must obey the tangency condition
\begin{equation}
  \{\phi_A,q_{\rm phys}\}\big|_{\phi=0}=0 .
  \label{eq:physical-q-preserve-constraints}
\end{equation}
This condition requires that, after the constraints are imposed, the Hamiltonian flow generated by $q_{\rm phys}$ remains tangent to the constrained surface.
We may therefore improve the bare charge by constrained generators,
\begin{equation}
  q_{\rm phys}
  =
  q_{\rm bare}
  +
  \lambda^A\phi_A
  +
  \lambda^{AB}\phi_A\phi_B
  +\cdots,
  \qquad
  q_{\rm phys}\big|_{\phi=0}
  =
  q_{\rm bare}\big|_{\phi=0},
  \label{eq:q-constraint-improvement-main}
\end{equation}
so the value of the charge on the reduced surface is unchanged.
Higher powers are allowed because the constrained generators need not commute.
The compensator is fixed by imposing\footnote{For higher powers in \eqref{eq:q-constraint-improvement-main}, the full condition is $\{\phi_A,q_{\rm phys}\}$ lying in the ideal generated by the $\phi_A$'s.  Expanding order by order in $\phi$ gives the higher-order Poisson-bracket conditions; the displayed equation is the leading one.}
\begin{equation}
  \{\phi_A,q_{\rm bare}\}\big|_{\phi=0}
  +
  \lambda^B\{\phi_A,\phi_B\}\big|_{\phi=0}
  =
  0 .
  \label{eq:constraint-compensator-equation-main}
\end{equation}
The nonzero bracket of the positive fermionic pair, discussed in \eqref{eq:positive-fermion-right-algebra}, is precisely what makes this cancellation possible at fixed non-zero $\mathfrak q$.

The same construction has a direct group-theoretic interpretation.
Let $\mathbf Z(z)$ denote the embedding of the boundary superline into super-$\mathrm{AdS}_2$ superspace, using the same boldface superspace notation as in Section~\ref{sec:superjt-boundary-particle}.
The group representative is then $\mathcal G(\mathbf Z(z))$, namely the boundary representative evaluated at the target superspace point.
The embedding $\mathbf Z(z)$ is not arbitrary, because $\mathcal G(\mathbf Z)$ must obey the fixed-source constraints on $\mathcal G^{-1}d\mathcal G$.
A physical supertranslation must map one allowed embedding to another,
$\mathbf Z(z)\mapsto\mathbf Z(z)+\epsilon\,\delta\mathbf Z(z)$.\footnote{Here the embedding is varied in a fixed super-$\mathrm{AdS}_2$ background with fixed simple boundary metric, as in the standard Schwarzian derivation from a fluctuating cutoff curve; see \cite{Mertens:2022irh}.}
A naive left action by $\mathsf F_{-,p}$ fails to do this by itself, because it generally refactorizes as
\begin{equation}
  e^{\epsilon\mathsf F_{-,p}}\mathcal G(\mathbf Z)
  =
  \mathcal G(\mathbf Z+\epsilon\,\delta\mathbf Z)\,
  e^{\epsilon\alpha_+(\mathbf Z)}
  +O(\epsilon^2),
  \qquad
  \alpha_+(\mathbf Z)\in\mathfrak n_+^{\rm susy}.
  \label{eq:minus-supertranslation-refactorization-main}
\end{equation}
The extra positive factor is precisely the piece which moves the configuration away from the fixed-source representative.
The compensator in $q_{\rm phys}$ absorbs $e^{\epsilon\alpha_+(\mathbf Z)}$ and restores the fixed-source representative.
The $\mathcal N=2$ and $\mathcal N=4$ examples below provide explicit checks of this construction.

\paragraph{The \texorpdfstring{$\mathcal N=2$}{N=2} case.}
In the $\mathcal N=2$ truncation the compensating construction takes a particularly simple form.
The physical representative whose bare piece is $\mathsf F_-$ is
\begin{equation}
  q_{\rm phys}
  =
  \mathsf F_-
  -
  \frac{\mathsf L_0+\mathsf R}{\mathsf L_+}\mathsf F_+ .
  \label{eq:n2-rational-q-main}
\end{equation}
The second term is the positive compensator.
It vanishes on the fixed positive surface, but its Hamiltonian flow cancels the component of the bare $\mathsf F_-$ flow which would leave the fixed-source slice.
The barred charge is
\begin{equation}
  \bar q_{\rm phys}
  =
  \bar{\mathsf F}_-
  +
  \frac{\mathsf L_0-\mathsf R}{\mathsf L_+}\bar{\mathsf F}_+ .
  \label{eq:n2-rational-qbar-main}
\end{equation}
In the one-complex-fermion case, no higher positive-fermion correction can be constructed.

For later use, we introduce numerator-cleared versions of these rational representatives.
With the sign convention $Q=-q_{\rm phys}\mathsf L_+$ and $\bar Q=-\bar q_{\rm phys}\mathsf L_+$, one obtains
\begin{align}
  Q
  &=
  \left(\mathsf L_0+\mathsf R\right)\mathsf F_+
  -
  \mathsf F_-\mathsf L_+,
  \nonumber\\
  \bar Q
  &=
  -\left(\mathsf L_0-\mathsf R\right)\bar{\mathsf F}_+
  -
  \bar{\mathsf F}_-\mathsf L_+ .
  \label{eq:n2-polynomial-q-main}
\end{align}
These charges obey the fixed-parabolic normalization
\begin{equation}
  Q^2=\bar Q^2=0,
  \qquad
  \{Q,\bar Q\}
  =
  \mathcal C_{\mathcal N=2}\mathsf L_+,
  \label{eq:n2-q-closure-main}
\end{equation}
where the ordered quadratic $\mathfrak{psu}(1,1|1)$ Casimir is $\mathcal C_{\mathcal N=2}=\mathsf L_0^2+\mathsf L_-\mathsf L_+-\mathsf R^2+\mathsf F_-\bar{\mathsf F}_+-\bar{\mathsf F}_-\mathsf F_+$.

\paragraph{The \texorpdfstring{$\mathcal N=4$}{N=4} case.}
In the $\mathcal N=4$ theory the same correction contains one additional feature.
The positive fermion constraints form a doublet, which allows the nilpotent bilinear $\mathsf F_{+,1}\mathsf F_{+,2}$.
We use $a=1,2$ for the $SU(2)_{\mathfrak R}$ doublet index and $\mathsf R_\pm=\mathsf R_1\pm i\mathsf R_2$.
For the charge whose bare piece is $\mathsf F_{-,1}$, the physical representative is
\begin{equation}
  q_{\rm phys}^1
  =
  \mathsf F_{-,1}
  -
  \frac{\mathsf L_0+\mathsf R_3}{\mathsf L_+}\mathsf F_{+,1}
  -
  \frac{\mathsf R_+}{\mathsf L_+}\mathsf F_{+,2}
  +
  \frac{1}{(\mathsf L_+)^2}
  \mathsf F_{+,1}\bar{\mathsf F}_{+}^{2}\mathsf F_{+,2} .
  \label{eq:n4-q1-physical-rational-main}
\end{equation}
The first two compensating terms are the direct analogues of the $\mathcal N=2$ correction.
The cubic term is allowed because of the nilpotent positive bilinear.
In the normalization of Appendix~\ref{app:supergroup-conventions}, nilpotency and closure fix its coefficient to be $1$.

The corresponding numerator-cleared charge is $Q^1=-(\mathsf L_+)^2q_{\rm phys}^1$, namely
\begin{equation}
  Q^1
  =
  (\mathsf L_0+\mathsf R_3)\mathsf F_{+,1}\mathsf L_+
  +
  \mathsf R_+\mathsf F_{+,2}\mathsf L_+
  -
  \mathsf F_{-,1}(\mathsf L_+)^2
  -
  \mathsf F_{+,1}\bar{\mathsf F}_{+}^{2}\mathsf F_{+,2} .
  \label{eq:n4-q1-main}
\end{equation}
The remaining three numerator-cleared charges are obtained by the same compensation, together with conjugation and $SU(2)_{\mathfrak R}$:
\begin{align}
  Q^2
  &=
  (\mathsf L_0-\mathsf R_3)\mathsf F_{+,2}\mathsf L_+
  +
  \mathsf R_-\mathsf F_{+,1}\mathsf L_+
  -
  \mathsf F_{-,2}(\mathsf L_+)^2
  -
  \mathsf F_{+,2}\bar{\mathsf F}_{+}^{1}\mathsf F_{+,1},
  \nonumber\\[1mm]
  \bar Q_1
  &=
  -(\mathsf L_0-\mathsf R_3)\bar{\mathsf F}_{+}^{1}\mathsf L_+
  +
  \mathsf R_-\bar{\mathsf F}_{+}^{2}\mathsf L_+
  -
  \bar{\mathsf F}_{-}^{1}(\mathsf L_+)^2
  +
  \bar{\mathsf F}_{+}^{1}\mathsf F_{+,2}\bar{\mathsf F}_{+}^{2},
  \nonumber\\[1mm]
  \bar Q_2
  &=
  -(\mathsf L_0+\mathsf R_3)\bar{\mathsf F}_{+}^{2}\mathsf L_+
  +
  \mathsf R_+\bar{\mathsf F}_{+}^{1}\mathsf L_+
  -
  \bar{\mathsf F}_{-}^{2}(\mathsf L_+)^2
  +
  \bar{\mathsf F}_{+}^{2}\mathsf F_{+,1}\bar{\mathsf F}_{+}^{1} .
  \label{eq:n4-cubic-q-main}
\end{align}
The terms proportional to $\mathsf F_{-,a}$ or $\bar{\mathsf F}_{-}^{a}$ are the original minus supertranslations.
The remaining terms are precisely the positive compensators required to stay inside the fixed-source slice.
The numerator-cleared $\mathcal N=4$ charges satisfy
\begin{equation}
  \{Q^a,\bar Q_b\}
  =
  \delta^a{}_{b}\mathcal C_{\mathcal N=4}(\mathsf L_+)^3,
  \qquad
  \{Q^a,Q^b\}
  =
  \{\bar Q_a,\bar Q_b\}
  =
  0 .
  \label{eq:n4-q-closure-main}
\end{equation}
Here the quadratic $\mathfrak{psu}(1,1|2)$ Casimir is
\begin{equation}
  \mathcal C_{\mathcal N=4}=\mathsf L_0^2+\mathsf L_-\mathsf L_+-\mathsf L_0-\mathsf R_3^2-\mathsf R_-\mathsf R_+-\mathsf R_3+\mathsf F_{-,a}\bar{\mathsf F}_{+}^{a}-\bar{\mathsf F}_{-}^{a}\mathsf F_{+,a}.
  \label{eq:n4-casimir-main}
\end{equation}
and this Casimir becomes the reduced Hamiltonian.

\subsection{The supersymmetric propagator}
\label{subsec:supersymmetric-propagator}

We next construct the propagator for the particle on a group.
The supercharges obtained above generate the boundary super-Poincare algebra, and the particle on a group must furnish a representation of this algebra.
Consequently the propagator is fixed by charge conditions: it carries the appropriate $SU(2)_{\mathfrak R}$ quantum numbers and satisfies the differential equations generated by the supercharges.

The two-endpoint propagator is a function of the group variables at both endpoints.
Following \cite{Lin:2022zxd}, we first use the symmetry under diagonal left multiplication to organize this dependence into invariant variables.
Once these variables are chosen, the construction proceeds in three steps.
We determine the action of the supercharges on the invariants, write the most general invariant ansatz, and impose the resulting charge equations.

\paragraph{Construction of invariants.}
We begin with the group coordinates before eliminating the positive fermions.
An endpoint of the particle on a group carries $x,\rho,U,y,\theta_-^p,\bar\theta_{-,p},\theta_+^p,\bar\theta_{+,p}$.
These are the coordinates of the Gauss representative of the particle on a group.
At this stage the $\theta_+^p,\bar\theta_{+,p}$ variables are still present; their elimination and the introduction of the $\chi^p$ polarization occur only after the fixed positive data are imposed.
The invariant variables are defined as in \cite{Lin:2022zxd}.
Before any positive variables are eliminated, a two-endpoint function is an invariant if it is annihilated by the diagonal left action of the global supergroup.\footnote{This follows from the definition of the endpoint states and propagator.
The group-labeled states are defined by $|\mathcal G,i\rangle=U(\mathcal G)|i\rangle$, so a physical left action sends $|\mathcal G,i\rangle$ to $|a\mathcal G,i\rangle$.
Since the Hamiltonian is the Casimir, the propagator is invariant under simultaneous left multiplication, as in \eqref{eq:left-invariance-propagator}.
Right multiplication is different: it gives equivariance, not invariance.
See Section~\ref{sec:n0-jt-particle-main} and Appendix~\ref{subsec:n0-bf-particle},
especially \eqref{eq:right-equivariant-wavefunction-state}--\eqref{eq:right-covariance-propagator-state},
for details.}
Writing $\mathcal D_{\mathsf T,i}^{L}$ and $\mathcal D_{\mathsf T,f}^{L}$ for the left differential operator generated by $\mathsf T\in\mathfrak{psu}(1,1|2)$ at the initial and final endpoints, this condition is
\begin{equation}
  \left(
    \mathcal D_{\mathsf T,i}^{L}
    +
    \mathcal D_{\mathsf T,f}^{L}
  \right)I
  =
  0,
  \qquad
  \mathsf T\in\mathfrak{psu}(1,1|2).
  \label{eq:diagonal-left-invariant-condition-main}
\end{equation}
Thus an invariant is unchanged when both endpoints are moved by the same global superisometry.
Before eliminating the positive fermions, we introduce the auxiliary quantities
\[
  \widetilde X_{if}=x_f-x_i-\theta_{-,i}^a\bar\theta_{-,f,a}-\bar\theta_{-,i,a}\theta_{-,f}^a,
  \qquad
  \theta_{if,-}^{a}=\theta_{-,f}^{a}-\theta_{-,i}^{a},
  \qquad
  \bar\theta_{if,-,a}=\bar\theta_{-,f,a}-\bar\theta_{-,i,a}.
\]
The first terms in the invariant variables are
\begin{equation}
\begin{aligned}
  L
  &=
  \frac{e^{-\frac12(\rho_i+\rho_f)}}{\widetilde X_{if}}
  +O(\theta^4),
  \\
  W^a{}_{b}
  &=
  \left[
  U_i
  \begin{pmatrix}
    1-\dfrac{\varepsilon_{cd}\theta_{if,-}^c\bar\theta_{if,-,d}}{\widetilde X_{if}}
    &
    \dfrac{(\bar\theta_{if,-})^2}{\widetilde X_{if}}
    \\[1.1em]
    \dfrac{(\theta_{if,-})^2}{\widetilde X_{if}}
    &
    1+\dfrac{\varepsilon_{cd}\theta_{if,-}^c\bar\theta_{if,-,d}}{\widetilde X_{if}}
  \end{pmatrix}
  U_f^{-1}
  \right]^a{}_{b}
  +O(\theta^4),
  \\
  Y_i-Y_f
  &=
  y_i-y_f+\frac{e^{-\rho_i}+e^{-\rho_f}}{\widetilde X_{if}}
  +\left(\theta_{+,i}^a\bar f_{i,a}-f_i^a\bar\theta_{+,i,a}\right)
  -\left(\theta_{+,f}^a\bar f_{f,a}-f_f^a\bar\theta_{+,f,a}\right)
  +O(\theta^4).
  \\
  f_{\alpha}^{a}
  &=
  \theta_{+,\alpha}^{a}
  +
  \frac{e^{-\rho_\alpha/2}(U_\alpha)^a{}_{b}\,
  \bar\theta_{if,-}^{b}}{\widetilde X_{if}}
  +O(\theta^3),
  \\
  \bar f_{\alpha,a}
  &=
  \bar\theta_{+,\alpha,a}
  +
  \frac{e^{-\rho_\alpha/2}(U_\alpha^{-1})_{a}{}^{b}\,
  \theta_{if,-}^{b}}{\widetilde X_{if}}
  +O(\theta^3).
\end{aligned}
\label{eq:unreduced-invariants-low-order-main}
\end{equation}
Here $\alpha=i,f$ and $\varepsilon_{12}=1$.
The complete all-fermion expressions are recorded in the Mathematica supplement.

We now impose the final positive constraints of Section~\ref{subsec:susy-hamiltonian-reduction}.
The fermionic constraints eliminate the independent $\theta_+^a,\bar\theta_{+,a}$ dependence and replace it by the $\chi^a$ polarization.
This step also fixes the treatment of the positive-fermion terms in $Y_i-Y_f$ in \eqref{eq:unreduced-invariants-low-order-main}.
Using the local constraints \eqref{eq:positive-sector-coordinate-constraints}, the $\theta_+,\bar\theta_+$ dependence is part of the positive fiber; after passing to the $\chi$-polarization we choose the reduced representative $\theta_{+,\alpha}^a=\bar\theta_{+,\alpha,a}=0$, so $(Y_i-Y_f)_{\rm red}=y_i-y_f+(e^{-\rho_i}+e^{-\rho_f})/\widetilde X_{if}+O(\theta^4)$.
The fermionic data which originally sat in the positive variables are not discarded.
They are carried by the $\eta_\alpha^a$ variables below and by the $\chi_\alpha^a$-derivative representation of the conjugate variables.
In this polarization the reduced fermionic invariants are
\begin{equation}
  \eta_\alpha^a
  \equiv
  f_\alpha^a\big|_{\rm red}
  =
  \chi_\alpha^a
  +
  \frac{e^{-\rho_\alpha/2}(U_\alpha)^a{}_{b}\,
  \bar\theta_{if,-}^{b}}{\widetilde X_{if}}
  +O(\theta^3),
  \qquad
  \alpha=i,f .
  \label{eq:reduced-eta-invariants-main}
\end{equation}
In this polarization, the conjugate $\bar f_{\alpha,a}$ variables are represented by derivatives with respect to $\chi_\alpha^a$.
The bosonic right constraint fixes the $Y$-dependence, and therefore the two-endpoint wavefunction has the fixed-charge form
\begin{equation}
  P_{if}
  =
  e^{-\mathfrak q(Y_i-Y_f)}
  \Psi_{\rm WdW}(L,W,\eta_i,\eta_f).
  \label{eq:fixed-charge-invariant-wavefunction-main}
\end{equation}
The reduced factor $\Psi_{\rm WdW}$ will be identified below with the Wheeler--DeWitt wavefunction.

\paragraph{Action of the supercharges.}
The differential operators used below are obtained directly from the supercharges in \eqref{eq:n4-q1-main}--\eqref{eq:n4-cubic-q-main}.
No additional ansatz is required.
Since a left action on an endpoint group element is represented in coordinates by a right-invariant vector field, each algebra generator $\mathsf T$ in the polynomial charge is replaced by the corresponding endpoint operator $\mathcal D^R_{\mathsf T}$.
For example, \eqref{eq:n4-q1-main} becomes
\begin{equation}
  Q^1
  =
  \left(\mathcal D^R_{\mathsf L_0}+\mathcal D^R_{\mathsf R_3}\right)
  \mathcal D^R_{\mathsf F_{+,1}}\mathcal D^R_{\mathsf L_+}
  +
  \mathcal D^R_{\mathsf R_+}\mathcal D^R_{\mathsf F_{+,2}}\mathcal D^R_{\mathsf L_+}
  -
  \mathcal D^R_{\mathsf F_{-,1}}\left(\mathcal D^R_{\mathsf L_+}\right)^2
  -
  \mathcal D^R_{\mathsf F_{+,1}}\mathcal D^R_{\bar{\mathsf F}_{+}^{2}}
  \mathcal D^R_{\mathsf F_{+,2}}.
  \label{eq:q1-final-generator-substitution-main}
\end{equation}
This operator acts on a function of the original endpoint variables through the invariant variables $L,Y,W,\eta_i,\eta_f$.
Equivalently, if $F_{\rm inv}=F(L,Y,W,\eta_i,\eta_f)$, then
\begin{equation}
  Q^aF_{\rm inv}
  =
  \left.
  Q^a
  F\!\left(
    L(x,\rho,U,\theta_-,\bar\theta_-,\chi),\,
    Y(x,\rho,U,\theta_-,\bar\theta_-,\chi),\,
    W(x,\rho,U,\theta_-,\bar\theta_-,\chi),\,
    \eta_i,\eta_f
  \right)
  \right|_{\rm inv},
  \label{eq:q-action-chain-rule-main}
\end{equation}
where the last step means that the result is rewritten again in the invariant variables.
This procedure gives the $Q$ charge as a differential operator on the invariant variables.
Let $\mathsf R_{f,3},\mathsf R_{f,\pm}$ denote the $SU(2)_{\mathfrak R}$ differential operators on $W$ generated from the final endpoint, and let $\mathsf R_{i,3},\mathsf R_{i,\pm}$ denote the corresponding initial-endpoint operators.
We also write the opposite-orientation Wilson line as $\bar W$, defined by $W^a{}_{c}\bar W^c{}_{b}=\bar W^a{}_{c}W^c{}_{b}=\delta^a{}_{b}$.
The endpoint label on $\partial_{Y_f}$ or $\partial_{Y_i}$ records which endpoint positive coordinate is differentiated before the fixed-charge phase is imposed.
For the final endpoint, we obtain
\begin{align}
  Q_f^1
  &=
  \left[
    \frac12(L\partial_L-2\mathsf R_{f,3})\partial_{\eta_f^1}
    +\partial_{\eta_f^2}\mathsf R_{f,+}
    -L\left(W^1{}_1\eta_i^1+W^1{}_2\eta_i^2\right)\partial_{Y_f}^2
    +\frac12\partial_{\eta_f^1}\partial_{\eta_f^2}\eta_f^2
  \right]\partial_{Y_f},
  \nonumber\\
  \bar Q_{f,1}
  &=
  \left[
    -\frac12(L\partial_L+2\mathsf R_{f,3})\eta_f^1
    +\eta_f^2\mathsf R_{f,-}
    -L\left(W^2{}_2\partial_{\eta_i^1}-W^2{}_1\partial_{\eta_i^2}\right)
    -\frac12\eta_f^1\eta_f^2\partial_{\eta_f^2}
  \right]\partial_{Y_f}^2,
  \nonumber\\
  Q_f^2
  &=
  \left[
    -\frac12(L\partial_L+2\mathsf R_{f,3})\partial_{\eta_f^2}
    -\partial_{\eta_f^1}\mathsf R_{f,-}
    -L\left(W^2{}_1\eta_i^1+W^2{}_2\eta_i^2\right)\partial_{Y_f}^2
    -\frac12\partial_{\eta_f^2}\partial_{\eta_f^1}\eta_f^1
  \right]\partial_{Y_f},
  \nonumber\\
  \bar Q_{f,2}
  &=
  \left[
    \frac12(L\partial_L-2\mathsf R_{f,3})\eta_f^2
    -\eta_f^1\mathsf R_{f,+}
    -L\left(W^1{}_1\partial_{\eta_i^2}-W^1{}_2\partial_{\eta_i^1}\right)
    +\frac12\eta_f^2\eta_f^1\partial_{\eta_f^1}
  \right]\partial_{Y_f}^2 .
  \label{eq:main-final-q-operators}
\end{align}
Repeating the same calculation at the initial endpoint, with $Y_i-Y_f$ and $W$ kept fixed, gives
\begin{align}
  Q_i^1
  &=
  \left[
    \frac12(L\partial_L-2\mathsf R_{i,3})\partial_{\eta_i^1}
    +\partial_{\eta_i^2}\mathsf R_{i,+}
    -L\left(\bar W^1{}_1\eta_f^1+\bar W^1{}_2\eta_f^2\right)\partial_{Y_i}^2
    +\frac12\partial_{\eta_i^1}\partial_{\eta_i^2}\eta_i^2
  \right]\partial_{Y_i},
  \nonumber\\
  \bar Q_{i,1}
  &=
  \left[
    -\frac12(L\partial_L+2\mathsf R_{i,3})\eta_i^1
    +\eta_i^2\mathsf R_{i,-}
    -L\left(\bar W^2{}_2\partial_{\eta_f^1}-\bar W^2{}_1\partial_{\eta_f^2}\right)
    -\frac12\eta_i^1\eta_i^2\partial_{\eta_i^2}
  \right]\partial_{Y_i}^2,
  \nonumber\\
  Q_i^2
  &=
  \left[
    -\frac12(L\partial_L+2\mathsf R_{i,3})\partial_{\eta_i^2}
    -\partial_{\eta_i^1}\mathsf R_{i,-}
    -L\left(\bar W^2{}_1\eta_f^1+\bar W^2{}_2\eta_f^2\right)\partial_{Y_i}^2
    -\frac12\partial_{\eta_i^2}\partial_{\eta_i^1}\eta_i^1
  \right]\partial_{Y_i},
  \nonumber\\
  \bar Q_{i,2}
  &=
  \left[
    \frac12(L\partial_L-2\mathsf R_{i,3})\eta_i^2
    -\eta_i^1\mathsf R_{i,+}
    -L\left(\bar W^1{}_1\partial_{\eta_f^2}-\bar W^1{}_2\partial_{\eta_f^1}\right)
    +\frac12\eta_i^2\eta_i^1\partial_{\eta_i^1}
  \right]\partial_{Y_i}^2 .
  \label{eq:main-initial-q-operators}
\end{align}
On the fixed-charge sector \eqref{eq:fixed-charge-invariant-wavefunction-main}, these expressions reproduce the Liouville supercharges of \cite{Lin:2025wof} after replacing the endpoint derivatives by $\partial_{Y_i}\to-\mathfrak q$ and $\partial_{Y_f}\to+\mathfrak q$.

\paragraph{BPS propagator.}
The BPS propagator is annihilated by the supercharges at both endpoints and obeys the $SU(2)_{\mathfrak R}$-singlet condition at each endpoint,
\begin{equation}
  Q_i^aP_{if}=\bar Q_{i,a}P_{if}=0,
  \qquad
  Q_f^aP_{if}=\bar Q_{f,a}P_{if}=0,
  \qquad
  \mathcal J_iP_{if}=\mathcal J_fP_{if}=0,
  \label{eq:bps-propagator-q-constraints-main}
\end{equation}
where $\mathcal J_i$ and $\mathcal J_f$ denote the $SU(2)_{\mathfrak R}$-singlet conditions at the two endpoints.

In the reduced $\eta$-variable convention, the two $SU(2)_{\mathfrak R}$-singlet fermion structures are
\[
  \eta_f^a W_a{}^b\eta_{i,b},
  \qquad
  \eta_i^1\eta_i^2+\eta_f^1\eta_f^2 .
\]
Therefore the invariant ansatz is
\begin{equation}
  P^{\rm BPS}
  =
  e^{-\mathfrak q(Y_i-Y_f)}
  \left[
    A(L)\,
    \eta_f^a W_a{}^b\eta_{i,b}
    +
    B(L)
    \left(
      \eta_i^1\eta_i^2
      +
      \eta_f^1\eta_f^2
    \right)
  \right].
  \label{eq:bps-propagator-ansatz-main}
\end{equation}
The first structure contracts the two endpoint fermions through the $SU(2)_{\mathfrak R}$ Wilson line, while the second is the endpoint singlet in the $\chi$-polarization.
Acting with the reduced supercharges on \eqref{eq:bps-propagator-ansatz-main} gives two coupled first-order radial equations for $A(L)$ and $B(L)$.
In the convention used here they can be written as
\begin{equation}
  \left(L\partial_L-1\right)A(L)
  +
  2\mathfrak q L\,B(L)
  =
  0,
  \qquad
  L\partial_L B(L)
  +
  2\mathfrak q L\,A(L)
  =
  0 .
  \label{eq:bps-radial-equations-main}
\end{equation}
The solution is
\begin{equation}
  A(L)=L K_0(2\mathfrak q L),
  \qquad
  B(L)=L K_1(2\mathfrak q L).
  \label{eq:bps-radial-functions-main}
\end{equation}
The resulting BPS interval propagator is
\begin{equation}
  P^{\rm BPS}
  =
  e^{-\mathfrak q(Y_i-Y_f)}
  \left[
    L K_0(2\mathfrak q L)\,
    \eta_f^a W_a{}^b\eta_{i,b}
    +
    L K_1(2\mathfrak q L)
    \left(
      \eta_i^1\eta_i^2
      +
      \eta_f^1\eta_f^2
    \right)
  \right].
  \label{eq:bps-propagator-main}
\end{equation}

\paragraph{Non-BPS propagator.}
Away from the BPS sector, the supermultiplet contains several components, and the propagator becomes correspondingly matrix-valued.
For fixed energy $E$, equivalently fixed Casimir eigenvalue, and spin $j$, we use the non-BPS multiplet.
For a more detailed discussion of these multiplets and their wavefunctions, see \cite{Heydeman:2020hhw,Lin:2025wof}.
\begin{equation}
  \mathcal M_{E,j}
  =
  |\mathrm H;E,j\rangle
  \oplus
  |\Psi;E,j-\tfrac12\rangle
  \oplus
  |\chi;E,j-\tfrac12\rangle
  \oplus
  |\mathrm L;E,j-1\rangle .
  \label{eq:nonbps-supermultiplet-main}
\end{equation}
Here $\mathrm H,\Psi,\chi,\mathrm L$ are component labels, and $\mathrm L$ should not be confused with the invariant $L$.
Schematically, the multiplet has the diamond structure
\begin{equation}
\begin{array}{ccccc}
  && |\mathrm H;E,j\rangle && \\[0.4em]
  & \swarrow Q && Q\searrow & \\[0.4em]
  |\Psi;E,j-\tfrac12\rangle &&&& |\chi;E,j-\tfrac12\rangle \\[0.4em]
  & Q\searrow && \swarrow Q & \\[0.4em]
  && |\mathrm L;E,j-1\rangle &&
\end{array}
  \label{eq:nonbps-supermultiplet-diamond-main}
\end{equation}
where each arrow denotes an appropriate component of the supercharge.
The bottom state is the spin-$j-1$ component obtained after two supercharge actions.\footnote{The naive second supercharge action also contains pieces in the higher-spin channel, so obtaining the $\mathrm L$ state includes the usual orthogonalization.}
For the non-BPS multiplet with highest spin $j=\frac12$, the $\mathrm L$ component is absent.
The continuous part of the spectrum is labelled by $s>0$.
In the $\mathcal N=4$ convention of \cite{Lin:2025wof}, a non-BPS multiplet of highest spin $j$ has $E=E_{\rm gap}(j)+s^2$, with $E_{\rm gap}(j)=j^2$ up to the overall Schwarzian energy normalization.
It is useful to package a boundary-segment state as
\begin{equation}
  \mathbf s=(s,j,\Phi),
  \label{eq:boundary-segment-state-label}
\end{equation}
where $s$ is the continuous energy label, $j$ is the highest spin of the worldline supermultiplet, and $\Phi$ labels the component inside the multiplet, such as $\mathrm H,\Psi,\chi,\mathrm L$.

Let us display the $H$-component for the highest-spin $j=\frac12$ multiplet, which is produced most directly by the invariant calculation.
In the endpoint ordering used here, introduce
\begin{equation}
  \eta_i\cdot W\cdot\eta_f
  \equiv
  \varepsilon_{ab}\eta_i^a W^b{}_{c}\eta_f^c,
  \qquad
  \mathcal U_1=W^1{}_{1}\,\eta_i\cdot W\cdot\eta_f,
  \qquad
  \mathcal U_2=W^1{}_{1}\left(\eta_i^1\eta_i^2+\eta_f^1\eta_f^2\right),
  \qquad
  \mathcal U_3=\eta_i^1\eta_f^2 .
  \label{eq:nonbps-h-angular-structures-main}
\end{equation}
The endpoint-orientation sign is fixed to be $+1$ throughout this paragraph.
Before imposing the supercharge equations, the invariant ansatz is
\begin{equation}
  P^{1/2}_{HH}
  =
  e^{-\mathfrak q(Y_i-Y_f)}
  \left[
    \mathfrak q A(L)\mathcal U_1
    +
    B(L)\mathcal U_2
    +
    C(L)\mathcal U_3
  \right].
  \label{eq:nonbps-h-ansatz-main}
\end{equation}
The factor multiplying $A(L)$ is part of the normalization convention for this component.
The functions $A,B,C$ are arbitrary at this stage.
The first-order charge equations, for example the pair generated by $Q_f^1$ and $\bar Q_{i,1}$, fix $B$ and $C$ in terms of $A$:
\begin{equation}
  B(L)
  =
  \frac{A(L)-\frac12 L A'(L)}{L},
  \qquad
  C(L)
  =
  \frac{
    2\left(\mathfrak q^2L^2-2\right)A(L)
    +
    L\left(3A'(L)-LA''(L)\right)
  }{
    4\mathfrak q L^2
  } .
  \label{eq:nonbps-h-charge-relations-main}
\end{equation}
On the fixed-charge sector, the endpoint supercharge algebra acting on the propagator is
\begin{equation}
  \left\{Q_\alpha^a,\bar Q_{\alpha,b}\right\}P
  =
  \delta^a{}_{b}\,\mathfrak q^3\,\widehat H\,P,
  \qquad
  \alpha=i,f .
  \label{eq:nonbps-endpoint-hamiltonian-algebra-main}
\end{equation}
Using this equation, the Hamiltonian equation on this component becomes the radial equation
\begin{equation}
  \left[
    (L\partial_L-1)^2
    -
    (2\mathfrak q L)^2
    +
    4s^2
  \right]A(L)=0 .
  \label{eq:nonbps-h-radial-equation-main}
\end{equation}
The normalizable solution is
\begin{equation}
  A_s(L)=L K_{2is}(2\mathfrak q L).
  \label{eq:nonbps-h-radial-solution-main}
\end{equation}
Therefore the $H$-component propagator is
\begin{equation}
  P^{1/2}_{HH}=e^{-\mathfrak q(Y_i-Y_f)}
  \left[
    \mathfrak q A_s(L)\mathcal U_1
    +\frac{A_s(L)-\frac12 L A_s'(L)}{L}\mathcal U_2
    +\frac{2\left(\mathfrak q^2L^2-2\right)A_s(L)+L\left(3A_s'(L)-LA_s''(L)\right)}{4\mathfrak q L^2}\mathcal U_3
  \right].
  \label{eq:nonbps-h-propagator-main}
\end{equation}
In this multiplet, $\Psi$ and $\chi$ are the two scalar components.
Their diagonal propagators are obtained by applying endpoint supercharges to $P^{1/2}_{HH}$:
\begin{equation}
  P^{1/2}_{\Psi\Psi}\propto Q_i^2\bar Q_{f,2}P^{1/2}_{HH},
  \qquad
  P^{1/2}_{\chi\chi}\propto \bar Q_{i,1}Q_f^1P^{1/2}_{HH},
  \label{eq:nonbps-scalar-propagators-from-h-main}
\end{equation}
with the proportionality constants fixed by the normalization of the component states.

\paragraph{Measure and relation to Liouville quantum mechanics.}

The gluing measure is the quotient Berezin--Haar measure from the Hamiltonian reduction.
In the coordinate convention of Section~\ref{subsec:susy-hamiltonian-reduction}, the $\mathcal N=4$ measure contains
\begin{equation}
  d\mu_{\rm phys}
  =
  e^{-\rho}
  d\rho\,dx\,d\mu_{SU(2)}
  d\theta_-^1d\theta_-^2
  d\bar\theta_{-,1}d\bar\theta_{-,2}
  d\chi^1d\chi^2 ,
  \label{eq:n4-reduced-measure-main}
\end{equation}
up to an overall normalization constant.
What needs to be noticed in the $\mathcal N=4$ discussion is the factor $e^{-\rho}$: this factor plays an important role when translating the group propagator to Liouville quantum mechanics.
There one inserts $P(1,2)P(2,1)$, divides by the simultaneous global $G$ action on the two endpoints, and uses the symmetry to fix one representative of the two-point configuration.
In the present $\mathcal N=4$ theory $G=\mathrm{PSU}(1,1|2)$, while in the bosonic detour it is $SL(2,\mathbb R)$.
The invariant variables then reduce to the length, the Euler angle, and the two remaining $\chi$ variables, and the final answer is written with the flat $d\ell$ Liouville measure.
In our variables, after the same kind of gauge fixing, only one radial integration remains.
Suppressing the $SU(2)_{\mathfrak R}$ and fermionic contractions, the radial part has the schematic form
\begin{equation}
  \int
  \frac{d\mu_{\rm phys}(A)d\mu_{\rm phys}(B)}
       {\operatorname{Vol}(G)}\,
  P_{AB}P_{BA}
  \longrightarrow
  \int d\rho_B\,e^{-\rho_B}\,
  \mathcal R(\rho_B)^2 .
  \label{eq:gauge-fixed-radial-norm-main}
\end{equation}
The arrow means that a representative of the diagonal orbit has been chosen.
The same quotient-measure mechanism is the one used below in the endpoint formulas for three- and four-point functions.
In the same gauge the invariant length is $L=e^{-\ell/2}\propto e^{-\rho_B/2}$.
Thus
\begin{equation}
  \int d\rho_B\,e^{-\rho_B}\,
  \mathcal R(\rho_B)^2
  =
  \int d\ell\,
  \left[e^{-\rho_B/2}\mathcal R(\rho_B)\right]^2
  \sim
  \int d\ell\,
  \left[L\,\mathcal R(L)\right]^2 .
  \label{eq:lqm-norm-measure-factor-main}
\end{equation}
This is the useful way to remember the relation: the leftover measure factor is shared by the two propagators in the norm.
Each Liouville wavefunction therefore receives one extra factor $e^{-\rho_B/2}=L$ relative to the corresponding group propagator component.
For example, the $K_0$ component of the Liouville BPS wavefunction is $L^2K_0(2\mathfrak q L)$, with $L=e^{-\ell/2}$, rather than the $LK_0(2\mathfrak q L)$ appearing in the BPS propagator above.
With this measure dictionary, any diagonal propagator in the same supermultiplet can be obtained from the corresponding Liouville quantum mechanics wavefunction in \cite{Lin:2025wof} by the replacements
\begin{equation}
  e^{-\ell/2}\longrightarrow L,
  \qquad
  g^a{}_{b}\longrightarrow W^a{}_{b},
  \qquad
  (\psi_l,\bar\psi_l;\psi_r,\bar\psi_r)
  \longrightarrow
  (\eta_i,\partial_{\eta_i};\partial_{\eta_f},\eta_f),
  \label{eq:lqm-to-group-diagonal-dictionary-main}
\end{equation}
with the fixed charge restored in the Bessel function argument as $2e^{-\ell/2}\to2\mathfrak q L$, and with the same fermion polarization as in \eqref{eq:main-final-q-operators}--\eqref{eq:main-initial-q-operators}.
The composition of propagators will be discussed more structurally in Section~\ref{sec:correlators-length-kernel}.

\subsection{Bulk representations and Wilson-line operators}
\label{subsec:bulk-matter-boundary-operators}

The final ingredient needed for correlators is the operator inserted
between two endpoints of the particle on a group. We will not
repeat the bosonic derivation in Section \ref{sec:n0-jt-particle-main} here. The point of this
subsection is instead to isolate the new feature in the supersymmetric
case. A matter field carries a bulk $PSU(1,1|2)$ representation, while
the boundary dynamics is described by a reduced $\mathcal N=4$
super-Poincare representation. The Wilson line is the bridge between these two
languages, but this difference leads to a few subtleties that we will
try to resolve in this subsection.

\paragraph{Symmetries and representations.}
There are two representation spaces which should be kept separate. The
first one is the representation of $PSU(1,1|2)$ in which the bulk
matter operator transforms. We denote it by $\mathbf R$, and write
its vectors as $\mathbf v\in\mathbf R$; we sometimes also call these
vectors polarizations.
Concretely, the matter operator can be packaged as a superfield in
this representation, ${\cal O}_{\mathbf R}(X,\theta_\pm,\bar\theta_\pm)$.
Expanding it as a polynomial in the fermionic variables defines the
component fields, which are ordinary fields of $X$. The odd
generators $\mathsf F_{\pm,a},\bar{\mathsf F}_{\pm}^{a}$ act on the
same representation space and move one component field into another.
These component fields should still be distinguished from the vectors $\mathbf v$.
Each component field contains a tower of modes
transforming under the bosonic subalgebra
$SL(2,\mathbb R)\times SU(2)_{\mathfrak R}$. Choosing a component
and a mode in this tower is what gives a vector
$\mathbf v\in\mathbf R$. All such modes of all component fields
together form the $PSU(1,1|2)$ representation $\mathbf R$.\footnote{We also comment on terminology related to the fermionic generators. When we refer to a superconformal primary, we mean the vector killed by the positive generators, in particular by $\mathsf F_{+,a},\bar{\mathsf F}_{+}^{a}$. This is a mode of the leading component field. Other component fields in the same multiplet often contain $\mathsf L_+$-primary modes, but they need not contain an $\mathsf F_+$-primary mode.}

The second representation is the one carried by the particle on a
group, whose propagator was discussed above. This Hilbert space carries
a $\mathcal N=4$ super-Poincare representation. We write a state
label schematically as $\mathbf s=(s,j,\Phi)$. Although
this representation is closely related to $PSU(1,1|2)$, the particle
on a group must also obey the fixed positive-charge constraints, so not
every group-element configuration is allowed. Therefore $\mathbf s$
is acted on by the physical worldline supercharges of
Section~\ref{subsec:physical-worldline-supercharges}, rather than by
the bare $\mathsf F_{\pm,a},\bar{\mathsf F}_{\pm}^{a}$ generators.

\paragraph{Component bilocals: primaries, compensators, and descendants.}

We now spell out what the preceding covariance statement means for actual component
two-point functions.  In fixed Euclidean $\mathrm{AdS}_2$, before the boundary mode is made
dynamical, a component matter two-point function is fixed by the isometries.  If the
component carries $SL(2)$ weight $h$ and transforms in an $SU(2)_{\mathfrak R}$ representation
of spin $j$, its boundary limit has the schematic form
\begin{equation}
  G^{(h,j)}{}_{m}{}^{n}(i,j)
  =
  C_{h,j}\,
  L_{ij}^{\,2h}\,
  \bigl[D^{(j)}(U_{ij})\bigr]_{m}{}^{n},
  \label{eq:component-primary-green}
\end{equation}
where $L_{ij}$ is the invariant boundary length, $U_{ij}\in SU(2)_{\mathfrak R}$ is the
R-symmetry Wilson-line group element between the two endpoints, and
$D^{(j)}(U_{ij})$ is its matrix in the spin-$j$ representation.  For $j=0$
this reduces to a single power of $L_{ij}$.  For the fundamental representation,
$j=\frac12$, it is proportional to $L_{ij}(U_{ij})^a{}_b$.  The corresponding
full matter Wilson-line matrix element is
\begin{equation}
  W^{\mathbf R}_{ij}(\mathbf v_i,\mathbf v_j)
  =
  \langle \mathbf v_i|\mathbf R(\mathcal G_i^{-1}\mathcal G_j)|\mathbf v_j\rangle .
  \label{eq:component-wilson-line}
\end{equation}
Here $\mathcal G_i$ and $\mathcal G_j$ are the reduced endpoint group elements, and
$\mathbf v_i,\mathbf v_j\in \mathbf R$ are matter polarizations.  They select the component field and
the mode of that component at the two endpoints.  After the Hamiltonian reduction this
fixed-background statement is not automatic, so let us first focus on the simple cases
where it survives in the reduced variables.  The simple one-term form in
\eqref{eq:component-primary-green} is guaranteed only
for polarizations on which the positive generators used in the Hamiltonian reduction
act trivially.  More explicitly, suppose the endpoint polarization is annihilated by
the positive subgroup,
\begin{equation}
  \mathsf L_+\mathbf v=0,
  \qquad
  \mathsf F_{+,p}\mathbf v=0,
  \qquad
  \bar{\mathsf F}_{+}^{p}\mathbf v=0,
  \label{eq:positive-primary-condition}
\end{equation}
with the corresponding dual condition at the bra endpoint.  Then the compensating
factor $h_A(X;\epsilon)\in \mathbf{N}_+^{\rm susy}$ in
\begin{equation}
  e^{\epsilon A}\mathcal G(X)
  =
  \mathcal G(X+\epsilon\delta_A X)\,h_A(X;\epsilon)
  +
  O(\epsilon^2)
  \label{eq:positive-compensator-refactorization}
\end{equation}
acts trivially on the endpoint polarizations.  In that case the component bilocal is
obtained by simply evaluating the Wilson line on the reduced invariants.
For the bottom $SU(2)_{\mathfrak R}$-singlet matter component, this is the direct supersymmetric analogue of \eqref{eq:main-n0-schwarzian-bilocal}:
\begin{equation}
  \begin{aligned}
  \mathcal O^{\rm scalar}_\Delta(i,j)
  &=
  \left[
    \frac{\Pi_i\Pi_j}{\widetilde X_{ij}^{\,2}}
  \right]^\Delta
  =
  L_{ij}^{2\Delta},
  \\
  \widetilde X_{ij}
  &=
  x_j-x_i-\theta^a_{-,i}\bar\theta_{-,j,a}
  -\bar\theta_{-,i,a}\theta^a_{-,j},
  \qquad
  L_{ij}
  =
  \frac{e^{-\frac12(\rho_i+\rho_j)}}{\widetilde X_{ij}} .
  \end{aligned}
  \label{eq:scalar-bilocal-operator}
\end{equation}
Here the source condition has been used to replace the endpoint super-Jacobians by the reduced length $L_{ij}$.
The scalar insertion used in Section~\ref{sec:correlators-length-kernel} is this simplest case.

This simplification should not be extended to arbitrary components of a matter
multiplet.  A component obtained from another one by a fermionic generator is still
computed from the same Wilson line, but the generator has to be represented on the
reduced variables.  The point is that the bare action of $\mathsf F_{-,p}$, or of its conjugate,
does not preserve the fixed-source representative.  It must be completed by the same
positive compensator which appeared in the construction of the physical worldline
supercharges.  Thus, at the level of reduced bilocals, one should not write
``act with the bare $\mathsf F_-$ on the old answer.''  The descendant bilocal is
therefore the result of the same compensated worldline action used in
Section~\ref{subsec:physical-worldline-supercharges}.

A simple short-multiplet example makes the distinction explicit.  Let the reduced
throat multiplet contain a fermionic component whose primary two-point function is
\begin{equation}
  \mathcal O^{\rm moduli}_{\psi}{}^a{}_b(i,f)
  =
  L_{if}\,W^a{}_b(i,f).
  \label{eq:fermion-primary-bilocal}
\end{equation}
This is the one-term answer for a primary polarization with $h=\frac12$ and
$j=\frac12$.  The bosonic supercharge descendant in the same short multiplet is
obtained by applying the corresponding reduced supercharges to the Wilson line.  In
the normalization where the purely bosonic term has coefficient one, the useful model
operator is
\begin{equation}
  \mathcal O^{\rm moduli}_{\rm marg}(i,f)
  =
  L_{if}^{\,2}
  +
  i\,L_{if}\,
  \eta_f^a W_a{}^b(i,f)\eta_{i,b}.
  \label{eq:berry-marginal-model-bilocal}
\end{equation}
Equation \eqref{eq:berry-marginal-model-bilocal} is not a new primary formula.  It is the
descendant of \eqref{eq:fermion-primary-bilocal}.  The second term is the finite
effect of the compensating positive transformation in the physical worldline
supercharge.

Interestingly, this also matches the Liouville quantum mechanics derivation.
In the $\mathcal N=4$ Liouville variables of \cite{Lin:2025wof}, the
spin-$\frac12$ Wilson-line component is
\begin{equation}
  \mathcal O_{1/2}
  =
  e^{-\frac12\ell}\,
  D^{1/2}_{\frac12,\frac12}(g).
  \label{eq:n4-lqm-spin-half-operator}
\end{equation}
Acting with the supercharges which give the bosonic descendant gives
\begin{equation}
  -i\{\bar Q_{l,1},[Q_r^2,\mathcal O_{1/2}]\}
  =
  e^{-\ell}
  +
  i e^{-\frac12\ell}\,
  \psi_l^a D^{1/2}_{a}{}^{b}(g)\psi_{r,b}.
  \label{eq:n4-lqm-descendant-normalization-check}
\end{equation}
This is the Liouville version of \eqref{eq:berry-marginal-model-bilocal}.\footnote{The
$\mathcal N=2$ submultiplet gives the same one-line check:
for $\mathcal O_{1/2}=e^{-\frac12\ell-ia}$, the reduced supercharges give
$-i\{\bar Q_l,[Q_r,\mathcal O_{1/2}]\}
=e^{-\ell}+i e^{-\frac12\ell-ia}\bar\psi_l\psi_r$.}
This example will be useful in the Berry-phase discussion of
Section~\ref{sec:discussion-future}.

\section{Correlators from the supersymmetric length kernel}
\label{sec:correlators-length-kernel}

We now turn the particle-on-a-group formalism into an algorithm for computing correlators.
The basic strategy is to use the propagators and Wilson-line operators constructed in Section~\ref{sec:worldline-supercharges-propagator}, glue them using the reduced particle measure reviewed in Section~\ref{sec:n0-jt-particle-main}, and rewrite the result in terms of invariant length variables.
In this language the fundamental object is the three-length composition kernel $I_3$.
Section~\ref{subsec:I3-from-propagator-gluing} explains how this kernel arises from gluing two propagators at a common endpoint.
The bosonic kernel is the familiar JT $I_3$ kernel, while its $\mathcal N=2$ and $\mathcal N=4$ analogues are obtained by applying the same gluing procedure to the supersymmetric propagators.
Once this kernel is known, scalar three-point functions follows almost directly from length integrals: we compute the zero-energy $\mathcal N=2$ scalar three-point function in closed Gamma-function form, and the $\mathcal N=4$ BPS scalar three-point function as a finite sum of balanced ${}_4F_3(1)$'s.

Section~\ref{subsec:scalar-four-point-function} applies the same length-space algorithm to four-point functions.
The time-ordered channel is obtained by iterating the three-point kernel, while the crossed channel gives the out-of-time-ordered correlator.
We derive the zero-energy scalar OTOC for the crossed bilocal pairing $\mathcal O_\Delta(1,3)\mathcal O_\Delta(2,4)$ and then study its large-$\Delta$ approximation.
In the $\mathcal N=2$ case the resulting length integral can be evaluated explicitly as a finite hypergeometric sum, while in the $\mathcal N=4$ BPS case the same method gives a controlled saddle approximation.
The integral manipulations used in this section are collected in Appendix~\ref{app:IBP-four-point-kernels}, and Appendix~\ref{app:N0-I3-gamma-Wilson} shows that the identical length-integral method reproduces the classic $\mathcal N=0$ JT OTOC.

\subsection{The composition kernel and scalar three-point functions}
\label{subsec:I3-from-propagator-gluing}

\paragraph{The composition kernel.}
It is useful to start from the bosonic $\mathcal N=0$ problem.  Let $P^{\mathcal N=0}_{ij}$ be the propagator for
an interval with endpoints $i,j$.  In the fixed-charge sector it has the form
$P^{\mathcal N=0}_{ij}=e^{-qY_{ij}}\Psi_{\rm WdW}^{\mathcal N=0}(L_{ij})$.
The composition law is
\begin{equation}
  \int d\mu_{\rm phys}(j)\,
  P^{\mathcal N=0}_{ij}P^{\mathcal N=0}_{jk}
  =
  P^{\mathcal N=0}_{ik}.
  \label{eq:section4-composition-recall}
\end{equation}
To extract the length kernel, we quotient the common global symmetry and use
$L_{ij},L_{jk}$ as the two integration variables, with $L_{ik}$ fixed by the
external endpoints.  In these variables the product of phases can be written as
\begin{equation}
  e^{-q Y_{ij}}e^{-q Y_{jk}}
  =
  e^{-q Y_{ik}}\,
  \exp\!\left[
    -q
    \left(
      \frac{L_{ij}L_{jk}}{L_{ik}}
      +
      \frac{L_{jk}L_{ik}}{L_{ij}}
      +
      \frac{L_{ik}L_{ij}}{L_{jk}}
    \right)
  \right]
	  ,
	  \label{eq:section4-phase-exponent}
	\end{equation}
	The left-hand side is the phase part of the two propagators being composed.
	On the right-hand side, the endpoint phase $e^{-qY_{ik}}$
	combines with $\Psi_{\rm WdW}^{\mathcal N=0}(L_{ik})=K_{2is}(2q L_{ik})$
	to give the bosonic JT propagator $P_{ik}^{\mathcal N=0}$.
	The remaining exponential on the right-hand side is the phase contribution that
	stays inside the length integral; together with the reduced length measure, it
	defines the three-length composition kernel.  In other words, after the quotient the measure and phase
	part takes the form\footnote{Equation~\eqref{eq:section4-I3-extraction} should be read as the result of the two-fold integral over the intermediate endpoint.  Passing to the two length variables $L_{ij},L_{jk}$ also brings in the Jacobian of this change of variables, which is included in the displayed $I_3$ normalization.}
\begin{equation}
  d\mu_{\rm phys}(j)\,
  e^{-q Y_{ij}}e^{-q Y_{jk}}
  =
  \frac{dL_{ij}}{L_{ij}}\frac{dL_{jk}}{L_{jk}}\,
  e^{-q Y_{ik}}\,
  I_3^{\mathcal N=0}(L_{ij},L_{jk},L_{ik}),
  \label{eq:section4-I3-extraction}
\end{equation}
The bosonic kernel is
\begin{equation}
  I_3^{\mathcal N=0}(L_{ij},L_{jk},L_{ik})
  =
  \frac{1}{\pi}\,
  \exp\!\left[
    -q
    \left(
      \frac{L_{ij}L_{jk}}{L_{ik}}
      +
      \frac{L_{jk}L_{ik}}{L_{ij}}
      +
      \frac{L_{ik}L_{ij}}{L_{jk}}
    \right)
  \right].
  \label{eq:section4-bosonic-I3}
\end{equation}
This is the bosonic Schwarzian $I_3$ kernel
\cite{Mertens:2017mtv,Blommaert:2018oro}.

The supersymmetric calculation follows the same logic, but the factor multiplying
the WdW wavefunction contains fermions and R-symmetry data.  The simpler
$\mathcal N=2$ zero-energy calculation makes this point explicit.  Translated
to the present length convention, the $j=0$ propagator of
\cite{Lin:2022zxd} is
\begin{equation}
  P^{\mathcal N=2}_{0,ij}
  =
  e^{-\mathfrak qY_{ij}}
  \left(
    \eta_j e^{-\frac{i}{2}\Sigma_{ij}}
    +
    \eta_i e^{\frac{i}{2}\Sigma_{ij}}
  \right)
  \Psi_0^{\mathcal N=2}(L_{ij}),
  \label{eq:section4-N2-j0-propagator}
\end{equation}
where the $j=0$ normalization is fixed by the norm convention described above, and $\Psi_0^{\mathcal N=2}(L)=\frac{2\mathfrak q L}{\sqrt{\pi}}K_{1/2}(2\mathfrak q L)=\sqrt{\mathfrak q L}\,e^{-2\mathfrak q L}$.
The analogue of \eqref{eq:section4-I3-extraction} is obtained by stripping off
the two radial factors $\Psi_0^{\mathcal N=2}$ and doing the $U(1)_{\mathfrak R}$
and Grassmann part of the intermediate-endpoint integral:
\begin{equation}
  \begin{aligned}
  &\int_{U(1)_{\mathfrak R},{\rm Gr.}}\!\!
  d\mu_{\rm phys}^{\mathcal N=2}(j)\,
  e^{-\mathfrak qY_{ij}}
  \left(
    \eta_j e^{-\frac{i}{2}\Sigma_{ij}}
    +
    \eta_i e^{\frac{i}{2}\Sigma_{ij}}
  \right)
  e^{-\mathfrak qY_{jk}}
  \left(
    \eta_k e^{-\frac{i}{2}\Sigma_{jk}}
    +
    \eta_j e^{\frac{i}{2}\Sigma_{jk}}
  \right)
  \\
  &\qquad =
  \frac{dL_{ij}}{L_{ij}}\frac{dL_{jk}}{L_{jk}}\,
  e^{-\mathfrak qY_{ik}}
  \left(
    \eta_k e^{-\frac{i}{2}\Sigma_{ik}}
    +
    \eta_i e^{\frac{i}{2}\Sigma_{ik}}
  \right)
  I_3^{\mathcal N=2}(L_{ij},L_{jk},L_{ik}) .
  \end{aligned}
  \label{eq:section4-N2-I3-extraction}
\end{equation}
This defines the corresponding three-length kernel.  Its value is
\begin{equation}
  \begin{aligned}
  I_3^{\mathcal N=2}(L_{ij},L_{jk},L_{ik})
  &=
  \exp\!\left[
    -\mathfrak q
    \left(
      \frac{L_{ij}L_{jk}}{L_{ik}}
      +
      \frac{L_{jk}L_{ik}}{L_{ij}}
      +
      \frac{L_{ik}L_{ij}}{L_{jk}}
    \right)
  \right]
  \\
  &\quad\times
  \left[
    -1
    +
    \frac{2\mathfrak q}{L_{ij}L_{jk}L_{ik}}
    \left(
      L_{ij}L_{jk}
      +
      L_{jk}L_{ik}
      +
      L_{ik}L_{ij}
    \right)^2
  \right].
  \end{aligned}
  \label{eq:section4-N2-I3}
\end{equation}
This uses the same length convention as the $\mathcal N=4$ formula below:
the Bessel function argument is $2\mathfrak q L$.

For $\mathcal N=4$, the kernel contains two component matrices,
\begin{align}
  \left(\mathbb I^{(0)}_{ijk}\right)_{00}
  &=
  -\frac{\mathfrak q^2}{4}
  \left(
    \frac{L_{ij}L_{jk}}{L_{ik}}
    +\frac{L_{jk}L_{ik}}{L_{ij}}
    +\frac{L_{ik}L_{ij}}{L_{jk}}
  \right)^2
  +\frac{5\mathfrak q}{4}
  \left(
    \frac{L_{ij}L_{jk}}{L_{ik}}
    +\frac{L_{jk}L_{ik}}{L_{ij}}
    +\frac{L_{ik}L_{ij}}{L_{jk}}
  \right)
  -1
  \nonumber\\
  &\quad
  -\mathfrak q^2
  \left(
    L_{ij}^2+L_{jk}^2+L_{ik}^2
  \right),
  \nonumber\\
  \left(\mathbb I^{(0)}_{ijk}\right)_{01}
  &=
  \mathfrak q L_{jk}
  \left[
  2-\mathfrak q
  \left(
  3\frac{L_{ij}L_{ik}}{L_{jk}}
  +\frac{L_{jk}L_{ik}}{L_{ij}}
  +\frac{L_{ij}L_{jk}}{L_{ik}}
  \right)
  \right],
  \nonumber\\
  \left(\mathbb I^{(0)}_{ijk}\right)_{10}
  &=
  \mathfrak q L_{ij}
  \left[
  2-\mathfrak q
  \left(
  3\frac{L_{jk}L_{ik}}{L_{ij}}
  +\frac{L_{ij}L_{ik}}{L_{jk}}
  +\frac{L_{ij}L_{jk}}{L_{ik}}
  \right)
  \right],
  \nonumber\\
  \left(\mathbb I^{(0)}_{ijk}\right)_{11}
  &=
  \mathfrak q L_{ik}
  \left[
  1-\mathfrak q
  \left(
  3\frac{L_{ij}L_{jk}}{L_{ik}}
  +\frac{L_{jk}L_{ik}}{L_{ij}}
  +\frac{L_{ik}L_{ij}}{L_{jk}}
  \right)
  \right].
  \label{eq:section4-I3-matrix-zero}
\end{align}
and
\begin{align}
  \left(\mathbb I^{(1)}_{ijk}\right)_{00}
  &=
  \mathfrak q L_{ik}
  \left[
  -2+\mathfrak q
  \left(
  3\frac{L_{ij}L_{jk}}{L_{ik}}
  +\frac{L_{jk}L_{ik}}{L_{ij}}
  +\frac{L_{ik}L_{ij}}{L_{jk}}
  \right)
  \right],
  \nonumber\\
  \left(\mathbb I^{(1)}_{ijk}\right)_{01}
  &=
  \mathfrak q L_{ij}
  \left[
  -1+\mathfrak q
  \left(
  3\frac{L_{jk}L_{ik}}{L_{ij}}
  +\frac{L_{ij}L_{ik}}{L_{jk}}
  +\frac{L_{ij}L_{jk}}{L_{ik}}
  \right)
  \right],
  \nonumber\\
  \left(\mathbb I^{(1)}_{ijk}\right)_{10}
  &=
  \mathfrak q L_{jk}
  \left[
  -1+\mathfrak q
  \left(
  3\frac{L_{ij}L_{ik}}{L_{jk}}
  +\frac{L_{jk}L_{ik}}{L_{ij}}
  +\frac{L_{ij}L_{jk}}{L_{ik}}
  \right)
  \right],
  \nonumber\\
  \left(\mathbb I^{(1)}_{ijk}\right)_{11}
  &=
  \frac{\mathfrak q^2}{4}
  \left(
    \frac{L_{ij}L_{jk}}{L_{ik}}
    +\frac{L_{jk}L_{ik}}{L_{ij}}
    +\frac{L_{ik}L_{ij}}{L_{jk}}
  \right)^2
  -\frac{\mathfrak q}{4}
  \left(
    \frac{L_{ij}L_{jk}}{L_{ik}}
    +\frac{L_{jk}L_{ik}}{L_{ij}}
    +\frac{L_{ik}L_{ij}}{L_{jk}}
  \right)
  -\frac14
  \nonumber\\
  &\quad
  +\mathfrak q^2
  \left(
    L_{ij}^2+L_{jk}^2+L_{ik}^2
  \right).
  \label{eq:section4-I3-matrix-one}
\end{align}
Combining these component matrices with the common exponential factor gives the $\mathcal N=4$ length kernel,
\begin{equation}
  I_3^{\mathcal N=4}(L_{ij},L_{jk},L_{ik})
  =
  \frac{
  \exp\!\left[-\mathfrak q
  \left(
    \frac{L_{ij}L_{jk}}{L_{ik}}
    +
    \frac{L_{jk}L_{ik}}{L_{ij}}
    +
    \frac{L_{ik}L_{ij}}{L_{jk}}
  \right)\right]
  }{2\mathfrak q L_{ik}}
  \left(\mathbb I^{(0)}_{ijk},\mathbb I^{(1)}_{ijk}\right).
  \label{eq:section4-N4-I3-pair}
\end{equation}

\paragraph{Normalization, and three-point functions.}
The first place the kernel appears is the composition law.  Schematically, after quotienting the common symmetry and passing to length variables,
\[
  \int d\mu_{\rm phys}(j)\,
  P_{ij}^{\rm BPS}P_{jk}^{\rm BPS}
  \;\longrightarrow\;
  I_3^{\mathcal N=4}(L_{ij},L_{jk},L_{ik})\,
  P_{ik}^{\rm BPS}.
\]
Since the BPS propagator has two fermionic structures, this single composition statement splits into two radial projections.  With the normalization in \eqref{eq:section4-N4-I3-pair}, these projections are
\begingroup
\small
\setlength{\arraycolsep}{1pt}
\begin{align}
  &\int_0^\infty \frac{dL_{ij}}{L_{ij}}\frac{dL_{jk}}{L_{jk}}\,
  \frac{e^{-\mathfrak q(\cdots)}}{2\mathfrak q L_{ik}}\!
  \begin{pmatrix}(2\mathfrak q L_{ij})^2K_0(2\mathfrak q L_{ij})&(2\mathfrak q L_{ij})^2K_1(2\mathfrak q L_{ij})\end{pmatrix}
  \!\mathbb I^{(0)}_{ijk}\!
  \begin{pmatrix}(2\mathfrak q L_{jk})^2K_0(2\mathfrak q L_{jk})\\ (2\mathfrak q L_{jk})^2K_1(2\mathfrak q L_{jk})\end{pmatrix}
  =
  -2\mathfrak q L_{ik}K_0(2\mathfrak q L_{ik}),
  \label{eq:section4-I3-zero-action}
  \\
  &\int_0^\infty \frac{dL_{ij}}{L_{ij}}\frac{dL_{jk}}{L_{jk}}\,
  \frac{e^{-\mathfrak q(\cdots)}}{2\mathfrak q L_{ik}}\!
  \begin{pmatrix}(2\mathfrak q L_{ij})^2K_0(2\mathfrak q L_{ij})&(2\mathfrak q L_{ij})^2K_1(2\mathfrak q L_{ij})\end{pmatrix}
  \!\mathbb I^{(1)}_{ijk}\!
  \begin{pmatrix}(2\mathfrak q L_{jk})^2K_0(2\mathfrak q L_{jk})\\ (2\mathfrak q L_{jk})^2K_1(2\mathfrak q L_{jk})\end{pmatrix}
  =
  2\mathfrak q L_{ik}K_1(2\mathfrak q L_{ik}).
  \label{eq:section4-I3-one-action}
\end{align}
\endgroup
Equations \eqref{eq:section4-I3-zero-action} and \eqref{eq:section4-I3-one-action} are the practical check of the normalization.

We now turn from the composition law to correlators.
We first focus on BPS three-point functions.
For a scalar three-point function, the matter input is the SCFT three-point
structure.
Before reducing to length variables, the corresponding gluing formula has the same form as the four-point formula below; one should first quotient by the common diagonal orbit, or equivalently choose a representative and include the quotient Jacobian in the reduced length measure:
\begin{equation}
  \mathcal A_3^{\Delta_1,\Delta_2,\Delta_3}
  =
  \int
  \frac{\prod_{a=1}^{3}d\mu_{\rm phys}(a)}
       {\operatorname{Vol}(G)}\,
  P_{12}^{\rm BPS}
  P_{23}^{\rm BPS}
  P_{31}^{\rm BPS}\,
  \mathcal O_{\Delta_1,\Delta_2,\Delta_3}(1,2,3).
  \label{eq:section4-direct-three-point-gluing}
\end{equation}
Equivalently, one can fix one endpoint and include the quotient Jacobian in the reduced length measure.
The object $\mathcal O_{\Delta_1,\Delta_2,\Delta_3}(1,2,3)$ is the matter three-point structure, which should be read from the boundary SCFT.
For three scalar dimensions $\Delta_1,\Delta_2,\Delta_3$, set $\alpha_{12}=\Delta_1+\Delta_2-\Delta_3$, $\alpha_{23}=\Delta_2+\Delta_3-\Delta_1$, and $\alpha_{31}=\Delta_3+\Delta_1-\Delta_2$.
For the scalar component considered here, using the scalar boundary data of Section~\ref{subsec:bulk-matter-boundary-operators}, this SCFT structure becomes a length-space monomial,
\begin{equation}
  \mathcal O_{\Delta_1,\Delta_2,\Delta_3}(1,2,3)
  =
  C_{123}\,
  \frac{\Pi_1^{\Delta_1}\Pi_2^{\Delta_2}\Pi_3^{\Delta_3}}
       {\widetilde X_{12}^{\alpha_{12}}
        \widetilde X_{23}^{\alpha_{23}}
        \widetilde X_{31}^{\alpha_{31}}}
  =
  C_{123}
  L_{12}^{\alpha_{12}}
  L_{23}^{\alpha_{23}}
  L_{31}^{\alpha_{31}} .
  \label{eq:section4-scalar-three-point-operator}
\end{equation}
The integrated three-point function is then obtained by multiplying this length-space matter structure by one composition kernel and integrating over the three lengths.  For the $\mathcal N=2$ zero-energy example above, the integral is
\begin{equation}
  \begin{aligned}
  \mathcal A_3^{\mathcal N=2}
  =
  C_{123}\int
  \frac{dL_{12}}{L_{12}}
  \frac{dL_{23}}{L_{23}}
  \frac{dL_{31}}{L_{31}}\,
  \Psi_0^{\mathcal N=2}(L_{12})
  \Psi_0^{\mathcal N=2}(L_{23})
  \Psi_0^{\mathcal N=2}(L_{31})\,
  I_3^{\mathcal N=2}(L_{12},L_{23},L_{31})\,
  L_{12}^{\alpha_{12}}
  L_{23}^{\alpha_{23}}
  L_{31}^{\alpha_{31}} .
  \end{aligned}
  \label{eq:section4-N2-integrated-three-point}
\end{equation}
The same Bessel and Gamma-integral techniques reviewed in
Appendix~\ref{app:N0-I3-gamma-Wilson} evaluate this integral directly.\footnote{The $K_{1/2}$ wavefunction is special:
$K_{1/2}(z)=\sqrt{\pi/(2z)}e^{-z}$.  This turns the intermediate
length integrals into a particularly simple Gamma integral.}
The answer is
\begin{equation}
  \mathcal A_3^{\mathcal N=2}
  =
  C_{123}\,
  \frac{\sqrt{\pi}}{2^{2(\Delta_1+\Delta_2+\Delta_3)+1}}\,
  \mathfrak q^{-(\Delta_1+\Delta_2+\Delta_3)}
  \frac{
  \Gamma(2\Delta_1+1)
  \Gamma(2\Delta_2+1)
  \Gamma(2\Delta_3+1)}
  {\Gamma(\Delta_1+\Delta_2+\Delta_3+1)} .
  \label{eq:section4-N2-three-point-answer}
\end{equation}
For the $\mathcal N=4$ BPS sector, the same statement keeps the two $K_0/K_1$ components:
\begingroup
\small
\begin{equation}
  \mathcal A_3^{\mathcal N=4}
  =
  C_{123}\int
  \frac{dL_{12}}{L_{12}}\frac{dL_{23}}{L_{23}}\frac{dL_{31}}{L_{31}}\,
  \left[
  \vec\Psi_{s_1}(L_{12})^T\otimes
  \vec\Psi_{s_2}(L_{23})^T\otimes
  \vec\Psi_{s_3}(L_{31})^T
  \right]\cdot
  I_3^{\mathcal N=4}(L_{12},L_{23},L_{31})\,
  L_{12}^{\alpha_{12}}L_{23}^{\alpha_{23}}L_{31}^{\alpha_{31}} .
  \label{eq:section4-N4-integrated-three-point}
\end{equation}
\endgroup
The dot denotes the finite component contraction fixed by the BPS wavefunctions and by the two entries of $I_3^{\mathcal N=4}$.
Using the same Bessel and Gamma-integral techniques as in Appendix~\ref{app:N0-I3-gamma-Wilson}, each term in \eqref{eq:section4-N4-integrated-three-point} reduces to a six-pair Gamma integral.
After the final spectral integral the answer is a finite sum of balanced ${}_4F_3(1)$ functions,
\begin{equation}
  \mathcal A_3^{\mathcal N=4}
  =
  C_{123}\,
  \sum_{\ell=1}^{24}
  \mathcal C_\ell(\alpha_{12},\alpha_{23},\alpha_{31})\,
  {}_4F_3\!\left(\vec a_\ell(\alpha_{12},\alpha_{23},\alpha_{31});
  \vec b_\ell(\alpha_{12},\alpha_{23},\alpha_{31});1\right),
  \label{eq:section4-N4-three-point-answer}
\end{equation}
where $\vec a_\ell$, $\vec b_\ell$, and $\mathcal C_\ell$ are explicit linear and Gamma-function expressions in $(\alpha_{12},\alpha_{23},\alpha_{31})$.
The full expression is recorded in the Mathematica supplement.

\subsection{The scalar four-point function}
\label{subsec:otoc-longer-form}
\label{subsec:scalar-four-point-function}

We now spell out the four-point calculation in a form which will be useful for comparison between $\mathcal N=0$, $\mathcal N=2$, and $\mathcal N=4$.
For equal scaling dimensions, the time-ordered four-point correlator is
\begin{equation}
  \mathcal A_{4,\mathrm{TOC}}^\Delta
  =
  \int
  \frac{\prod_{a=1}^{4}d\mu_{\rm phys}(a)}
       {\operatorname{Vol}(G)}\,
  P_{12}
  P_{23}
  P_{34}
  P_{41}\,
  \mathcal O_\Delta(1,2)\mathcal O_\Delta(3,4).
  \label{eq:section4-long-toc-gluing}
\end{equation}
Since the two bilocals do not cross, this is just the product of the two corresponding two-point functions.  The hard part is the crossed pairing, which gives the OTOC.
The OTOC expression is
\begin{equation}
  \mathcal A_{4,\mathrm{OTOC}}^\Delta
  =
  \int
  \frac{\prod_{a=1}^{4}d\mu_{\rm phys}(a)}
       {\operatorname{Vol}(G)}\,
  P_{12}
  P_{23}
  P_{34}
  P_{41}\,
  \mathcal O_\Delta(1,3)\mathcal O_\Delta(2,4).
  \label{eq:section4-long-endpoint-gluing}
\end{equation}
From here on we focus on the OTOC; later in this subsection we specialize to the equal-scaling-dimension case.
We use hyperbolic geometry to determine the two diagonal lengths $L_{13}$ and $L_{24}$:
\begin{equation}
  \frac{1}{L_{13}L_{24}}
  =
  \frac{1}{L_{12}L_{34}}
  +
  \frac{1}{L_{23}L_{41}},
  \qquad
  L_{13}L_{24}
  =
  \frac{L_{12}L_{23}L_{34}L_{41}}
       {L_{12}L_{34}+L_{23}L_{41}} .
  \label{eq:section4-long-diagonal-relation}
\end{equation}
These diagonal lengths appear when we write the two scalar operators in the length basis.  For two equal scalar dimensions,
\begin{equation}
  \mathcal O_\Delta(1,3)\mathcal O_\Delta(2,4)
  =
  (L_{13}L_{24})^{2\Delta}
  =
  \left(
    \frac{L_{12}L_{23}L_{34}L_{41}}
         {L_{12}L_{34}+L_{23}L_{41}}
  \right)^{2\Delta},
  \label{eq:section4-long-spectator}
\end{equation}
The useful feature of the equal-dimension scalar case is that the product of scalar bilocals is independent of the gluing length $L_{13}$.

\paragraph{\texorpdfstring{$\mathcal N=0$}{N=0} detour.}
In ordinary JT gravity the three-length kernel is the scalar kernel
$I_3^{\mathcal N=0}$ in \eqref{eq:section4-bosonic-I3}.
In dimensionless lengths, gluing two copies along $L_{13}$ gives
\begin{equation}
  \begin{aligned}
  I_4(L_{12},L_{23},L_{34},L_{41})
  &=
  \int_0^\infty\frac{dL_{13}}{L_{13}}\,
  I_3^{\mathcal N=0}(L_{12},L_{23},L_{13})
  I_3^{\mathcal N=0}(L_{34},L_{41},L_{13})
  \\
  &=
  \frac{2}{\pi^2}
  K_0\!\left(2\mathcal R_{1234}\right),
  \label{eq:section4-long-bosonic-I4}
  \\
  \mathcal R_{1234}
  &=
  \sqrt{
    \frac{
    (L_{12}L_{23}+L_{34}L_{41})
    (L_{12}L_{34}+L_{23}L_{41})
    (L_{12}L_{41}+L_{23}L_{34})}
    {L_{12}L_{23}L_{34}L_{41}} } .
  \end{aligned}
\end{equation}
so that
\begin{equation}
  \begin{aligned}
  \mathcal A_4^{\mathcal N=0}
  &=
  \int
  \frac{dL_{12}}{L_{12}}
  \frac{dL_{23}}{L_{23}}
  \frac{dL_{34}}{L_{34}}
  \frac{dL_{41}}{L_{41}}\,
  \Psi_{s_1}(L_{12})
  \Psi_{s_2}(L_{23})
  \Psi_{s_3}(L_{34})
  \Psi_{s_4}(L_{41})
  \\
  &\hspace{1.0cm}\times
  \mathcal O_\Delta(1,3)\mathcal O_\Delta(2,4)\,
  I_4(L_{12},L_{23},L_{34},L_{41}) .
  \end{aligned}
  \label{eq:section4-long-N0-length-block}
\end{equation}
The details of the length calculation are given in Appendix~\ref{app:N0-I3-gamma-Wilson}.
After the length integrals and some standard manipulations, this reproduces the JT OTOC kernel with a constant prefactor,
\begin{equation}
  \mathcal A_4^{\mathcal N=0}
  \sim
  \mathcal A_4^{\rm JT},
  \label{eq:section4-long-N0-final}
\end{equation}
where $\mathcal A_4^{\rm JT}$ is the usual Schwarzian OTOC \cite{Stanford:2017thb,Mertens:2017mtv,Blommaert:2018oro}.
The supersymmetric cases use the same length geometry, with the scalar bosonic kernel replaced by the $K_0/K_1$ matrix structure produced by the supercharges.

\paragraph{\texorpdfstring{$\mathcal N=2$}{N=2} analogue.}
The zero-energy $\mathcal N=2$ OTOC is given by
\begin{equation}
  \begin{aligned}
  \mathcal A_4^{\mathcal N=2}
  &=
  \int
  \frac{dL_{12}}{L_{12}}
  \frac{dL_{23}}{L_{23}}
  \frac{dL_{34}}{L_{34}}
  \frac{dL_{41}}{L_{41}}\,
  \mathcal O_\Delta(1,3)\mathcal O_\Delta(2,4)
  \prod_{(ab)=(12),(23),(34),(41)}
  \frac{2L_{ab}}{\pi}\,
  K_{1/2}(2L_{ab})
  \\
  &\quad\times
  \int_0^\infty\frac{dL_{13}}{L_{13}}\,
  I_3^{\mathcal N=2}(L_{12},L_{23},L_{13})\,
  I_3^{\mathcal N=2}(L_{34},L_{41},L_{13}) .
  \end{aligned}
  \label{eq:section4-N2-pre-IBP-two-I3}
\end{equation}
Compared with the bosonic case, the rational factor in $I_3^{\mathcal N=2}$ given in \eqref{eq:section4-N2-I3} produces powers of the diagonal length $L_{13}$ and of the adjacent lengths $L_{12},L_{23},L_{34},L_{41}$.
To cast the result in a form familiar from the bosonic calculation, we perform an IBP reduction which is explained in Appendix~\ref{app:I3-derivation-normalization}.
The result is a finite numerator multiplying the same scalar-bilocal product and the same $K_0$ four-length kernel:
\begin{equation}
  \begin{aligned}
  \mathcal A_4^{\mathcal N=2}
  &=
  \int
  \frac{dL_{12}}{L_{12}}
  \frac{dL_{23}}{L_{23}}
  \frac{dL_{34}}{L_{34}}
  \frac{dL_{41}}{L_{41}}\,
  \left(
    T_1+T_2+T_3+T_4+T_5
  \right)
  \mathcal O_\Delta(1,3)\mathcal O_\Delta(2,4)
  \\
  &\quad\times
  K_0\!\left(2\mathcal R_{1234}\right)
  \prod_{(ab)=(12),(23),(34),(41)}
  \frac{2L_{ab}}{\pi}\,
  K_{1/2}(2L_{ab}) .
  \end{aligned}
  \label{eq:section4-N2-IBP-integrand}
\end{equation}
Here $\mathcal R_{1234}$ is the length combination in \eqref{eq:section4-long-bosonic-I4}, and
\begin{align}
  T_1
  &=
  4\Delta(2\Delta+1)
  \frac{L_{12}L_{23}L_{34}L_{41}}
       {(L_{12}L_{34}+L_{23}L_{41})^2},
  \label{eq:section4-N2-T1}
  \\
  T_2
  &=
  4\Delta(2\Delta+1)
  \frac{L_{23}L_{34}+L_{12}L_{41}}
       {L_{12}L_{34}+L_{23}L_{41}},
  \label{eq:section4-N2-T2}
  \\
  T_3
  &=
  2\Delta(2\Delta+1),
  \label{eq:section4-N2-T3}
  \\
  T_4
  &=
  -\frac{2(2\Delta+1)}
        {L_{12}L_{34}+L_{23}L_{41}}
  \left[
    L_{12}L_{23}L_{34}
    +
    L_{23}L_{34}L_{41}
    +
    L_{12}(L_{23}+L_{34})L_{41}
  \right],
  \label{eq:section4-N2-T4}
  \\
  T_5
  &=
  \frac{2}
       {L_{12}L_{34}+L_{23}L_{41}}
  \left[
    L_{12}L_{34}(L_{12}+L_{34})
    +
    L_{23}L_{41}(L_{23}+L_{41})
  \right].
  \label{eq:section4-N2-T5}
\end{align}

The resulting length integral can be evaluated completely.
The steps are explained in Appendix~\ref{app:N0-I3-gamma-Wilson}; here we only note that the $K_{1/2}$ structure simplifies the calculation substantially.
Writing
\begin{equation}
  \mathcal A_4^{\mathcal N=2}
  =
  \sum_{r=1}^{5}A_r(\Delta),
  \label{eq:section4-N2-sum}
\end{equation}
one finds
\begin{align}
  \pi\,2^{4\Delta} A_1(\Delta)
  &=
  \frac{\Delta\,\Gamma(1+2\Delta)^4}{4\,\Gamma(2+4\Delta)}
  \,{}_2F_1\!\left(
    \begin{matrix}
      2+2\Delta,\;2+2\Delta\\
      3+4\Delta
    \end{matrix}
    ;-1
  \right),
  \label{eq:section4-N2-A1}
  \\
  \pi\,2^{4\Delta} A_2(\Delta)
  &=
  \frac{\Gamma(1+2\Delta)^3\Gamma(2+2\Delta)}
       {4\,\Gamma(2+4\Delta)}
  \,{}_2F_1\!\left(
    \begin{matrix}
      1+2\Delta,\;1+2\Delta\\
      3+4\Delta
    \end{matrix}
    ;-1
  \right),
  \label{eq:section4-N2-A2}
  \\
  \pi\,2^{4\Delta} A_3(\Delta)
  &=
  \frac{\Delta(1+2\Delta)\Gamma(1+2\Delta)^4\Gamma(1+4\Delta)}
       {2\,\Gamma(2+4\Delta)^2}
  \,{}_3F_2\!\left(
    \begin{matrix}
      1+2\Delta,\;1+2\Delta,\;1+4\Delta\\
      2+4\Delta,\;2+4\Delta
    \end{matrix}
    ;-1
  \right),
  \label{eq:section4-N2-A3}
  \\
  \pi\,2^{4\Delta} A_4(\Delta)
  &=
  -\frac{(1+2\Delta)\Gamma(1+2\Delta)^2\Gamma(2+2\Delta)^2\Gamma(2+4\Delta)}
        {4\,\Gamma(3+4\Delta)^2}
  \,{}_3F_2\!\left(
    \begin{matrix}
      1+2\Delta,\;1+2\Delta,\;2+4\Delta\\
      3+4\Delta,\;3+4\Delta
    \end{matrix}
    ;-1
  \right),
  \label{eq:section4-N2-A4}
  \\
  \pi\,2^{4\Delta} A_5(\Delta)
  &=
  \frac{\Gamma(1+2\Delta)^2\Gamma(2+2\Delta)^2\Gamma(1+4\Delta)}
       {4\,\Gamma(2+4\Delta)\Gamma(3+4\Delta)}
  \,{}_3F_2\!\left(
    \begin{matrix}
      1+2\Delta,\;1+2\Delta,\;1+4\Delta\\
      2+4\Delta,\;3+4\Delta
    \end{matrix}
    ;-1
  \right).
  \label{eq:section4-N2-A5}
\end{align}

\begin{figure}[t]
  \centering
  \includegraphics[width=0.94\textwidth]{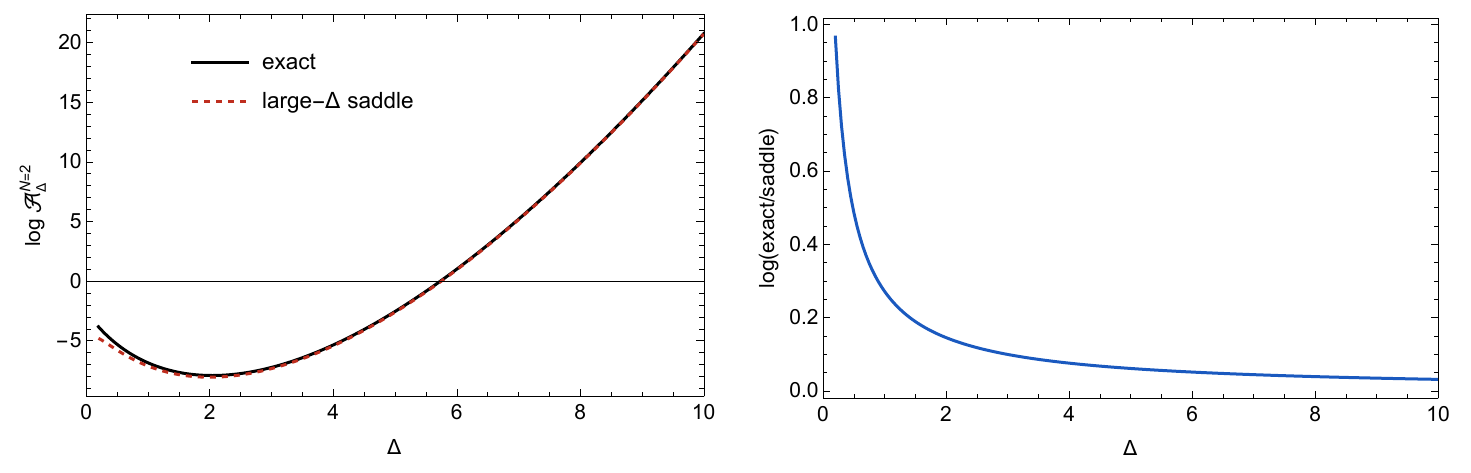}
  \caption{The zero-energy $\mathcal N=2$ OTOC as a function of $\Delta$, plotted on a logarithmic scale. Left: $\log\mathcal A_4^{\mathcal N=2}$, where the solid curve is the exact finite sum $A_1+\cdots+A_5$, and the dashed curve is the logarithm of the large-$\Delta$ saddle estimate in \eqref{eq:section4-N2-large-delta}. Right: the logarithmic residual $\log \mathcal A_4^{\mathcal N=2}-\log \mathcal A_{4,\rm 1-loop}^{\mathcal N=2}$, which is small on the scale of the left panel but visible after zooming in.}
  \label{fig:n2-otoc-delta}
\end{figure}

This exact expression can be used as a benchmark: when an analytic integral as precise as this is hard to obtain, it tells us how large the deviation is under a given approximation.
A useful approximation is to study a large-$\Delta$ saddle with equal lengths.
The dominant configuration is the symmetric one,
\[
  L_{12}=L_{23}=L_{34}=L_{41}=\lambda .
\]
The leading saddle sits at $\lambda_\ast=(2-\sqrt2)\Delta$, and the Gaussian fluctuations around this symmetric configuration give the large-$\Delta$ estimate
\begin{equation}
  \mathcal A_{4,\rm 1-loop}^{\mathcal N=2}
  \sim
  2^{-1/4}\sqrt\pi\,\Delta^{3/2}
  \left(
    \frac{(2-\sqrt2)\Delta}{\sqrt2\,e}
  \right)^{4\Delta}.
  \label{eq:section4-N2-large-delta}
\end{equation}
Figure \ref{fig:n2-otoc-delta} compares the logarithm of the exact answer \eqref{eq:section4-N2-sum} with the logarithm of this saddle approximation.
The right panel resolves the logarithmic residual $\log \mathcal A_4^{\mathcal N=2}-\log \mathcal A_{4,\rm 1-loop}^{\mathcal N=2}$.
Thus the two curves can nearly overlap in the left panel while the right panel still shows a visible residual: the left panel has the much larger vertical range of the logarithmic amplitude, whereas the right panel zooms in on the log-ratio itself.
Near $\Delta=1$, the one-loop saddle estimate is only about $30\%$ away from the exact answer which shows the large $\Delta$ approximation, with one-loop corrections, is actually not a bad approximation, and the agreement improves quickly as $\Delta$ is increased.

\paragraph{\texorpdfstring{$\mathcal N=4$}{N=4} BPS length integral and saddle.}
We now return to the $\mathcal N=4$ BPS sector.
The direct analogue of \eqref{eq:section4-N2-pre-IBP-two-I3} keeps the two components of the $I_3^{\mathcal N=4}$ kernel.
Suppressing the external $K_0/K_1$ indices of the BPS wavefunctions, the pre-IBP length integral is
\begin{equation}
  \begin{aligned}
  \mathcal A_4^{\mathcal N=4}
  &=
  \int
  \frac{dL_{12}}{L_{12}}
  \frac{dL_{23}}{L_{23}}
  \frac{dL_{34}}{L_{34}}
  \frac{dL_{41}}{L_{41}}\,
  \left[
    \vec\Psi_{s_1}(L_{12})^T
    \otimes
    \vec\Psi_{s_2}(L_{23})^T
    \otimes
    \vec\Psi_{s_3}(L_{34})^T
    \otimes
    \vec\Psi_{s_4}(L_{41})^T
  \right]\!\cdot
  \\
  &\quad\times
  \mathcal O_\Delta(1,3)\mathcal O_\Delta(2,4)
  \int_0^\infty \frac{dL_{13}}{L_{13}}\,
  I_3^{\mathcal N=4}(L_{12},L_{23},L_{13})^T\,
  I_3^{\mathcal N=4}(L_{34},L_{41},L_{13}) .
  \end{aligned}
  \label{eq:section4-N4-pre-IBP-two-I3}
\end{equation}
Here $I_3^{\mathcal N=4}$ is the two-component kernel in \eqref{eq:section4-N4-I3-pair}.

The expression \eqref{eq:section4-N4-pre-IBP-two-I3} is not yet a useful final form.
As in the $\mathcal N=2$ calculation, one can simplify it by integration by parts; the needed identities are collected in Appendix~\ref{app:I3-derivation-normalization}.
A useful check is obtained by setting the scalar-bilocal product to one.
Then the IBP identities reduce the two $I_3^{\mathcal N=4}$ kernels to the composition law of the BPS propagator, with the $K_0$ and $K_1$ pieces treated separately and with the boundary terms accounted for as in Appendix~\ref{app:I3-derivation-normalization}.
For the OTOC, the same IBP operation also differentiates the scalar-bilocal product $(L_{12}L_{23}L_{34}L_{41}/(L_{12}L_{34}+L_{23}L_{41}))^{2\Delta}$.
The result is a finite sum of ordinary length integrals with the same exponential factors and Bessel functions, but with shifted powers of the lengths and with a finite $K_0/K_1$ combination structure.
From this form one can either try to do the full analytic length integral term by term,\footnote{\label{fn:n4-witten-integrals}A typical term after the IBP has the schematic form
\[
  \int_0^\infty\frac{dL_{13}}{L_{13}}\,
  E_{123}E_{341}\,
  L_{13}^{m_{13}}
  \prod_{a=12,23,34,41}
  L_a^{m_a}K_0(2L_a),
\]
with $E_{ijk}$ defined in \eqref{eq:app-I3-exponential-factor}; other terms replace some external $K_0$'s by $K_1$'s or shift the integer powers $m_a$.
After the $L_{13}$-integral this gives ordinary JT Witten-diagram type integrals, schematically
\[
  \prod_{a=12,23,34,41}L_a^{m_a}K_0(2L_a)\,
  K_0(2\mathcal R_{1234})
\]
or the corresponding $\mathcal R_{1234}^{-1}K_1(2\mathcal R_{1234})$ form.  In this sense the Witten diagram is of the following crossed shape,
\[
\begin{tikzpicture}[baseline=-0.6ex, line cap=round, line join=round, scale=1.28]
  \def\r{1.0}
  \draw[thick] (90:\r) arc (90:0:\r);
  \draw[thick] (0:\r) arc (0:-90:\r);
  \draw[thick] (-90:\r) arc (-90:-180:\r);
  \draw[thick] (180:\r) arc (180:90:\r);
  \draw[line width=1.32pt] (180:\r) -- (0:\r);
  \draw[thick] (-90:\r) -- (90:\r);
  \node[font=\scriptsize] at (0.45,0.22) {$\Delta+m_{13}$};
  \node[font=\scriptsize] at (-0.18,-0.42) {$\Delta$};
  \draw[thick] (90:\r) -- (180:\r);
  \draw[thick] (-90:\r) -- (0:\r);
  \node[font=\scriptsize] at (-0.55,0.50) {$m_{41}$};
  \node[font=\scriptsize] at (0.55,-0.50) {$m_{23}$};
  \node[font=\scriptsize] at (45:1.35) {$K_0(2L_{12})$};
  \node[font=\scriptsize] at (-45:1.35) {$K_0(2L_{23})$};
  \node[font=\scriptsize] at (-135:1.35) {$K_0(2L_{34})$};
  \node[font=\scriptsize] at (135:1.35) {$K_0(2L_{41})$};
\end{tikzpicture}
\]
Following the Appendix~\ref{app:N0-I3-gamma-Wilson} route opens these integrals into Mellin--Barnes variables, but for the generic $\mathcal N=4$ terms we do not find a collapse to a single-fold Mellin--Barnes integral.  Therefore the term-by-term answer is not simply a single ${}_pF_q$.  A better IBP basis may still combine the terms more efficiently, but we have not found such a basis yet.}
or extract the large-$\Delta$ saddle.
Below we take the second route.

The full length integral has not yet been reduced to a closed energy-space crossing kernel, but the large-$\Delta$ saddle can already be extracted from this length representation.
The dominant configuration is the same symmetric one,
\[
  L_{12}=L_{23}=L_{34}=L_{41}=\lambda,
  \qquad
  \lambda_\ast=(2-\sqrt2)\Delta .
\]
Thus the exponential part of the answer is unchanged from the $\mathcal N=2$ calculation.
The difference between the $\mathcal N=4$ and $\mathcal N=2$ integrands is then contained in the finite polynomial in the lengths multiplying the same saddle exponent.
Evaluating this polynomial at the symmetric saddle, and including the Gaussian factors, gives the one-loop saddle estimate
\begin{equation}
  \mathcal A_{4,\rm 1-loop}^{\mathcal N=4}
  \sim
  \frac{2^{3/4}\pi^{9/2}}{16}
  (111-78\sqrt2)\,
  \Delta^{15/2}
  \left(
    \frac{(2-\sqrt2)\Delta}{\sqrt2\,e}
  \right)^{4\Delta}.
  \label{eq:section4-N4-large-delta}
\end{equation}
This gives a useful comparison with the zero-energy $\mathcal N=2$ answer,
\begin{equation}
  \frac{
  \mathcal A_{4,\rm 1-loop}^{\mathcal N=4}}
  {\mathcal A_{4,\rm 1-loop}^{\mathcal N=2}}
  \sim
  \frac{\pi^4}{8}(111-78\sqrt2)\,\Delta^6 .
  \label{eq:section4-N4-over-N2-large-delta}
\end{equation}
This $\Delta^6$ enhancement is an interesting fact, and we believe it comes from a shorter effective wormhole length in the higher supersymmetry case.

\section{Discussion and future directions}
\label{sec:discussion-future}

\paragraph{An OPE/representation-theoretic way to derive the OTOC.}\hfill\par\noindent

We now describe an alternative representation-theoretic route to the four-point
function, following the discussion of Wilson functions and polynomials in
\cite{Jafferis:2022wez,Jafferis:2022uhu}.  The reason to bring this up is that
our formalism supplies one of the main ingredients, which is the three-point
functions.  This suggests that a closed-form expression for the
super-JT OTOC may be within reach.

The time-ordered channel is the cleanest place to see what data are needed.
We write an OPE representation of the time-ordered four-point function: in
the $12|34$ channel it takes the form
\begin{equation}
  \mathcal A_{4,\mathrm{TOC}}^{12|34}
  =
  \sum_{n=0}^{\infty}
  \mathcal A_{3,n}^{12}(s_1,s_2,s_3)\,
  \eta_n\,
  \bigl(\Gamma_{13}^{h_n}\bigr)^{-1}\,
  \mathcal A_{3,n}^{34}(s_1,s_4,s_3),
  \label{eq:discussion-raw-A3-toc}
\end{equation}
Equation~\eqref{eq:discussion-raw-A3-toc} should be read as requiring three
pieces of information.
\begin{enumerate}
\item \emph{The channel sum.}  One must know which exchanged representations
appear in the OPE.  For identical bosonic scalars this is
\begin{equation}
  D^+_{\Delta}\otimes D^+_{\Delta}
  =
  \bigoplus_{n=0}^{\infty}D^+_{2\Delta+n},
  \qquad h_n=2\Delta+n .
  \label{eq:discussion-bosonic-sl2-decomposition}
\end{equation}
In the notation of \cite{Jafferis:2022wez,Jafferis:2022uhu},
this is the same $n$-sum appearing in the Wilson-polynomial identity
resolution $\sum_n P_n^{\Delta,\Delta}(s_2;s_1,s_3)
P_n^{\Delta,\Delta}(s_4;s_1,s_3)$.

\item \emph{The gluing kernel.}  In our raw $\mathcal A_3$ normalization, the
object sitting between the two three-point vertices is
\[
  \eta_n\,\bigl(\Gamma_{13}^{h_n}\bigr)^{-1},
  \qquad
  \Gamma_{13}^{h}
  :=
  \frac{\Gamma(h\pm i s_1\pm i s_3)}{\Gamma(2h)} .
\]
Here $\Gamma_{13}^{h}$ is the Schwarzian two-point leg of the exchanged
operator.  The scalar part of the gluing kernel is
\begin{equation}
  \eta_n
  =
  \frac{\Gamma(h_n)^2}{n!\,(4\Delta+n-1)_n}
  =
  \frac{\Gamma(2\Delta+n)^2}{n!\,(4\Delta+n-1)_n}.
  \label{eq:discussion-raw-A3-eta}
\end{equation}
The denominator is the norm of the ordered double-trace primary
$[O_\Delta O_\Delta]_n$\footnote{\[
  [O_\Delta O_\Delta]_n^{12}
  =
  \sum_{k=0}^n
  (-1)^k \binom{n}{k}
  \frac{(2\Delta+n-k)_k}{(2\Delta)_k}
  \partial^k O_1\,\partial^{n-k}O_2 .
\]}, while the two powers of $\Gamma(h_n)$ match the
normalization of the raw length-integral three-point vertex.  In
\cite{Jafferis:2022wez,Jafferis:2022uhu} this gluing kernel is hidden by
the normalized Wilson-polynomial basis:
their $P_n$'s include the Wilson norm, so the TOC identity resolution has no
visible $\eta_n(\Gamma_{13}^{h_n})^{-1}$ between the two factors.

\item \emph{The three-point function.}  In our notation the three-point block in
\eqref{eq:discussion-raw-A3-toc} is the fixed-energy three-point function for
unit-normalized operators
\begin{equation}
  \begin{aligned}
  \mathcal A_{3,n}^{12}(s_1,s_2,s_3)
  =
  &\int
  \frac{dL_{12}}{L_{12}}
  \frac{dL_{23}}{L_{23}}
  \frac{dL_{31}}{L_{31}}\,
  \Psi_{s_1}(L_{12})
  \Psi_{s_2}(L_{23})
  \Psi_{s_3}(L_{31})\\
  &\quad\times
  I_3^{\mathcal N=0}(L_{12},L_{23},L_{31})\,
  V_n(L_{12},L_{23},L_{31}) ,
  \end{aligned}
  \label{eq:discussion-bosonic-A3n}
\end{equation}
with $V_n=L_{12}^{-n}L_{23}^{2\Delta+n}L_{31}^{2\Delta+n}$.  This is the
analog, with a different normalization, of the
Wilson-polynomial $P_n^{\Delta,\Delta}(s_2;s_1,s_3)$ in the notation of
\cite{Jafferis:2022wez,Jafferis:2022uhu}.
\end{enumerate}

The OTOC requires one more piece of data, namely the exchange action on the same
channel space:
\begin{equation}
  \mathcal A_{4,\mathrm{OTOC}}^{12|34}
  =
  \sum_{n=0}^{\infty}
  \mathcal A_{3,n}^{12}(s_1,s_2,s_3)\,
  \eta_n\,\lambda_n\,
  \bigl(\Gamma_{13}^{h_n}\bigr)^{-1}\,
  \mathcal A_{3,n}^{34}(s_1,s_4,s_3).
  \label{eq:discussion-raw-A3-otoc}
\end{equation}
For identical bosonic scalars this exchange coefficient is
\begin{equation}
  \lambda_n=(-1)^n .
  \label{eq:discussion-bosonic-exchange-sign}
\end{equation}
This comes from
$[O_\Delta O_\Delta]_n^{12}=(-1)^n[O_\Delta O_\Delta]_n^{21}$.

Now let us study the supersymmetric analogue using the same three ingredients.
The external insertions are components of graded matter multiplets, so a channel
label is no longer just a number $n$.  A convenient notation for small
$\mathcal N=4$ multiplets labels the superprimary by $[2J]_h$, where $h$
is its $SL(2)$ weight and $J$ is its highest $SU(2)_{\mathfrak R}$ spin;
we write $A[2J]_h$ for short multiplets and $L[2J]_h$ for long multiplets,
following \cite{Lee:2019uen}.  The analogue of
\eqref{eq:discussion-raw-A3-toc} has the schematic form
\begin{equation}
  \mathcal A_{4,\mathrm{TOC}}^{12|34}
  =
  \sum_{\Xi}\,
  \mathcal A_{3,\Xi,a}^{12}\,
  \mathcal G_{\Xi}^{ab}\,
  \mathcal A_{3,\Xi,b}^{34},
  \label{eq:discussion-susy-toc-schematic}
\end{equation}
where repeated tensor-structure indices are summed.  This formula should again
be read as requiring three pieces of information.
\begin{enumerate}
\item \emph{The channel sum.}  One must know which exchanged supermultiplets
$\Xi$ appear in the OPE of the two external multiplets, including possible
multiplicity spaces.  The representation-theory literature, in particular
\cite{Lee:2019uen}, supplies this part.  For example,
\[
  A[1]_{1/2}\otimes A[1]_{1/2}
  =
  A[2]_1\oplus \bigoplus_{m=0}^{\infty}L[0]_{1+m}.
\]
This example is multiplicity-free, but in general there can be multiplicities
from different tensor structures.

\item \emph{The gluing kernel.}  The factor
$\mathcal G_{\Xi}^{ab}$ is the supersymmetric analogue of
$\eta_n(\Gamma_{13}^{h_n})^{-1}$.  It contains the inverse two-point function
of the exchanged multiplet in the fixed-energy basis, together with an
energy-independent normalization factor.  In the supersymmetric case this factor
can become a matrix on the tensor-structure space.

\item \emph{The three-point function.}  The objects
$\mathcal A_{3,\Xi,a}^{12}$ and $\mathcal A_{3,\Xi,b}^{34}$ are the
fixed-energy three-point functions obtained from the length-space propagator
construction in this paper.  For genuine supermultiplets the three-point
function is more complicated than a product of powers of lengths, so the
composition kernel is more complicated than the scalar
$I_3^{\mathcal N=2,4}$ in
Section~\ref{subsec:I3-from-propagator-gluing}.  The logic of obtaining it from
propagators, however, is the same.
\end{enumerate}

The OTOC needs one further finite piece of data, the graded exchange map on the
same channel space:
\begin{equation}
  \mathcal A_{4,\mathrm{OTOC}}^{12|34}
  =
  \sum_{\Xi}\,
  \mathcal A_{3,\Xi,a}^{12}\,
  \mathcal G_{\Xi}^{ac}\,
  \Lambda_{\Xi,c}{}^{b}\,
  \mathcal A_{3,\Xi,b}^{34}.
  \label{eq:discussion-susy-otoc-schematic}
\end{equation}
Here $\Lambda_\Xi$ is a finite exchange matrix.  For component operators of
grading $\#_i=0,1$, it includes the ordinary grading sign
$(-1)^{\#_1\#_2}$.

To summarize, the problem of getting OTOC from this method is very concrete:
combine the two three-point functions computed by the present
propagator method with the supermultiplet decomposition and exchange data from
\cite{Lee:2019uen}, keeping careful track of grading and normalization.  That
data gives the supersymmetric TOC through
\eqref{eq:discussion-susy-toc-schematic} and the supersymmetric OTOC through
\eqref{eq:discussion-susy-otoc-schematic}.

\paragraph{D1--D5--P Berry curvature.}
Another direction is to study the Berry curvature of BPS black hole states.
Recently, Berry curvature was proposed as a diagnostic of chaos in degenerate BPS sectors \cite{Chen:2026vml}; see also related work \cite{deBoer:2008qe}.
These analyses are mostly formulated from the CFT point of view; here we study the same physics from gravity.
As mentioned in the introduction, our approach is to follow the flow from the UV region to the edge of the near-horizon throat, and then use dimensional reduction to map the calculation to a two-dimensional JT problem.

First recall the definition.
Let $\mathcal H_{\rm BPS}$ be the degenerate BPS Hilbert space at fixed charges, with projector $\Pi_{\rm BPS}=\sum_a|a\rangle\langle a|$.
The basic definition is that the non-Abelian Berry curvature is the curvature of this projector,
\begin{equation}
  (F_{IJ})_{ab}
  =
  \langle a|[\partial_{I}\Pi_{\rm BPS},
  \partial_{J}\Pi_{\rm BPS}]|b\rangle .
  \label{eq:discussion-berry-curvature-definition}
\end{equation}
Equivalently, after inserting a complete set of states outside $\mathcal H_{\rm BPS}$, the same curvature is
\begin{equation}
  (F_{IJ})_{ab}
  =
  \sum_{n\notin\mathcal H_{\rm BPS}}
  \frac{
  \langle a|V_{I}|n\rangle
  \langle n|V_{J}|b\rangle
  -
  \langle a|V_{J}|n\rangle
  \langle n|V_{I}|b\rangle}
  {(E_n-E_0)^2}.
  \label{eq:discussion-berry-spectral}
\end{equation}
Here $E_0$ is the common BPS energy and $V_{I}$ is the equal-time perturbing operator.
This is the form we use below; the equivalent Euclidean conformal-perturbation derivation of \cite{Baggio:2017aww} gives the same expression after spectral decomposition, with the two Euclidean time integrals producing the factor $(E_n-E_0)^{-2}$.

The system we have in mind is D1--D5--P.
Without the momentum charge, the bulk dual is $\mathrm{AdS}_3\times S^3\times M_4$, and the Berry curvature is a highly protected quantity determined by symmetry and the chiral-ring data \cite{deBoer:2008ss,Baggio:2012rr}.
With large momentum charge, the bulk dual is expected to be an extremal BTZ black hole times the same transverse dimensions \cite{Strominger:1996sh}.
In this case there is less symmetry, and the Berry curvature is not expected to be uniquely fixed.
We provide a gravity approach to computing it.
Since the low-energy spectrum of a (near-)BPS black hole is captured by super-JT gravity, we use the techniques developed in this paper to study the corresponding Berry curvature.
The index $I$ labels a direction in the D1--D5 conformal manifold, which has dimension $20$ for $M_4=T^4$ and $84$ for $M_4=K3$ \cite{LarsenMartinec:1999,David:1999ec}.
More concretely, the conformal-manifold deformation is represented by an exactly marginal operator
\begin{equation}
  \mathcal O_{I}^{\rm marg}
  \sim
  \left(
  \mathsf F^{(L)}_{-1/2}\mathsf F^{(R)}_{-1/2}\Phi_{I}
  \right),
  \qquad
  (h,\bar h)=(1,1),
  \label{eq:discussion-d1d5-marginal-descendant}
\end{equation}
where the superconformal primary obeys
\begin{equation}
  \Phi_{I}:
  \qquad
  (h,\bar h)=\left(\frac12,\frac12\right),
  \qquad
  (j,\bar j)=\left(\frac12,\frac12\right).
  \label{eq:discussion-d1d5-primary-quantum-numbers}
\end{equation}
At strong coupling $\mathcal O^{\rm marg}_{I}$ is dual to the neutral massless scalar $t_{I}$ \cite{Seiberg:1999xz,David:1999ec}.
\begin{equation}
  \delta S_{\rm CFT}
  =
  \sum_{I}\lambda^{I}
  \int d^2z\,\mathcal O^{\rm marg}_{I}(z,\bar z),
  \qquad
  V_{I}
  =
  \int d\sigma\,\mathcal O^{\rm marg}_{I}(0,\sigma).
  \label{eq:discussion-VI-def}
\end{equation}

We then propagate this boundary source to the edge of the $\mathrm{AdS}_2$ region.
This amounts to solving the massless scalar equation in the BTZ background
\[
  ds_3^2
  =
  -N^2dt^2+N^{-2}dr^2+r^2(d\varphi+N^\varphi dt)^2,
  \qquad
  N^2=
  \frac{(r^2-r_+^2)(r^2-r_-^2)}{r^2},
\]
with $N^\varphi=-r_+r_-/(r^2)$.
The source is a static zero mode, so we take $t_{I}(t,r,\varphi)=\lambda_{I}R(r)$, with
\[
  R(r)=c_1+\frac{c_2}{2(r_+^2-r_-^2)}
  \log\!\left(\frac{r^2-r_+^2}{r^2-r_-^2}\right)
\]
which solves\footnote{In the extremal limit the nonconstant solution becomes $(r^2-r_0^2)^{-1}$.}
\begin{equation}
  \partial_r\left[
  \frac{(r^2-r_+^2)(r^2-r_-^2)}{r}
  \partial_r R
  \right]=0 .
  \label{eq:discussion-btz-static-radial-equation}
\end{equation}
Regularity at the horizon removes the nonconstant branch, so the propagation to the throat is trivial.

We can now compute the moments of Berry curvature from $\mathcal N=4$ super-JT by a simple symmetry argument: with our convention the D1--D5--P momentum charge is carried by the left-moving sector, while the extremal BTZ geometry preserves the $PSU(1,1|2)$ superisometry in the right-moving sector.

In the JT description we should not expect to compute a prescribed matrix element of $F_{IJ}$ in a microscopic basis of $\mathcal H_{\rm BPS}$.
Constructing such a basis for individual black-hole microstates is difficult, despite substantial progress from cohomological constructions \cite{Grant:2008sk,Chang:2013fba,Chang:2022mjp,Choi:2022caq,Choi:2023znd,Budzik:2023vtr,Chang:2023zqk,Choi:2023vdm,Chang:2024zqi}.
The natural gravity observables are instead statistical and basis-independent quantities, such as traces of moments, and even the spectral statistics of the matrix $(F_{IJ})_{ab}$ as in \cite{Chen:2026vml}.
Here we fix two independent deformation directions $I,J$ and use $a,b$ to label BPS states.\footnote{In the specific D1--D5--P Berry curvature problem all directions in the moduli space are equivalent, so we can pick any two independent deformation directions.}
The simplest such basis-independent observable is
\begin{equation}
  \mathcal V_{IJ;IJ}
  =
  \frac{1}{D_{\rm BPS}}\,
  \overline{\operatorname{Tr}_{\rm BPS}
  \left(F_{IJ}
  F_{IJ}^{\dagger}\right)},
  \qquad
  D_{\rm BPS}=\dim\mathcal H_{\rm BPS},
  \label{eq:discussion-berry-variance}
\end{equation}
where the overline denotes the statistical average that the JT calculation is meant to approximate.

The next step is to translate this observable into explicit JT correlators.
Substituting \eqref{eq:discussion-berry-spectral} into \eqref{eq:discussion-berry-variance}, we obtain an average over four $V$ matrix elements, which translates into JT four-point functions.\footnote{See especially Eq.~(2.41) of \cite{Brown:2024ajk}, where a normalized trace with four energy projectors is identified with a product of four energy-basis matrix elements and the JT four-point function.}
Concretely, the two different orderings of $I,J$, namely $I,J,J,I$ and $I,J,I,J$, give
\begin{equation}
  \begin{aligned}
  \mathcal V_{IJ;IJ}
  &=
  \frac{2}{D_{\rm BPS}}
  \sum_{m,n\notin\mathcal H_{\rm BPS}}
  \sum_{a,b\in\mathcal H_{\rm BPS}}
  \frac{1}{(E_n-E_0)^2(E_m-E_0)^2}
  \\
  &\quad\times
  \left[
  \overline{
  \langle a|V_{I}|n\rangle
  \langle n|V_{J}|b\rangle
  \langle b|V_{J}|m\rangle
  \langle m|V_{I}|a\rangle}
  -
  \overline{
  \langle a|V_{I}|n\rangle
  \langle n|V_{J}|b\rangle
  \langle b|V_{I}|m\rangle
  \langle m|V_{J}|a\rangle}
  \right].
  \end{aligned}
  \label{eq:discussion-berry-moment-spectral}
\end{equation}
Since $V_{I}$ and $V_{J}$ correspond to two different deformation directions, there is no Wick contraction between them.
The two averages should therefore be translated into the following two JT gravity correlators:
\[
  \overline{
  \langle a|V_{I}|n\rangle
  \langle n|V_{J}|b\rangle
  \langle b|V_{J}|m\rangle
  \langle m|V_{I}|a\rangle}
  \quad\longleftrightarrow\quad
  \begin{tikzpicture}[baseline=-0.6ex, line cap=round, line join=round, scale=1.08]
  \def\r{1.0}
  \draw[thick] (90:\r) arc (90:0:\r);
  \draw[thick] (0:\r) arc (0:-90:\r);
  \draw[thick] (-90:\r) arc (-90:-180:\r);
  \draw[thick] (180:\r) arc (180:90:\r);
  \draw[thick] (90:\r) arc[start angle=180,end angle=270,radius=\r];
  \draw[thick] (180:\r) arc[start angle=90,end angle=0,radius=\r];
  \node[font=\scriptsize] at (90:1.28) {$V_{I}$};
  \node[font=\scriptsize] at (180:1.28) {$V_{J}$};
  \node[font=\scriptsize] at (-90:1.28) {$V_{J}$};
  \node[font=\scriptsize] at (0:1.28) {$V_{I}$};
  \node[font=\scriptsize] at (45:1.35) {$|{\rm BPS},a\rangle$};
  \node[font=\scriptsize] at (-45:1.35) {$|n\rangle$};
  \node[font=\scriptsize] at (-135:1.35) {$|{\rm BPS},b\rangle$};
  \node[font=\scriptsize] at (135:1.35) {$|m\rangle$};
  \end{tikzpicture}
\]
\[
  \overline{
  \langle a|V_{I}|n\rangle
  \langle n|V_{J}|b\rangle
  \langle b|V_{I}|m\rangle
  \langle m|V_{J}|a\rangle}
  \quad\longleftrightarrow\quad
  \begin{tikzpicture}[baseline=-0.6ex, line cap=round, line join=round, scale=1.08]
  \def\r{1.0}
  \draw[thick] (90:\r) arc (90:0:\r);
  \draw[thick] (0:\r) arc (0:-90:\r);
  \draw[thick] (-90:\r) arc (-90:-180:\r);
  \draw[thick] (180:\r) arc (180:90:\r);
  \draw[thick] (90:\r) -- (90:0.45);
  \draw[thick] (0:\r) -- (0:0.45);
  \draw[thick] (-90:\r) -- (-90:0.45);
  \draw[thick] (180:\r) -- (180:0.45);
  \draw[thick] (90:0.45) -- (-90:0.45);
  \draw[thick] (180:0.45) -- (0:0.45);
  \node[font=\scriptsize] at (90:1.28) {$V_{I}$};
  \node[font=\scriptsize] at (180:1.28) {$V_{J}$};
  \node[font=\scriptsize] at (-90:1.28) {$V_{I}$};
  \node[font=\scriptsize] at (0:1.28) {$V_{J}$};
  \node[font=\scriptsize] at (45:1.35) {$|{\rm BPS},a\rangle$};
  \node[font=\scriptsize] at (-45:1.35) {$|n\rangle$};
  \node[font=\scriptsize] at (-135:1.35) {$|{\rm BPS},b\rangle$};
  \node[font=\scriptsize] at (135:1.35) {$|m\rangle$};
  \end{tikzpicture}
\]
where the lines in the disk, denoted by $\mathcal O_{I}^{\rm marg}(i,j)$, are the Wilson-line two-point insertions for $V_{I}$ and $V_{J}$.
Following the propagator gluing method of Section~\ref{subsec:otoc-longer-form}, we obtain the corresponding four-point expressions
\begin{equation}
  \resizebox{0.93\textwidth}{!}{$\displaystyle
  \mathcal A_{4,\rm TOC}(E_n,E_m)=
  \int
  \frac{\prod_{a=1}^{4}d\mu_{\rm phys}(a)}
       {\operatorname{Vol}(G)}\,
  P_{12}^{\rm BPS}P_{23}^{\rm non\text{-}BPS}(E_n)
  P_{34}^{\rm BPS}P_{41}^{\rm non\text{-}BPS}(E_m)
  \bigl[\mathcal O_{I}^{\rm marg}(1,2)
  \mathcal O_{J}^{\rm marg}(3,4)
  +(I\leftrightarrow J)\bigr]$}
  \label{eq:discussion-berry-toc-gluing}
\end{equation}
\begin{equation}
  \resizebox{0.93\textwidth}{!}{$\displaystyle
  \mathcal A_{4,\rm OTOC}(E_n,E_m)=
  \int
  \frac{\prod_{a=1}^{4}d\mu_{\rm phys}(a)}
       {\operatorname{Vol}(G)}\,
  P_{12}^{\rm BPS}P_{23}^{\rm non\text{-}BPS}(E_n)
  P_{34}^{\rm BPS}P_{41}^{\rm non\text{-}BPS}(E_m)
  \bigl[\mathcal O_{I}^{\rm marg}(1,3)
  \mathcal O_{J}^{\rm marg}(2,4)
  +(I\leftrightarrow J)\bigr]$}
  \label{eq:discussion-berry-otoc-gluing}
\end{equation}
Here $P^{\rm non\text{-}BPS}(E)$ is the fixed-energy propagator for non-BPS states.

We will comment on the bilocal operator $\mathcal O_{I}^{\rm marg}$ below.
With the spectral normalization of \eqref{eq:discussion-berry-spectral}, the second moment is
\begin{equation}
  \mathcal V_{IJ;IJ}
  =
  \int dE_n\,dE_m\,
  \frac{
    \rho_{\rm non\text{-}BPS}(E_n)\rho_{\rm non\text{-}BPS}(E_m)}
  {D_{\rm BPS}(E_n-E_0)^2(E_m-E_0)^2}
  \left[
    \mathcal A_{4,\mathrm{TOC}}(E_n,E_m)
    -
    \mathcal A_{4,\mathrm{OTOC}}(E_n,E_m)
  \right].
  \label{eq:discussion-berry-second-moment-A4}
\end{equation}
The two BPS propagators in $\mathcal A_4$ include the sums over the two BPS labels $a,b$.\footnote{The explicit factor $D_{\rm BPS}=e^{S_0}$ in the denominator and the factor of $e^{S_0}$ hidden in $P$ combine to make $\mathcal V_{IJ;IJ}\sim e^{S_0}$.}

We now turn to the operator $\mathcal O^{\rm marg}$ and the supersymmetric selection rule.
The $\mathcal O_{I}^{\rm marg}$'s in \eqref{eq:discussion-berry-toc-gluing} and \eqref{eq:discussion-berry-otoc-gluing} are the $\mathrm{AdS}_2$ boundary-anchored two-point functions for the deformation $V_{I}$.
As discussed in Section~\ref{subsec:bulk-matter-boundary-operators}, for a primary component of weight $h$ and $SU(2)_{\mathfrak R}$ spin $j$, the invariant dependence is $L_{ij}^{2h}D^{(j)}(U_{ij})$.
The marginal operator is instead a supercharge descendant of the spin-$1/2$, dimension-$1/2$ primary, so its invariant expression is more complicated: the physical boundary supercharge includes the compensating transformation needed by the Hamiltonian reduction.
In the normalization of \eqref{eq:berry-marginal-model-bilocal}, this means $\mathcal O_{I}^{\rm marg}(i,j)=Q_iQ_j\bigl[L_{ij}D^{(1/2)}(U_{ij})\bigr]=L_{ij}^2+iL_{ij}\eta_j^aW_a{}^b(i,j)\eta_{i,b}$, where $Q_i$ is the physical boundary supercharge, not the naive bare $\mathsf F_-$ action.

This also tells us which non-BPS sector to use: there is a supersymmetry plus angular-momentum selection rule.
In any vertex $\langle a|V_{I}|n\rangle$, angular momentum conservation forces $n$ to have zero angular momentum.
There are two non-BPS $\mathcal N=4$ super-Poincare worldline multiplets in Section~\ref{subsec:supersymmetric-propagator} which contain such scalar states: one has maximum angular momentum $J=1/2$ and contains the $\Psi,\chi$ states, while the other has maximum angular momentum $J=1$ and contains the $\mathrm L$ state.
Since $V_{I}$ is a superdescendant of a spin-$1/2$ superprimary, the $\mathrm L$ state is ruled out, and we are left with
\begin{equation}
  \Pi_{\rm marg}
  =
  \Pi^{J=\frac12}_{\Psi,j=0}
  +
  \Pi^{J=\frac12}_{\chi,j=0}.
  \label{eq:discussion-marginal-projector}
\end{equation}

To summarize, the ingredients needed for the second moment are already present in this paper.
The BPS and $J=1/2$ supermultiplet propagators were derived in Section~\ref{subsec:supersymmetric-propagator}, while the marginal bilocal operator was discussed in Section~\ref{subsec:bulk-matter-boundary-operators}.
Combining these inputs with the four-point gluing formula \eqref{eq:discussion-berry-second-moment-A4} gives a concrete JT protocol for computing the second moment of the Berry curvature.
We emphasize that this is not limited to the D1--D5--P Berry curvature.
In principle the same procedure applies to any BPS black hole with a moduli space; D1--D5--P is just the concrete example for which we assemble the ingredients in this paper.

A more ambitious goal is indeed more chalenging, that is to quantify the exact chaotic behavior of $F_{IJ}$.
A natural diagnostic is the statistical distribution of its eigenvalues, but determining that distribution requires essentially all moments.
In the present formalism the $n$-th moment is a higher $2n$-point function of the same marginal bilocal.
It is not yet clear whether an all-$n$ formula is practical; if the length-integral representation admits a useful recursion relation, then there may be a systematic way forward.
This would provide a direct bridge between super-JT correlators and statistical diagnostics of chaos in the degenerate BPS Hilbert space.

\phantomsection
\section*{Acknowledgments}
I thank Ahmed Abdalla, Stefano Antonini, Anna Biggs, Andreas  Blommaert, Jan  Boruch, Yiming Chen, Gabriele Di Ubaldo, Elliott Gesteau, Luca Iliesiu, Henry Lin, Masamichi Miyaji, Arvin Shahbazi-Moghaddam, Elisabeth Tabor, Wayne Weng, Jiuci Xu, Cynthia Yan, and especially Sean Colin-Ellerin and Mykhaylo Usatyuk for encourangement, discussions and comments on the manuscript.
This work was supported in part by the Leinweber Institute for Theoretical Physics at UC Berkeley, and by the Department of Energy, Office of Science, Office of High Energy Physics through award DE-SC0025293 and QuantISED award DE-SC0019380.

\appendix

\section{Notations and conventions}
\label{app:notation}

This appendix collects notation used in the main text.
Bulk component fields are ordinary symbols such as $G_{\mu\nu}$,
$E_\mu{}^a$, and $\Psi_\mu$.
Bulk superfields are bold, for example $\mathbf E^A$ and $\mathbf \Phi$.
BF fields are calligraphic, for example $\mathcal A$ and $\mathcal X$, and
the full supergroup element is $\mathcal G$.

\subsection{(Super) Lie algebra and bilinear form}
\label{app:supergroup-conventions}

Algebra generators are sans-serif:
\[
  \mathsf L_-,\quad \mathsf L_0,\quad \mathsf L_+,\qquad
  \mathsf R,\quad \mathsf R_i,\qquad
  \mathsf F_{-,p},\quad \bar{\mathsf F}_{-}^{p},\quad
  \mathsf F_{+,p},\quad \bar{\mathsf F}_{+}^{p}.
\]
Here $\mathsf R$ is the $U(1)_{\mathfrak R}$ generator of the
$\mathcal N=2$ subalgebra.
For the $SU(2)_{\mathfrak R}$ part of the $\mathcal N=4$ algebra,
$p,q=1,2$ are fundamental indices and $i,j,k=1,2,3$ are adjoint indices.

\paragraph{\texorpdfstring{$SL(2)$}{SL(2)} basis and bracket.}
Throughout the paper, the bracket $\langle\,,\,\rangle$ is always the
gravity-normalized invariant bilinear form.
For the bosonic $SL(2)$ block we fix the basis by
\begin{equation}
  {}[\mathsf L_0,\mathsf L_\pm]=\pm\mathsf L_\pm,
  \qquad
  [\mathsf L_+,\mathsf L_-]=2\mathsf L_0 .
  \label{eq:sl2-commutators}
\end{equation}
The bosonic invariant form is normalized by
\begin{equation}
  \langle \mathsf L_0,\mathsf L_0\rangle=1,
  \qquad
  \langle \mathsf L_+,\mathsf L_-\rangle=2,
  \label{eq:sl2-gravity-pairing}
\end{equation}
with all other independent pairings vanishing.
In first-order JT variables we also use
\begin{equation}
  \mathsf P_{\hat u}=\frac{1}{2}(\mathsf L_-+\mathsf L_+),
  \qquad
  \mathsf P_{\hat r}=\mathsf L_0,
  \qquad
  \mathsf J=\frac{1}{2}(\mathsf L_- - \mathsf L_+),
  \label{eq:first-order-jt-basis}
\end{equation}
which gives
\begin{equation}
  \langle \mathsf P_a,\mathsf P_b\rangle=\delta_{ab},
  \qquad
  \langle \mathsf J,\mathsf J\rangle=-1 .
  \label{eq:first-order-jt-basis-pairing}
\end{equation}
This convention is twice the common representation-theory normalization of the
$SL(2)$ Casimir; it is the convention used everywhere below.

\paragraph{\texorpdfstring{$\mathcal N=2$}{N=2} algebra and pairing.}
The $\mathcal N=2$ subalgebra is generated by
\[
  \mathsf L_-,\quad \mathsf L_0,\quad \mathsf L_+,\qquad
  \mathsf R,\qquad
  \mathsf F_-,\quad \bar{\mathsf F}_-,\quad
  \mathsf F_+,\quad \bar{\mathsf F}_+ .
\]
The nonzero brackets and anticommutators are
\begin{equation}
\begin{gathered}
  {}[\mathsf L_0,\mathsf L_\pm]=\pm\mathsf L_\pm,
  \qquad
  [\mathsf L_+,\mathsf L_-]=2\mathsf L_0,
  \qquad
  [\mathsf R,\mathsf L_0]=[\mathsf R,\mathsf L_\pm]=0,\\
  [\mathsf L_0,\mathsf F_\pm]=\pm\frac12\mathsf F_\pm,
  \qquad
  [\mathsf L_0,\bar{\mathsf F}_\pm]=\pm\frac12\bar{\mathsf F}_\pm,
  \qquad
  [\mathsf R,\mathsf F_\pm]=\frac12\mathsf F_\pm,
  \qquad
  [\mathsf R,\bar{\mathsf F}_\pm]=-\frac12\bar{\mathsf F}_\pm,\\
  [\mathsf L_\pm,\mathsf F_\pm]
  =
  [\mathsf L_\pm,\bar{\mathsf F}_\pm]
  =
  0,
  \qquad
  [\mathsf L_\pm,\mathsf F_\mp]=-\mathsf F_\pm,
  \qquad
  [\mathsf L_\pm,\bar{\mathsf F}_\mp]=\bar{\mathsf F}_\pm,\\
  \{\mathsf F_\pm,\bar{\mathsf F}_\pm\}=\mathsf L_\pm,
  \qquad
  \{\mathsf F_+,\bar{\mathsf F}_-\}=\mathsf R-\mathsf L_0,
  \qquad
  \{\mathsf F_-,\bar{\mathsf F}_+\}=\mathsf R+\mathsf L_0,\\
  \{\mathsf F_\epsilon,\mathsf F_{\epsilon'}\}
  =
  \{\bar{\mathsf F}_\epsilon,\bar{\mathsf F}_{\epsilon'}\}
  =
  0,
  \qquad
  \epsilon,\epsilon'=\pm .
\end{gathered}
  \label{eq:n2-algebra}
\end{equation}

The invariant pairing is the gravity-normalized Casimir pairing.
In these conventions the nonzero pairings are
\begin{equation}
\begin{gathered}
  \langle \mathsf L_+,\mathsf L_-\rangle=2,
  \qquad
  \langle \mathsf L_0,\mathsf L_0\rangle=1,
  \qquad
  \langle \mathsf R,\mathsf R\rangle=-1,\\
  \langle \mathsf F_+,\bar{\mathsf F}_-\rangle=-2,
  \qquad
  \langle \bar{\mathsf F}_-,\mathsf F_+\rangle=2,
  \qquad
  \langle \mathsf F_-,\bar{\mathsf F}_+\rangle=-2,
  \qquad
  \langle \bar{\mathsf F}_+,\mathsf F_-\rangle=2 .
\end{gathered}
  \label{eq:n2-invariant-pairings}
\end{equation}
The signs in the odd pairings are the graded signs of the invariant form.
All other independent pairings vanish.

\paragraph{\texorpdfstring{$\mathcal N=4$}{N=4} algebra and pairing.}
For the full $\mathcal N=4$ particle we use the $PSU(1,1|2)$ algebra in an
$SU(2)_{\mathfrak R}$-covariant basis.
The even generators are
\[
  \mathsf L_-,\quad \mathsf L_0,\quad \mathsf L_+,\qquad
  \mathsf R_i\quad (i=1,2,3),
\]
and the odd generators are
\[
  \mathsf F_{\pm,p},\qquad
  \bar{\mathsf F}_{\pm}^{p},
  \qquad p=1,2 .
\]
The $SL(2)$ brackets are still \eqref{eq:sl2-commutators}.
The remaining nonzero brackets and anticommutators are
\begin{equation}
\begin{gathered}
  {}[\mathsf R_i,\mathsf R_j]
  =
  i\epsilon_{ijk}\mathsf R_k,
  \qquad
  [\mathsf R_i,\mathsf L_0]
  =
  [\mathsf R_i,\mathsf L_\pm]
  =
  0,\\
  [\mathsf L_0,\mathsf F_{\pm,p}]
  =
  \pm\frac12\mathsf F_{\pm,p},
  \qquad
  [\mathsf L_0,\bar{\mathsf F}_{\pm}^{p}]
  =
  \pm\frac12\bar{\mathsf F}_{\pm}^{p},
  \qquad
  [\mathsf R_i,\mathsf F_{\pm,p}]
  =
  \frac12(\sigma_i)^q{}_{p}\mathsf F_{\pm,q},
  \qquad
  [\mathsf R_i,\bar{\mathsf F}_{\pm}^{p}]
  =
  -\frac12(\sigma_i)^p{}_{q}\bar{\mathsf F}_{\pm}^{q},\\
  [\mathsf L_\pm,\mathsf F_{\pm,p}]
  =
  [\mathsf L_\pm,\bar{\mathsf F}_{\pm}^{p}]
  =
  0,
  \qquad
  [\mathsf L_\pm,\mathsf F_{\mp,p}]
  =
  -\mathsf F_{\pm,p},
  \qquad
  [\mathsf L_\pm,\bar{\mathsf F}_{\mp}^{p}]
  =
  \bar{\mathsf F}_{\pm}^{p}.
  \\[1mm]
  \{\mathsf F_{\pm,p},\bar{\mathsf F}_{\pm}^{q}\}
  =
  \delta_p{}^q\,\mathsf L_\pm,
  \qquad
  \{\mathsf F_{+,p},\bar{\mathsf F}_{-}^{q}\}
  =
  -\delta_p{}^q\,\mathsf L_0
  +(\sigma^i)^q{}_{p}\mathsf R_i,
  \qquad
  \{\mathsf F_{-,p},\bar{\mathsf F}_{+}^{q}\}
  =
  \delta_p{}^q\,\mathsf L_0
  +(\sigma^i)^q{}_{p}\mathsf R_i,
  \\[1mm]
  \{\mathsf F_{\epsilon,p},\mathsf F_{\epsilon',q}\}
  =
  \{\bar{\mathsf F}_{\epsilon}^{p},\bar{\mathsf F}_{\epsilon'}^{q}\}
  =
  0,
  \qquad
  \epsilon,\epsilon'=\pm .
\end{gathered}
  \label{eq:n4-algebra}
\end{equation}
Here $\sigma_i$ are the standard Pauli matrices.
The invariant bilinear form is the gravity-normalized Casimir pairing of the
projective algebra.\footnote{A useful warning is that one can easily write a
$3\times3$, or $2|1$, matrix realization for the
$\mathfrak{psu}(1,1|1)$ subalgebra, but one should not expect an analogous
honest $4\times4$, or $2|2$, matrix realization for
$\mathfrak{psu}(1,1|2)$.
This is why the small $\mathcal N=2$ check works, while the natural full $\mathcal N=4$ check is only projective.
After complexification the former is essentially $\mathfrak{sl}(2|1)$,
whereas the latter is
$\mathfrak{psl}(2|2)=\mathfrak{sl}(2|2)/\mathbb C\,\mathbf 1$.
In the $2|1$ case, $\operatorname{Str}\mathbf 1_{2|1}=2-1\neq0$, so the
identity is not an element of $\mathfrak{sl}(2|1)$.
The scalar-looking singlet in the odd anticommutator can therefore be absorbed
into the genuine $U(1)_{\mathfrak R}$ generator, giving literal identities
such as $\{\mathsf F_+,\bar{\mathsf F}_-\}=\mathsf R-\mathsf L_0$ and
$\{\mathsf F_-,\bar{\mathsf F}_+\}=\mathsf R+\mathsf L_0$.
In the $2|2$ case, $\operatorname{Str}\mathbf 1_{2|2}=2-2=0$, so the identity
lies inside $\mathfrak{sl}(2|2)$ and is central.
The simple algebra is the projective quotient, and the elementary $2|2$
matrices close only up to this central term:
$\{\mathsf F_{+,p},\bar{\mathsf F}_{-}^{q}\}
=-\delta_p{}^q\mathsf L_0+(\sigma^i)^q{}_{p}\mathsf R_i
+\frac12\delta_p{}^q\mathbf 1_{2|2}$.
The last term vanishes in $\mathfrak{psl}(2|2)$, but not as an ordinary $4\times4$ matrix.
Thus the obstruction is the special $m=n$ feature of $\mathfrak{sl}(m|n)$, not a bad basis choice.
Accordingly the pairing below is defined from the projective Casimir, not from an ordinary $4\times4$ trace.}
In these conventions the nonzero pairings are
\begin{equation}
\begin{gathered}
  \langle \mathsf L_+,\mathsf L_-\rangle=2,
  \qquad
  \langle \mathsf L_0,\mathsf L_0\rangle=1,
  \qquad
  \langle\mathsf R_i,\mathsf R_j\rangle=-\delta_{ij},
  \qquad
  \langle\mathsf F_{+,p},\bar{\mathsf F}_{-}^{q}\rangle=-2\delta_p{}^q,
  \qquad
  \langle\bar{\mathsf F}_{-}^{q},\mathsf F_{+,p}\rangle=2\delta_p{}^q,\\
  \langle\mathsf F_{-,p},\bar{\mathsf F}_{+}^{q}\rangle=-2\delta_p{}^q,
  \qquad
  \langle\bar{\mathsf F}_{+}^{q},\mathsf F_{-,p}\rangle=2\delta_p{}^q .
\end{gathered}
  \label{eq:n4-pairings}
\end{equation}
All other independent pairings vanish.

\paragraph{Ordered coordinates and left/right derivatives.}
For the $PSU(1,1|2)$ group element we use the ordered coordinates
\begin{equation}
  \mathcal G
  =
  e^{x\mathsf L_-}
  e^{\theta_-^p\mathsf F_{-,p}
     +\bar\theta_{-,p}\bar{\mathsf F}_{-}^{p}}
  e^{\rho\mathsf L_0}
  U
  e^{\theta_+^p\mathsf F_{+,p}
     +\bar\theta_{+,p}\bar{\mathsf F}_{+}^{p}}
  e^{y\mathsf L_+}.
  \label{eq:app-ordered-group-coordinates}
\end{equation}
For a function $f(\mathcal G)$, the left and right vector fields are defined by
\begin{equation}
  \mathcal D_T^L f(\mathcal G)
  =
  \left.\frac{d}{d\epsilon}f(e^{\epsilon T}\mathcal G)\right|_{\epsilon=0},
  \qquad
  \mathcal D_T^R f(\mathcal G)
  =
  \left.\frac{d}{d\epsilon}f(\mathcal G e^{\epsilon T})\right|_{\epsilon=0}.
  \label{eq:app-left-right-vector-field-definition}
\end{equation}
It is useful to separate the fermionic number operators
\[
  N_-
  =
  \theta_-^p\partial_{\theta_-^p}
  +\bar\theta_{-,p}\partial_{\bar\theta_{-,p}},
  \qquad
  N_+
  =
  \theta_+^p\partial_{\theta_+^p}
  +\bar\theta_{+,p}\partial_{\bar\theta_{+,p}} .
\]
The simple left derivatives are
\begin{align}
  \mathcal D_{\mathsf L_-}^L
  &=
  \partial_x,
  &
  \mathcal D_{\mathsf L_0}^L
  &=
  -x\partial_x+\partial_\rho-\frac12N_-,
  &
  \mathcal D_{\mathsf F_{-,p}}^L
  &=
  \partial_{\theta_-^p}
  -\frac12\bar\theta_{-,p}\partial_x,
  &
  \mathcal D_{\bar{\mathsf F}_{-}^{p}}^L
  &=
  \partial_{\bar\theta_{-,p}}
  -\frac12\theta_-^p\partial_x,
  \qquad p=1,2.
  \label{eq:app-left-derivatives}
\end{align}
The simple right derivatives are
\begin{align}
  \mathcal D_{\mathsf L_+}^R
  &=
  \partial_y,
  &
  \mathcal D_{\mathsf L_0}^R
  &=
  -y\partial_y+\partial_\rho-\frac12N_+,
  &
  \mathcal D_{\mathsf F_{+,p}}^R
  &=
  \partial_{\theta_+^p}
  +\frac12\bar\theta_{+,p}\partial_y,
  &
  \mathcal D_{\bar{\mathsf F}_{+}^{p}}^R
  &=
  \partial_{\bar\theta_{+,p}}
  +\frac12\theta_+^p\partial_y,
  \qquad p=1,2.
  \label{eq:app-right-derivatives}
\end{align}
The remaining left-plus and right-minus vector fields are longer and will be
provided in the supplemental Mathematica code.

\subsection{Superspace, superfield, and super-Schwarzian}
\label{app:superspace-notation}

\paragraph{Bulk and boundary supercoordinates.}
Bulk spacetime indices are $\mu,\nu=r,u$, tangent-space indices are
$a,b=\hat r,\hat u$, and spinor indices are $\alpha,\beta$.
For superspace bookkeeping, the bulk and boundary supercoordinates are
\[
  \mathbf Z^M=(X^\mu=(r,u);\Theta^{p\alpha},\bar\Theta_p{}^\alpha),
  \qquad
  z^m=(u,\vartheta^p,\bar\vartheta_p).
\]
Here $\mathbf Z^M$ are the bulk supercoordinates, while $z^m$ are the boundary superline coordinates.

On the flat boundary superline we use the basis one-forms
\[
  \Delta u
  =
  du+i\,d\vartheta^p\bar\vartheta_p
     +i\,d\bar\vartheta_p\,\vartheta^p,
  \qquad
  d\vartheta^p,
  \qquad
  d\bar\vartheta_p .
\]
The corresponding flat superspace derivatives are
\[
  D_p=\partial_{\vartheta^p}-i\bar\vartheta_p\,\partial_u,\qquad
  \bar D^p=\partial_{\bar\vartheta_p}-i\vartheta^p\,\partial_u,\qquad
  \{D_p,\bar D^q\}=-2i\delta_p{}^q\partial_u .
\]
For any boundary superfield $F(u,\vartheta,\bar\vartheta)$,
\[
  dF
  =
  \Delta u\,\partial_uF
  +d\vartheta^pD_pF
  +d\bar\vartheta_p\bar D^pF .
\]
\paragraph{Supergravity multiplets.}

The bulk component fields are grouped into a supergravity multiplet and a dilaton multiplet.
We use the component diamonds
\[
\mathfrak G_{\mathcal N}=
\begin{array}{ccccc}
&& E_\mu{}^a && \\[-1mm]
& \Psi_\mu{}^p && \bar\Psi_{\mu,p} & \\[-1mm]
&& \omega_\mu,\ A_\mu{}^I &&
\end{array}
\qquad
\mathfrak D_{\mathcal N}=
\begin{array}{ccccc}
&& \Phi && \\[-1mm]
& \zeta_\alpha{}^p && \bar\zeta_{\alpha,p} & \\[-1mm]
&& X^a,\ b^I &&
\end{array} .
\]
The index ranges are
\[
p=
\begin{cases}
1, & \mathcal N=2,\\
1,2, & \mathcal N=4,
\end{cases}
\qquad
I=
\begin{cases}
\mathfrak R, & \mathcal N=2,\quad \mathfrak R=U(1)_{\mathfrak R},\\
i=1,2,3, & \mathcal N=4,\quad \mathfrak R=SU(2)_{\mathfrak R}.
\end{cases}
\]
The fields $X^a$ and $b^I$ are the bosonic auxiliary, or multiplier,
components of the superdilaton multiplet: $X^a$ imposes the supertorsion
constraint, while $b^I$ is conjugate to the $\mathfrak R$-symmetry field
strength.
For $\mathcal N=2$, $b^I=b$ is the $U(1)_{\mathfrak R}$ multiplier; for
$\mathcal N=4$, $b^I=b^i$ is an $SU(2)_{\mathfrak R}$ triplet
\cite{Turiaci:2023jfa,Iliesiu:2021are}.

These component fields are encoded in the superfields as follows.
In Wess--Zumino gauge, the bulk supervielbein begins schematically as
\[
  \mathbf E^a
  =
  dX^\mu
  \bigl[
    \ubtwo{E_\mu{}^a}{\mathrm{frame}}{\mathrm{field}}
    +i\,\bar\Theta_p\gamma^a
      \ubone{\Psi_\mu^p}{\mathrm{gravitino}}
    +i\,
      \ubbarone{\bar\Psi_{\mu,p}}{\mathrm{gravitino}}
      \gamma^a\Theta^p
    +O(\Theta^2)
  \bigr]
  +i\,d\bar\Theta_p\,\gamma^a\Theta^p
  +i\,\bar\Theta_p\gamma^a d\Theta^p
  +O(\Theta^2 d\Theta),
\]
and the odd components begin as
\[
  \mathbf E^{p\alpha}
  =
  d\Theta^{p\alpha}
  +dX^\mu
  \bigl[
    \ubone{\Psi_\mu^{p\alpha}}{\mathrm{gravitino}}
    +O(\Theta)
  \bigr],
  \qquad
  \bar{\mathbf E}_{p}{}^\alpha
  =
  d\bar\Theta_p{}^\alpha
  +dX^\mu
  \bigl[
    \ubbarone{\bar\Psi_{\mu,p}{}^\alpha}{\mathrm{gravitino}}
    +O(\Theta)
  \bigr].
\]
The bulk superdilaton has component expansion
\[
  \mathbf \Phi(\mathbf Z)
  =
    \ubone{\Phi(X)}{\mathrm{dilaton}}
  +\Theta^{p\alpha}
    \ubbarone{\bar\zeta_{p\alpha}(X)}{\mathrm{dilatino}}
  +\bar\Theta_p{}^\alpha
    \ubone{\zeta^p{}_\alpha(X)}{\mathrm{dilatino}}
  +O(\Theta^2).
\]

\paragraph{First-order fields.}

The first-order fields are packaged into a $\mathfrak{psu}(1,1|2)$-valued connection
\[
  \mathcal A
  =
  \mathcal A^-\,\mathsf L_-
  +\mathcal A^0\,\mathsf L_0
  +\mathcal A^+\,\mathsf L_+
  +\ubtwo{\mathsf A^i}{\mathrm{SU(2)}_{\mathfrak R}}{\mathrm{connection}}\,
    \mathsf R_i
  +\ubone{\Psi_-^p}{\mathrm{gravitino}}\,
    \mathsf F_{-,p}
  +\ubbarone{\bar\Psi_{-,p}}{\mathrm{gravitino}}\,
    \bar{\mathsf F}_{-}^{p}
  +\ubone{\Psi_+^p}{\mathrm{gravitino}}\,
    \mathsf F_{+,p}
  +\ubbarone{\bar\Psi_{+,p}}{\mathrm{gravitino}}\,
    \bar{\mathsf F}_{+}^{p}.
\]
The adjoint scalar, equivalently the first-order dilaton multiplet, is
\[
  \mathcal X
  =
  X_-\,\mathsf L_-
  +X_0\,\mathsf L_0
  +X_+\,\mathsf L_+
  +\ubtwo{X_{\mathcal R}^i}{\mathrm{SU(2)}_{\mathfrak R}}{\mathrm{multiplier}}\,
    \mathsf R_i
  +\ubone{\zeta_-^p}{\mathrm{dilatino}}\,
    \mathsf F_{-,p}
  +\ubbarone{\bar\zeta_{-,p}}{\mathrm{dilatino}}\,
    \bar{\mathsf F}_{-}^{p}
  +\ubone{\zeta_+^p}{\mathrm{dilatino}}\,
    \mathsf F_{+,p}
  +\ubbarone{\bar\zeta_{+,p}}{\mathrm{dilatino}}\,
    \bar{\mathsf F}_{+}^{p}.
\]

The radial asymptotic form of the boundary connection is
\[
  \mathcal A_u
  =
  e^r
  \ubtwo{e_u^{(0)}}{\mathrm{metric}}{\mathrm{source}}
  \mathsf L_-
  +e^{r/2}
  \ubtwo{v_u^{(0)p}}{\mathrm{gravitino}}{\mathrm{source}}
  \mathsf F_{-,p}
  +e^{r/2}
  \ubbartwo{\bar v_{u,p}^{(0)}}{\mathrm{gravitino}}{\mathrm{source}}
  \bar{\mathsf F}_{-}^{p}
  +\ubtwo{a_u^{(0)i}}{\mathrm{SU(2)}_{\mathfrak R}}{\mathrm{source}}
  \mathsf R_i
\]
\[
  \hspace{1.2cm}
  +e^{-r/2}
  \ubone{\mathcal S^p}{\mathrm{supercurrent}}
  \mathsf F_{+,p}
  +e^{-r/2}
  \ubbarone{\bar{\mathcal S}_p}{\mathrm{supercurrent}}
  \bar{\mathsf F}_{+}^{p}
  +\ubtwo{\mathcal R^i}{\mathrm{SU(2)}_{\mathfrak R}}{\mathrm{current}}
  \mathsf R_i
  +e^{-r}
  \ubtwo{\mathcal L}{\mathrm{stress}}{\mathrm{tensor}}
  \mathsf L_+
  +\cdots .
\]
This identifies the boundary super-Schwarzian multiplet as
\[
  \bigl(
    \mathcal L,\mathcal S^p,\bar{\mathcal S}_p,\mathcal R^i
  \bigr).
\]

\paragraph{Super-Schwarzian.}
\phantomsection\label{app:n2-superspace-super-schwarzian}
In the $\mathcal N=2$ truncation one may repackage the component vev multiplet
into the standard superspace super-Schwarzian.
We use the flat boundary superline coordinates
\[
  z=(u,\vartheta,\bar\vartheta),
  \qquad
  D=\partial_\vartheta-i\bar\vartheta\,\partial_u,
  \qquad
  \bar D=\partial_{\bar\vartheta}-i\vartheta\,\partial_u,
  \qquad
  \{D,\bar D\}=-2i\partial_u .
\]
A super-reparametrization is described by target supercoordinates
\[
  Z(z)=\bigl(X(z),\Theta(z),\bar\Theta(z)\bigr)
\]
obeying the $\mathcal N=2$ superconformal constraints
\begin{equation}
  D\bar\Theta=0,\qquad
  \bar D\Theta=0,\qquad
  DX+i\bar\Theta D\Theta=0,\qquad
  \bar D X+i\Theta\bar D\bar\Theta=0 .
  \label{eq:n2-superconformal-constraints-app}
\end{equation}
In these conventions the $\mathcal N=2$ super-Schwarzian used in
\cite{Fu:2016vas,Heydeman:2020hhw} is written as
\begin{equation}
  {\cal S}_{\mathcal N=2}(Z;z)
  =
  \frac{\partial_u(\bar D\bar\Theta)}{\bar D\bar\Theta}
  -
  \frac{\partial_u(D\Theta)}{D\Theta}
  +2i\,
  \frac{\partial_u\Theta\,\partial_u\bar\Theta}
  {(D\Theta)(\bar D\bar\Theta)} .
  \label{eq:n2-superspace-super-schwarzian-app}
\end{equation}
The factor of $i$ in the last term is a convention effect: it converts the more
common convention $\{D,\bar D\}=2\partial_\tau$ to our convention
$\{D,\bar D\}=-2i\partial_u$.
The corresponding superspace action is therefore
\begin{equation}
  I_{\mathcal N=2}
  =
  -C\int du\,d\vartheta\,d\bar\vartheta\,
  {\cal S}_{\mathcal N=2}(Z;z),
  \label{eq:n2-superspace-action-app}
\end{equation}
up to the overall normalization $C$.
The component variables used in the particle-on-a-group calculation are the
lowest components of the super-reparametrization,
\begin{equation}
  X|=x,\qquad
  \Theta|=\theta_-,\qquad
  \bar\Theta|=\bar\theta_-,\qquad
  (D\Theta)(\bar D\bar\Theta)|=\Pi,\qquad
  \frac{D\Theta}{\bar D\bar\Theta}\bigg|=e^{-2a},
  \label{eq:n2-superspace-component-dictionary-app}
\end{equation}
where
\[
  \Pi=\dot x+\bar\theta_-\dot\theta_-+\theta_-\dot{\bar\theta}_-.
\]
With this identification, the $\vartheta\bar\vartheta$ component of
${\cal S}_{\mathcal N=2}$ reproduces the component super-Schwarzian multiplet
$(\mathcal L,\mathcal S,\bar{\mathcal S},\mathcal R)$ obtained from the
particle on a group, with $\mathcal L$ understood as the improved stress-tensor
component.
Thus the component calculation in the main text is equivalently the all-orders
superspace $\mathcal N=2$ super-Schwarzian written in the derivative convention
of this appendix.

The $\mathcal N=4$ case is parallel, except that the spinor derivative is an
$SU(2)_{\mathfrak R}$ doublet.  We use
\[
  z=(u,\vartheta^p,\bar\vartheta_p),
  \qquad
  D_p=\partial_{\vartheta^p}-i\bar\vartheta_p\partial_u,
  \qquad
  \bar D^p=\partial_{\bar\vartheta_p}-i\vartheta^p\partial_u,
  \qquad
  \{D_p,\bar D^q\}=-2i\delta_p{}^q\partial_u .
\]
A super-reparametrization
\[
  Z(z)=\bigl(X(z),\Theta^p(z),\bar\Theta_p(z)\bigr)
\]
obeys
\begin{equation}
  \bar D^p\Theta^q=0,\qquad
  D_p\bar\Theta_q=0,\qquad
  D_pX+i\bar\Theta_qD_p\Theta^q=0,\qquad
  \bar D^pX+i\Theta^q\bar D^p\bar\Theta_q=0 .
  \label{eq:n4-superconformal-constraints-app}
\end{equation}
In the old $\mathcal N=4$ superspace construction
\cite{Matsuda:1989kp,Aoyama:2018lfc}, the super-Schwarzian object is the
logarithm of the spinor Jacobian,
\begin{equation}
  \Sigma_{\mathcal N=4}(Z;z)
  =
  \log\det\bigl(D_p\Theta^q\bigr).
  \label{eq:n4-log-jacobian-app}
\end{equation}
The constraints imply the equivalent barred and bosonic Jacobian forms, but we
will not need them explicitly.  The corresponding superspace action is
\begin{equation}
  I_{\mathcal N=4}
  =
  -C_4
  \int du\,d^2\vartheta\,d^2\bar\vartheta\,
  \Sigma_{\mathcal N=4}(Z;z),
  \label{eq:n4-superspace-action-app}
\end{equation}
up to the overall normalization $C_4$, with
$d^2\vartheta\,d^2\bar\vartheta
=d\vartheta^1d\vartheta^2d\bar\vartheta_1d\bar\vartheta_2$.
The component variables used in the particle-on-a-group calculation are obtained
from the lowest components of the super-reparametrization and of the spinor
Jacobian,
\begin{equation}
  X|=x,\qquad
  \Theta^p|=\theta_-^p,\qquad
  \bar\Theta_p|=\bar\theta_{-,p},\qquad
  (D_p\Theta^q)|=\Pi^{1/2}U^q{}_{p},\qquad
  (\bar D^p\bar\Theta_q)|=\Pi^{1/2}(U^{-1})^p{}_{q},
  \label{eq:n4-superspace-component-dictionary-app}
\end{equation}
where $U(u)\in SU(2)_{\mathfrak R}$ and
\[
  \Pi=\dot x+\bar\theta_{-,p}\dot\theta_-^p
  +\theta_-^p\dot{\bar\theta}_{-,p}.
\]
Thus $\det(D_p\Theta^q)|=\Pi$, in direct analogy with
$(D\Theta)(\bar D\bar\Theta)|=\Pi$ in
\eqref{eq:n2-superspace-component-dictionary-app}.  The same
$\mathcal N=4$ data can also be packaged as an
$SU(2)_{\mathfrak R}$ triplet superfield \cite{Heydeman:2020hhw},
\begin{equation}
  \begin{aligned}
  \mathcal J_{\rm Sch}^i
  &\sim
  -2D_p(\sigma^i)^p{}_{q}\bar D^q\,\Sigma_{\mathcal N=4}  \\
  &=
  2\mathcal R^i
  +\vartheta^p(\sigma^i)_p{}^q\bar{\mathcal S}_q
  +\mathcal S^p(\sigma^i)_p{}^q\bar\vartheta_q
  -\vartheta^p(\sigma^i)_p{}^q\bar\vartheta_q\,\mathcal L
  +\cdots ,
  \end{aligned}
  \label{eq:n4-triplet-current-app}
\end{equation}
up to an overall convention-dependent normalization.

\section{\texorpdfstring{$\mathcal N=0$}{N=0} JT, BF Theory, and Particle on a Group}
\label{app:n0-jt-bf-particle}

This appendix gives expansion of Section \ref{sec:n0-jt-particle-main}.
It is supposed to be a careful and hopefully self-contained and pedegogical derivation of the bosonic particle-on-a-group description of JT gravity. 

\subsection{From metric JT gravity to first-order BF variables}
\label{subsec:n0-jt-first-order}
In this subsection we review the translation from metric formalism to the BF formalism of JT gravity, and the reason is that from the latter one can easily pass to the particle on a group description.

\paragraph{Metric variables.}
We begin with the metric formalism, or say the second-order description of Euclidean JT gravity.
The fundamental fields are a two-dimensional metric $G_{\mu\nu}$ and a dilaton $\Phi$.
The spacetime $M$ is regulated by an asymptotic curve $\partial M$, with coordinate $u$ and induced line element $ds_{\partial}=\sqrt{h_{uu}}\,du$.
The nearly-AdS$_2$ Dirichlet boundary condition fixes the leading proper length of this curve together with the leading dilaton:\footnote{Many JT conventions set the boundary einbein to $1$ and put the physical thermal length in the range of $u$, for example $u\sim u+\beta$.
We keep the bosonic source $q$ of Section~\ref{sec:n0-jt-particle-main} explicit because the same number becomes the parabolic charge of \cite{Lin:2022zxd} in the particle reduction.}
\begin{equation}
  h_{uu}\big|_{\partial M}
  =
  \left(\frac{q}{\epsilon}\right)^2,
  \qquad
  \Phi\big|_{\partial M}
  =
  \frac{\phi_r^{(0)}(u)}{\epsilon},
  \qquad
  \epsilon\to 0 .
  \label{eq:jt-second-order-bc}
\end{equation}
We take the Euclidean action to be
\begin{equation}
  I[G,\Phi]
  =
  -\frac{1}{2}\int_M d^2X\,\sqrt G\,\Phi\,(R[G]+2)
  -\int_{\partial M} ds\,\Phi_{\partial}(K-1).
  \label{eq:jt-second-order-action-compact}
\end{equation}

{The variational problem is the standard gravitational Dirichlet problem for the regulated data in \eqref{eq:jt-second-order-bc}.}
The dilaton variation imposes the constant curvature condition
\begin{equation}
  R+2=0 .
  \label{eq:jt-dilaton-eom}
\end{equation}
while the metric variation gives the dilaton equation\footnote{Tracing the equation displayed below in two dimensions gives $\nabla^2\Phi=2\Phi$.
Substitution then gives the stronger Hessian equation $\nabla_\mu\nabla_\nu\Phi=G_{\mu\nu}\Phi$.
This equation has an immediate Killing-vector interpretation: if $k_\mu=\epsilon_\mu{}^\nu\nabla_\nu\Phi$, then $\nabla_{(\mu}k_{\nu)}=0$, because the Hessian of $\Phi$ is proportional to the metric.
Conversely, on a constant-curvature AdS$_2$ background the three Killing vectors can be generated this way from three independent solutions of the Hessian equation.
The existence of these three modes is special to the maximally symmetric background: commuting derivatives on $\nabla_\rho\Phi$ in the Hessian equation requires the curvature tensor of unit Euclidean AdS$_2$, equivalently $R=-2$ in our convention.
The three independent Hessian modes may be expanded with three coefficients, and these coefficients furnish the adjoint representation of $\mathfrak{sl}(2,\mathbb R)$.}
\begin{equation}
  \nabla_\mu\nabla_\nu\Phi
  -
  G_{\mu\nu}\nabla^2\Phi
  +
  G_{\mu\nu}\Phi
  =
  0 .
  \label{eq:jt-metric-eom}
\end{equation}

\paragraph{First-order variables.}
We now rewrite the same theory in first-order form in order to expose the BF structure.
The independent fields are the zweibein, the Euclidean spin connection, two auxiliary zero-forms, and the physical dilaton.
It is convenient to use the plus/minus frame basis
\[
  E^\pm
  =
  E^{\hat u}\pm iE^{\hat r},
  \qquad
  X^\pm
  =
  X_{\hat u}\pm iX_{\hat r},
  \qquad
  \omega=\omega_\mu dX^\mu,
  \qquad
  \Phi .
\]
The metric is recovered from
\begin{equation}
  G_{\mu\nu}=E_\mu{}^a E_\nu{}^b\eta_{ab}
  =
  \frac{1}{2}
  \left(
    E_\mu{}^+E_\nu{}^-+E_\mu{}^-E_\nu{}^+
  \right).
  \label{eq:first-order-metric}
\end{equation}
We choose $\epsilon_{\hat r\hat u}=+1$.\footnote{The minimal dictionary is $T^+=dE^+ + i\omega\wedge E^+$, $T^-=dE^- - i\omega\wedge E^-$, $R^a{}_{b}=d\omega^a{}_{b}+\omega^a{}_{c}\wedge\omega^c{}_{b}$, and $d\omega=\frac{R}{4i}E^+\wedge E^-$.}

In these variables the bulk action equivalent to \eqref{eq:jt-second-order-action-compact} is
\begin{equation}
  I
  =
  -\int_M
  \left[
    \frac{1}{2}X^-
    \left(
      dE^+ + i\omega\wedge E^+
    \right)
    +
    \frac{1}{2}X^+
    \left(
      dE^- - i\omega\wedge E^-
    \right)
    +
    \Phi
    \left(
      d\omega
      +\frac{1}{2i}E^+\wedge E^-
    \right)
  \right]
  +I_{\partial}.
  \label{eq:first-order-action}
	\end{equation}
and we comment on $I_\partial$ later.

A nice way to repackage the first order variables is to use the BF theory language, which is a gauge theory of the group $SL(2,\mathbb R)$.
We introduce two gauge fields: a connection one form which encodes the einbein
\footnote{This is the $\mathsf L_\pm,\mathsf L_0$ version of the standard first-order parametrization $\mathcal A=E^{\hat u}\mathsf P_{\hat u}+E^{\hat r}\mathsf P_{\hat r}+\omega\mathsf J$ and $\mathcal X=X_{\hat u}\mathsf P_{\hat u}+X_{\hat r}\mathsf P_{\hat r}+\Phi\mathsf J$. The $\mathsf P,\mathsf J$ basis is defined in \eqref{eq:first-order-jt-basis}; we also use $E^\pm=E^{\hat u}\pm iE^{\hat r}$ and $X^\pm=X_{\hat u}\pm iX_{\hat r}$.}
\begin{equation}
  \mathcal A
  =
  \left[
    \frac{1}{4}(E^++E^-)+\frac{1}{2}\omega
  \right]\mathsf L_-
  +
  \frac{1}{2i}(E^+-E^-)\,\mathsf L_0
  +
  \left[
    \frac{1}{4}(E^++E^-)-\frac{1}{2}\omega
  \right]\mathsf L_+ .
  \label{eq:bf-connection-first-order-fields}
\end{equation}
and an adjoint zero-form which encodes the dilaton
\begin{equation}
  \mathcal X
  =
  \left(
    \frac{\Phi}{2}+\frac{X^++X^-}{4}
  \right)\mathsf L_-
  +
  \frac{X^+-X^-}{2i}\,\mathsf L_0
  +
  \left(
    -\frac{\Phi}{2}+\frac{X^++X^-}{4}
  \right)\mathsf L_+ .
	  \label{eq:bf-scalar-first-order-fields}
	\end{equation}
Defining $\mathcal F[\mathcal A]=d\mathcal A+\mathcal A\wedge\mathcal A$, the bulk first-order action can be written as
	\begin{equation}
	  I_{\rm BF,bulk}[\mathcal A,\mathcal X]
	  =
	  -\int_M
	  \left\langle
	    \mathcal X,\mathcal F[\mathcal A]
	  \right\rangle.
	  \label{eq:bf-bulk-action-sign}
	\end{equation}

{At this stage the equivalence is only a bulk equivalence.
{And the equations of motion provide the first check of the equivalence.}
Varying $X^\pm$ imposes following two equations which in the gravity context is the torsion-free condition
\begin{equation}
  dE^+ + i\omega\wedge E^+ =0,
  \qquad
  dE^- - i\omega\wedge E^- =0 .
  \label{eq:first-order-torsion-eom}
\end{equation}
The variation of $\Phi$ gives
\begin{equation}
  d\omega+\frac{1}{2i}E^+\wedge E^-=0 .
  \label{eq:first-order-curvature-eom}
\end{equation}
{After combining with torsionlessness, this equation is precisely $R+2=0$.}

The variations of $\omega$ and $E^\pm$ give the remaining first-order equations,
\begin{equation}
  d\Phi
  +
  \frac{1}{2i}
  \left(
    X^+E^- - X^-E^+
  \right)
  =0 ,
  \label{eq:first-order-omega-eom}
\end{equation}
and
\begin{equation}
  dX^+ + i\omega X^+ + i\Phi E^+=0,
  \qquad
  dX^- - i\omega X^- - i\Phi E^-=0 .
  \label{eq:first-order-e-eom}
\end{equation}
{The equation obtained from $\omega$ shows that $X^\pm$ are NOT new local degrees of freedom.}
They are just the plus/minus frame components of the dilaton gradient, $X^-=2iE_+{}^\mu\partial_\mu\Phi$ and $X^+=-2iE_-{}^\mu\partial_\mu\Phi$.
Substituting these expressions into \eqref{eq:first-order-e-eom} reproduces the metric equation \eqref{eq:jt-metric-eom}.\footnote{Using torsionlessness, $E^\pm$ are covariantly constant, and the two $E^\pm$ equations become the frame components of $\nabla_\mu\nabla_\nu\Phi=G_{\mu\nu}\Phi$.
Taking the trace gives $\nabla^2\Phi=2\Phi$, hence \eqref{eq:jt-metric-eom}.} This verifies that the first-order system carries the same local content as second-order JT.

\paragraph{Gibbons--Hawking term and the path to the BF boundary action.}
{We now discuss the boundary condition and boundary term.}
The metric problem fixes the induced line element and the leading physical dilaton, while the BF variation uses the connection $\mathcal A$ and the full adjoint zero-form $\mathcal X$.
In a boundary-adapted orthonormal frame, the JT boundary data take the form
\[
  E^{\hat u}\big|_{\partial M}
  =
  ds
  =
  E_u{}^{\hat u}du
  \sim
  \frac{e_u^{(0)}(u)}{\epsilon}\,du,
  \qquad
  E^{\hat r}\big|_{\partial M}=0,
  \qquad
  \Phi_{\partial}\sim \frac{\phi_r^{(0)}(u)}{\epsilon}.
\]
These data fix the leading ratio $\Phi_{\partial}/E_u{}^{\hat u}$.
In Fefferman--Graham gauge the same boundary-adapted frame also fixes the boundary connection.
Using \eqref{eq:bf-connection-first-order-fields} and $E_u^+=E_u^-=E_u{}^{\hat u}$, the coefficients of $\mathcal A_u$ along $\mathsf L_-$, $\mathsf L_0$, and $\mathsf L_+$ are $(E_u{}^{\hat u}+\omega_u)/2$, $0$, and $(E_u{}^{\hat u}-\omega_u)/2$, respectively.\footnote{On the cutoff curve the spin connection pulls back as $\omega|_{\partial M}=K\,ds$, so $\omega_u=K E_u{}^{\hat u}$.}
For the zero-form, however, the metric condition fixes only the dilaton component $\Phi_\partial$.
It does not by itself specify the boundary values of $X^\pm$, which are auxiliary fields imposing torsion in the bulk rather than additional metric JT sources.
{This is the missing boundary input that the BF boundary action must supply.}

We can get some clue by looking at the boundary action. 
The JT boundary term is the Gibbons--Hawking--York term, and in first-order variables it becomes one part of the BF boundary term,
\begin{equation}
  -\int_{\partial M} ds\,\Phi K
  =
  -\int_{\partial M}\Phi\,\omega
  =
  -\int du\,\Phi\,\omega_u .
  \label{eq:first-order-Kminus1}
	\end{equation}
The full BF boundary term is suggested by the bulk variation
	\begin{equation}
	  \delta I_{\rm BF,bulk}
	  =
	  \text{bulk EoM}
	  -
	  \int_{\partial M}du\,
	  \left\langle
	    \mathcal X,\delta\mathcal A_u
	  \right\rangle,
	  \label{eq:bf-bulk-boundary-variation}
	\end{equation}
Since \eqref{eq:bf-bulk-boundary-variation} contains $-\int_{\partial M}\langle\mathcal X,\delta\mathcal A_u\rangle$, the natural BF boundary term is
	\begin{equation}
	  \int_{\partial M}\left\langle
	 \mathcal X,\mathcal A_u
	  \right\rangle
	  =\int_{\partial M} \left(
	  \frac{1}{2}X^-E_u^+
	  +
	  \frac{1}{2}X^+E_u^-
	  -
	  \Phi\,\omega_u \right) .
	  \label{eq:bf-boundary-pairing-components}
\end{equation}
It contains the Gibbons--Hawking--York term in \eqref{eq:first-order-Kminus1}, but also the auxiliary pieces $X^-E_u^+$ and $X^+E_u^-$; as emphasized above, the metric boundary condition does not fix $X^\pm$ as independent sources.
Moreover, although this term cancels the $\delta\mathcal A_u$ variation from the bulk, its own variation leaves $\int_{\partial M}\langle\mathcal A_u,\delta\mathcal X\rangle$.
This means $I_{\partial}=\int_{\partial M}\langle\mathcal A_u,\delta\mathcal X\rangle$ can not be the end of the story, because the $\delta \cal{X} = 0$ condition is just too strong.\footnote{If one literally solve $\delta \cal{X} = 0$, the out come is the same as $\partial_u {x,u}=0$ where ${x,u}$ is the Schwarzian derivative.}   

To avoid this, one can add an additional boundary term which only contains $\cal{X}$ like $H(\langle\cal{X},\cal{X} \rangle)$.
This is the remaining input for BF theory that we mentioned above. Therefore the boundary action is 
	\begin{equation}
	  I_{\partial}
	  =
	  \int_{\partial M}du\,
	  \left\langle
	    \mathcal X,\mathcal A_u
	  \right\rangle
	  +
	  \int_{\partial M}du\,
	  H\!\left(
	    \left\langle
	      \mathcal X,\mathcal X
	    \right\rangle
	  \right).
	  \label{eq:bf-boundary-action-schematic}
	\end{equation}
At this stage $H$ is not fixed by the Gibbons--Hawking--York term alone.
It is the additional boundary input which tells us how $\mathcal X$, including $X^\pm$, is tied to the boundary connection.
{Different BF boundary theories are specified by different boundary Hamiltonians.}
{Below we derive the JT choice, which is just $\frac{1}{2\phi_r}\langle\mathcal X,\mathcal X\rangle$, from the mixed boundary condition, obtaining \eqref{eq:jt-bf-boundary-hamiltonian}, and then use it to obtain the particle-on-a-group action.}

\subsection{From BF theory to the reduced particle on a group}
\label{subsec:n0-bf-particle}

To see what BF boundary term that we need to introduce, we want to dive into the boundary dynamics.
Here we follow the the standard AdS/CFT boundary paradigm: the coefficient of the growing piece, which is usually called source, is fixed, while the decaying piece, which is usually called vev, is allowed to fluctuate. 
To determine them, we need to solve the bulk equations of motion and find the radial solutions.

\paragraph{Bulk equations and radial solutions.}
The BF equations take the compact form
\begin{equation}
  \mathcal F[\mathcal A]=0,
  \qquad
  D_{\mathcal A}\mathcal X=0 .
  \label{eq:bf-bulk-eom}
\end{equation}
{Here $D_{\mathcal A}\mathcal X=d\mathcal X+[\mathcal A,\mathcal X]$.}
We impose radial gauge $\mathcal A_r=\mathsf L_0$, which is weaker than Fefferman--Graham gauge.
The $ru$ component of the flatness condition, $\partial_r\mathcal A_u+[\mathsf L_0,\mathcal A_u]=0$, then fixes the radial weights without fixing the boundary representative.
In the simple constant-source representative used below this gives
\begin{equation}
  \mathcal A
  =
  e^r e_u^{(0)}(u)\,\mathsf L_-\,du
  +
  \mathsf L_0\,dr
  +
  e^{-r}t(u)\,\mathsf L_+\,du .
  \label{eq:radial-gauge-A-solution}
\end{equation}
Using \eqref{eq:bf-connection-first-order-fields}, the same solution translates back to the first-order gravity fields as
\begin{equation}
  \omega
  =
  \left(e^r e_u^{(0)}-e^{-r}t\right)du,
  \qquad
  \frac{E^+ + E^-}{2}
  =
  \left(e^r e_u^{(0)}+e^{-r}t\right)du,
  \qquad
  \frac{E^+ - E^-}{2}
  =
  i\,dr .
  \label{eq:radial-gauge-E-omega-solution}
\end{equation}
The radial equation for the zero-form, obtained from $D_{\mathcal A}\mathcal X=0$, fixes the same $e^r,1,e^{-r}$ hierarchy for $\mathcal X$.
This hierarchy is the source/vev split: the leading $e^r\mathsf L_-$ coefficient is fixed source data, while the $e^{-r}\mathsf L_+$ coefficient is a normalizable vev mode.
In the constant-source JT ensemble we set $e_u^{(0)}=q$ and fix $\chi_-/e_u^{(0)}=\phi_r$, equivalently $\chi_-=q\phi_r$.
If we then use only the leading and finite parts of the $u$-component of $D_{\mathcal A}\mathcal X=0$, the same representative gives $\chi_0=0$ and $\chi_+=\phi_r t(u)$.
The off-shell asymptotic fields are therefore
\begin{equation}
  \mathcal A_u(r,u)
  =
  e^rq\,\mathsf L_-+e^{-r}t(u)\,\mathsf L_+,
  \qquad
  \mathcal X(r,u)
  =
  e^rq\phi_r\,\mathsf L_-+e^{-r}\phi_rt(u)\,\mathsf L_+ .
  \label{eq:constant-source-A-X}
\end{equation}
{We mention that Equation \eqref{eq:constant-source-A-X} represents off-shell asymptotic data, not a full solution of all BF equations.}
{The leading and finite parts of $D_u\mathcal X=0$ identify the fixed source and its conjugate vev data, while the remaining decaying $\mathsf L_+$ component gives the equation of motion for $t(u)$, which is actually the Schwarzian equation of motion $\dot t(u) = \partial_u 
 \{x,u\}=0$.}
{We do not impose this last equation in the off-shell Dirichlet problem; the vev coefficient $t(u)$ is path integrated over and its equation of motion is generated by the particle-on-a-group action.}\footnote{In the shifted radial representative $\mathcal A=e^r e_u^{(0)}\mathsf L_-du+\mathsf L_0(dr+s\,du)+e^{-r}t\mathsf L_+du$, the leading and finite parts of $D_u\mathcal X=0$ determine $\chi_0,\chi_+$.
For $e_u^{(0)}=q$ and $\chi_-=q\phi_r$, the result is $\chi_0=\phi_rs$ and $\chi_+=\phi_rt+(\phi_r/2q)\dot s$.}

\paragraph{JT mixed boundary condition and BF Hamiltonian.}
{We can now determine the boundary condition for $\mathcal X$.}
In the constant-source representative it is tied to the boundary connection by
\begin{equation}
  \left.\mathcal X(r,u)\right|_{\partial M}
  =
  \phi_r\,\left.\mathcal A_u(r,u)\right|_{\partial M},
  \qquad
  \left.\delta\mathcal X(r,u)\right|_{\partial M}
  =
  \phi_r\,\left.\delta\mathcal A_u(r,u)\right|_{\partial M}.
  \label{eq:jt-mixed-bc-bf}
\end{equation}
{The corresponding first-order BF boundary action should keep $\mathcal X$ as the boundary momentum:}
\begin{equation}
  I[\mathcal A,\mathcal X]
  =
  -\int_M\left\langle
    \mathcal X,\mathcal F[\mathcal A]
  \right\rangle
  +
  \int_{\partial M}du\,
  \left[
    \left\langle
      \mathcal X,\mathcal A_u
    \right\rangle
    -
    \frac{1}{2\phi_r}
    \left\langle
      \mathcal X,\mathcal X
    \right\rangle
  \right].
  \label{eq:complete-bf-jt-action}
\end{equation}
{The coefficient of the quadratic term is fixed by the variational principle.}
\begin{equation}
  \left.\delta I\right|_{\partial M}
  =
  \int_{\partial M}du\,
  \left\langle
    \mathcal A_u-\frac{1}{\phi_r}\mathcal X,
    \delta\mathcal X
  \right\rangle = 0.
  \label{eq:bf-boundary-variation-mixed}
\end{equation}
{This vanishes precisely on the mixed JT boundary condition $\mathcal X=\phi_r\mathcal A_u$.}
Equivalently,
\begin{equation}
  H\!\left(
    \left\langle
      \mathcal X,\mathcal X
    \right\rangle
  \right)
  =
  -\frac{1}{2\phi_r}
  \left\langle
    \mathcal X,\mathcal X
  \right\rangle
  \label{eq:jt-bf-boundary-hamiltonian}
\end{equation}
up to an additive constant.

{This form also clarifies the role of the auxiliary fields in the path integral.}
{The fields $X^\pm$ imposed torsion in the first-order gravity action; in the BF boundary action they remain auxiliary variables should be integrated over, not fixed as extra sources.}
{Since $\mathcal X$ appears without derivatives and only quadratically, it is a  Gaussian integral but with non-trivial peak centered at $\mathcal X=\phi_r\mathcal A_u$.}
Integrating it out gives the equivalent $\mathcal A$-only boundary term
\begin{equation}
  I_{\partial}
  =
  \frac{\phi_r}{2}
  \int_{\partial M}du\,
  \left\langle
    \mathcal A_u,\mathcal A_u
  \right\rangle.
  \label{eq:A-only-boundary-hamiltonian}
\end{equation}

\paragraph{Particle on a group I (path integral): action and constraints from Lagrange multiplier.}
{We now pass from the BF formulasion to the particle on a group formalism.}
{The bulk equation $\mathcal F[\mathcal A]=0$ is solved locally by writing the connection as a pure gauge.}
If one also imposed the zero-form equation completely, the solution would take the form
\begin{equation}
  \mathcal A
  =
  \mathcal G^{-1}d\mathcal G,
  \qquad
  \mathcal X
  =
  \mathcal G^{-1}\mathcal X_0\mathcal G,
  \qquad
  d\mathcal X_0=0 .
  \label{eq:fully-onshell-bf-group-solution}
\end{equation}
Here $\mathcal{G}$ is a group valued field on AdS$_2$ which will become the particle-on-a-group variable later.
{The above is true for any flat connection, but it does not yet define the JT boundary theory.}\footnote{The constant adjoint element $\mathcal X_0$ has three components.
A rigid $SL(2,\mathbb R)$ conjugation removes two of them on a generic orbit, leaving only the invariant Casimir $\langle\mathcal X_0,\mathcal X_0\rangle$.
This is why the on-shell phase space can be parametrized by the JT scale $\phi_r$ and its conjugate boundary variable, rather than by three independent zero-form sources.}
The JT boundary theory requires the additional asymptotic structure, meaning $\mathcal{A}$ can not be any flat conection, so that $\mathcal{G}$ can not be any group-valued funciton.

Now let us make it a bit more precise.
{For the off-shell path integral we therefore impose flatness and the JT source condition on $\mathcal A$, but not the final decaying component of $D_{\mathcal A}\mathcal X=0$.}
{Thus flatness allows us to write the boundary connection as $\mathcal A_u=\mathcal G^{-1}\partial_u\mathcal G$.}\footnote{The group current $\mathcal G^{-1}d\mathcal G$ is standard in particle actions on group manifolds and related coadjoint-orbit constructions; see \cite{Chu:1994hm,Alekseev:1990mp}.}
Substituting this into the first-order boundary action gives
\begin{equation}
  I_{\rm pG}[\mathcal G,\mathcal X]
  =
  \int du\,
  \left[
    \left\langle
      \mathcal X,\mathcal G^{-1}\partial_u\mathcal G
    \right\rangle
    -
    \frac{1}{2\phi_r}
    \left\langle
      \mathcal X,\mathcal X
    \right\rangle
  \right].
  \label{eq:first-order-particle-group-action-X}
\end{equation}
The algebraic equation for $\mathcal X$ is $\mathcal X=\phi_r\mathcal G^{-1}\partial_u\mathcal G$.
{Equivalently, in terms of the normalized particle momentum} $\mathcal P\equiv \frac{\mathcal X}{\phi_r}$,
the Schwarzian coupling $\phi_r$ multiplies the whole first-order particle action,
\begin{equation}
  I_{\rm pG}[\mathcal G,\mathcal P]
  =
  \phi_r
  \int du\,
  \left[
    \left\langle
      \mathcal P,\mathcal G^{-1}\partial_u\mathcal G
    \right\rangle
    -
    \frac{1}{2}
    \left\langle
      \mathcal P,\mathcal P
    \right\rangle
  \right].
  \label{eq:first-order-particle-group-action}
\end{equation}
Integrating out $\mathcal X$, or equivalently $\mathcal P$, gives the quadratic current action
\begin{equation}
  I_{\rm pG}[\mathcal G]
  =
  \frac{\phi_r}{2}
  \int du\,
  \left\langle
    \mathcal G^{-1}\dot{\mathcal G},
    \mathcal G^{-1}\dot{\mathcal G}
  \right\rangle.
  \label{eq:second-order-particle-group-action}
\end{equation}

Here there is no constraints on the particle yet, and the constraint comes from to impose the same asymptotic source condition on the group current.
In the constant-source FG representative, after removing the radial weights, the current is constrained to have the form $\mathcal G^{-1}\partial_u\mathcal G=q\mathsf L_-+t(u)\mathsf L_+$, with fixed source $q$ and fluctuating vev coefficient $t(u)$.
{In a more general gauge one writes $\mathcal G^{-1}\partial_u\mathcal G =e_u^{(0)}\mathsf L_-+s\mathsf L_0+t\mathsf L_+$.
The constant-source FG representative sets $e_u^{(0)}=q$ and $s=0$.
The scalar source fixes $\chi_-=q\phi_r$, while $t(u)$ remains off-shell vev data.} In invariant language, fixing $q$ is a genuine charge constraint, while $s=0$ is a choice of gauge slice rather than an additional boundary charge.

 It is useful to check that \eqref{eq:second-order-particle-group-action} contains the usual Schwarzian action.}
Use the Gauss parameterization
\begin{equation}
  \mathcal G(u)
  =
  e^{x(u)\mathsf L_-}e^{\rho(u)\mathsf L_0}e^{\gamma(u)\mathsf L_+}.
  \label{eq:gauss-param-group}
\end{equation}
The worldline group current is
\begin{equation}
  \mathcal G^{-1}\dot{\mathcal G}
  =
  e^\rho\dot x\,\mathsf L_-
  +\left(\dot\rho-2\gamma e^\rho\dot x\right)\mathsf L_0
  +\left(\dot\gamma+\gamma\dot\rho-\gamma^2e^\rho\dot x\right)\mathsf L_+ .
  \label{eq:gauss-group-current}
\end{equation}
Now impose the constant-source FG form of the group current reading from  \eqref{eq:constant-source-A-X},
\begin{equation}
  \mathscr J(u)
  \equiv
  \mathcal G^{-1}\dot{\mathcal G}
  =
  q\mathsf L_-+t(u)\mathsf L_+ .
  \label{eq:fg-current-group-constraint}
\end{equation}
Comparing the three components of \eqref{eq:gauss-group-current} gives
\begin{equation}
  e^\rho=\frac{q}{\dot x},
  \qquad
  \gamma=\frac{\dot\rho}{2q}
  =
  -\frac{\ddot x}{2q\dot x},
  \qquad
  t(u)
  =
  -\frac{1}{2q}
  \left[
    \frac{\dddot x}{\dot x}
    -
    \frac{3}{2}
    \left(\frac{\ddot x}{\dot x}\right)^2
  \right]
  =
  -\frac{1}{2q}\{x,u\}.
  \label{eq:fg-schwarzian-relation}
\end{equation}
Here $\{x,u\}$ is the Schwarzian derivative, and the sign in $e^\rho=q/\dot x$ follows from the orientation convention in \eqref{eq:gauss-param-group}.
Since
\[
  \left\langle
    q\mathsf L_-+t\mathsf L_+,
    q\mathsf L_-+t\mathsf L_+
  \right\rangle
  =
  4qt,
\]
the quadratic particle action becomes
\begin{equation}
  I_{\rm pG}
  =
  -\phi_r
  \int du\,\{x,u\},
  \label{eq:schwarzian-from-particle-group}
\end{equation}
with the Euclidean sign convention inherited from \eqref{eq:complete-bf-jt-action}.

Note that the above is at action level, now we quantize the theory. 
For the path integral, and later for canonical quantization, we use the Lagrange multiplier to impose teh following constraint
\begin{equation}
  \left\langle
    \mathcal P,\mathsf L_+
  \right\rangle
  =
  q,
  \qquad
  \text{or equivalently}\qquad
  \left\langle
    \mathcal X,\mathsf L_+
  \right\rangle
  =
  \phi_r q .
  \label{eq:normalized-source-constraint}
\end{equation}
We introduce a Lagrange multiplier, $\lambda(u)$ to the action and integrate over all $\lambda(u)$ 
\begin{equation}
  I[\mathcal G,\mathcal P,\lambda]
  =
  \phi_r
  \int du\,
  \left[
    \left\langle
      \mathcal P,
      \mathcal G^{-1}\partial_u\mathcal G-\lambda\mathsf L_+
    \right\rangle
    -
    \frac{1}{2}
    \left\langle
      \mathcal P,\mathcal P
    \right\rangle
    +
    q\lambda
  \right].
  \label{eq:right-nplus-lagrange-multiplier}
\end{equation}

Interestingly, the same Lagrange multiplier also keeps track of the endpoint character of the reduced wavefunction.
{Here $\mathcal G$ is still the unreduced $SL(2,\mathbb R)$ group variable.}
The subgroup being divided out is the right $N_+=\exp(\mathbb R\mathsf L_+)$ direction; the reduced configuration space is the right quotient $SL(2,\mathbb R)/N_+$, not a separate quotient group.
Local right $N_+$ transformations which vanish at the endpoints are gauge redundancies of \eqref{eq:right-nplus-lagrange-multiplier}.
If the right transformation has nonzero endpoint values, the action changes by a boundary term, which is precisely the character carried by the endpoint states.
Explicitly,
\[
  \mathcal G\to \mathcal G e^{\alpha\mathsf L_+},
  \qquad
  \mathcal P\to
  e^{-\alpha\mathsf L_+}\mathcal P e^{\alpha\mathsf L_+},
  \qquad
  \lambda\to\lambda+\dot\alpha ,
\]\footnote{Right and left multiplication are different here.  For right
multiplication $\delta\mathcal G=\mathcal G T$, so $\mathcal G^{-1}\delta\mathcal G=T$ and the normalized canonical pairing sees $\langle\mathcal P,T\rangle$.
For left multiplication $\delta\mathcal G=T\mathcal G$, so $\mathcal G^{-1}\delta\mathcal G=\mathcal G^{-1}T\mathcal G$ and the corresponding normalized charge is $\langle \mathcal G\mathcal P\mathcal G^{-1},T\rangle$.
The parabolic reduction of \cite{Lin:2022zxd} uses the right action generated by $\mathsf L_+$.} shifts the action only by the fixed charge term:
\begin{equation}
  I[\mathcal G e^{\alpha\mathsf L_+},
  e^{-\alpha\mathsf L_+}\mathcal P e^{\alpha\mathsf L_+},
  \lambda+\dot\alpha]
  =
  I[\mathcal G,\mathcal P,\lambda]
  +
  \phi_rq\bigl[\alpha_f-\alpha_i\bigr].
  \label{eq:right-nplus-action-character}
\end{equation}
{Equivalently, the bracketed normalized action shifts by $q[\alpha_f-\alpha_i]$; the full action is obtained by multiplying this normalized action by $\phi_r$.}
{This shows that the path integral produces an $N_+$-equivariant wavefunction on the original $SL(2,\mathbb R)$ variable, and we will see later this leads to a quotient by right $N_+$ imposed in the measure and inner product.}

Just from the equalvariance property, the propagator should obey
\begin{equation}
  K_q(\mathcal G_f e^{\alpha_f\mathsf L_+};
  \mathcal G_i e^{\alpha_i\mathsf L_+})
  =
  e^{-q\alpha_f}\,
  K_q(\mathcal G_f;\mathcal G_i)\,
  e^{q\alpha_i}.
  \label{eq:right-nplus-propagator-equivariance}
\end{equation}
Equivalently, a final-state wavefunction transforms as
\[
  \Psi(\mathcal G e^{\alpha\mathsf L_+})
  =
  e^{-q\alpha}\Psi(\mathcal G),
\]
{This is the Lagrangian origin of the constraint $L_+^R\Psi=-q\Psi$ later appearing in the Hamiltonian formalism.}

\paragraph{Particle on a group II (canonical quantization): reduced Hamiltonian and left invariant right equalvarient states}
{We now give the Hamiltonian version of the same reduction, following the conventions of \cite{Lin:2022zxd}.}
We keep the Gauss parameterization \eqref{eq:gauss-param-group} and the group current \eqref{eq:gauss-group-current}.
Substituting \eqref{eq:gauss-group-current} into the canonical one-form in \eqref{eq:first-order-particle-group-action}, we obtain
\begin{align}
  p_x
  &=
  e^\rho\left[
    \langle\mathcal P,\mathsf L_-\rangle
    -2\gamma\langle\mathcal P,\mathsf L_0\rangle
    -\gamma^2\langle\mathcal P,\mathsf L_+\rangle
  \right],
  \nonumber\\
  p_\rho
  &=
  \langle\mathcal P,\mathsf L_0\rangle
  +\gamma\langle\mathcal P,\mathsf L_+\rangle,
  \qquad
  p_\gamma
  =
  \langle\mathcal P,\mathsf L_+\rangle.
  \label{eq:gauss-canonical-momenta}
\end{align}
Solving these equations gives the right dual momenta in terms of the canonical coordinates:
\begin{equation}
  \langle\mathcal P,\mathsf L_+\rangle=p_\gamma,\qquad
  \langle\mathcal P,\mathsf L_0\rangle=p_\rho-\gamma p_\gamma,\qquad
  \langle\mathcal P,\mathsf L_-\rangle=e^{-\rho}p_x+2\gamma p_\rho-\gamma^2p_\gamma .
  \label{eq:right-dual-momenta-canonical}
\end{equation}
{This derivation uses the right covariant derivatives.} Meaning 
\begin{equation}
  p_\gamma \mathcal{G} = \mathcal{G} \mathsf L_+
\end{equation}
why the right not the left we will explain below when we talk about detailed definition of states.

{The Hamiltonian is obtained by an ordinary Legendre transform.}
The first term in the normalized particle Lagrangian, $\langle\mathcal P,\mathcal G^{-1}\dot{\mathcal G}\rangle$, is precisely the canonical one-form $p_x\dot x+p_\rho\dot\rho+p_\gamma\dot\gamma$.
Therefore
\begin{equation}
  H
  =
  p_x\dot x+p_\rho\dot\rho+p_\gamma\dot\gamma-L
  =
  \frac{1}{2}
  \left\langle
    \mathcal P,\mathcal P
  \right\rangle,
  \label{eq:canonical-jt-hamiltonian}
\end{equation}
and in canonical quantization, we promte everything to operators. 
Just to comment, the actual JT Hamiltonian  carries the overall Schwarzian coupling, $\widehat H_{\rm JT}=(2\phi_r)^{-1}\langle\widehat{\mathcal P}, \widehat{\mathcal P}\rangle$, with the ordering chosen so that the quadratic operator is the $SL(2,\mathbb R)$ Casimir.
For the present purpose we only need the ordered quadratic operator in the Gauss patch,
\begin{equation}
  \widehat C
  =
  L_0^2-\frac12\left(L_+L_-+L_-L_+\right)
  =
  \partial_\rho^2+\partial_\rho
  -e^{-\rho}\partial_x\partial_\gamma .
  \label{eq:lmrs-unreduced-casimir}
\end{equation}
{Up to the overall factor $1/(2\phi_r)$, and the usual possible additive constant, this is the Hamiltonian of the unreduced particle on $SL(2,\mathbb R)$.}

We now impose the constraint, which is $p_\gamma=\langle\mathcal P,\mathsf L_+\rangle$.
In the Euclidean conventions of \cite{Lin:2022zxd}, this is
\begin{equation}
  L_+^R\Psi=-q\Psi,
  \qquad
  \Psi(x,\rho,\gamma)=e^{-q\gamma}F(x,\rho).
  \label{eq:lmrs-right-charge-condition}
\end{equation}
Substitution of $\partial_\gamma\to -q$ in \eqref{eq:lmrs-unreduced-casimir} then gives the reduced operator
\begin{equation}
  \widehat C_q
  =
  \partial_\rho^2+\partial_\rho
  +
  q e^{-\rho}\partial_x .
  \label{eq:lmrs-reduced-casimir}
\end{equation}
The physical reduced JT Hamiltonian is $\widehat H_{{\rm JT},q}=(2\phi_r)^{-1}\widehat C_q$.
{This reproduces the operator of \cite{Lin:2022zxd} from the Lagrangian construction, with $\phi_r$ appearing only as the overall Schwarzian coupling.}

{We next fix the convention for states, since the same right action appears both in the wavefunction constraint and in the transformation of boundary propagators.}
Choose a reference state, or a set of reference states $|i\rangle$, and define
\[
  |\mathcal G,i\rangle=U(\mathcal G)|i\rangle .
\]
{This definition does not rely on any quotient or reduction; it is simply how the group-labeled states are introduced.}
Left multiplication acts directly on the group label,
\[
  U(a)|\mathcal G,i\rangle=|a\mathcal G,i\rangle .
\]
{Right multiplication is different because the factors multiply inside the same representation.}
If $h$ acts on the chosen reference states by $\rho(h)$ (namely $\rho$ a representation), then
\[
  |\mathcal G h,i\rangle
  =
  U(\mathcal G)U(h)|i\rangle
  =
  |\mathcal G,j\rangle\,\rho(h)^j{}_{i}.
\]
{Thus right multiplication should be intentionally separated from a left action such as $U(h)|\mathcal G,i\rangle=|\mathcal G h,i\rangle$.}
{It acts on the reference-state labels instead.}

The state transformation is reflected in wavefunction.
For example, take $h_\alpha=e^{\alpha\mathsf L_+}$ 
\[
  U(h_\alpha)|q\rangle=e^{q\alpha}|q\rangle,
  \qquad
  \langle q|U(h_\alpha)^{-1}=e^{-q\alpha}\langle q| .
\]
With $\langle \mathcal G;q|=\langle q|U(\mathcal G)^{-1}$, the wavefunction obeys\footnote{In general, wavefunctions transform as
\[
  \Psi_i(\mathcal G h)
  =
  \bigl(\rho(h)^{-1}\bigr)_i{}^j\Psi_j(\mathcal G).
\]
{This statement is more general than the JT example.}
{For example, reference states carrying $SU(2)_{\mathfrak R}$ indices transform this way under right $SU(2)_{\mathfrak R}$ multiplication even when $SU(2)_{\mathfrak R}$ is not being reduced.}}
\begin{equation}
  \Psi_q(\mathcal G e^{\alpha\mathsf L_+})
  =
  e^{-q\alpha}\Psi_q(\mathcal G),
  \qquad
  \Psi_q(\mathcal G)=\langle \mathcal G;q|\Psi\rangle ,
  \label{eq:right-equivariant-wavefunction-state}
\end{equation}
which is the finite form of $L_+^R\Psi_q=-q\Psi_q$ so that 
\begin{equation}
  \Psi_q(x,\rho,\gamma)=e^{-q\gamma}F(x,\rho).
\end{equation}
{These are the reduced wavefunctions for particle on a group.}

From the propagator, we can now understand the difference between left and right action better, and the slogan is that the propagator is left invariant and right equalvariant. 
The global $SL(2,\mathbb R)$ symmetry acts from the left on these states, $U(a)|\mathcal G;q\rangle=|a\mathcal G;q\rangle$.\footnote{Infinitesimal left multiplication in these coordinates gives $L_-=\partial_x$, $L_0=-x\partial_x+\partial_\rho$, and $L_+=x^2\partial_x-2x\partial_\rho+e^{-\rho}\partial_\gamma$, up to the overall sign convention.}
The Euclidean propagator is
\[
  K_q(\mathcal G_f,\mathcal G_i;T)
  =
  \langle \mathcal G_f;q|e^{-T\widehat H}|\mathcal G_i;q\rangle
  =
  \langle q|
  U(\mathcal G_f)^{-1}e^{-T\widehat H}U(\mathcal G_i)
  |q\rangle .
\]
Because the Hamiltonian is the Casimir, it commutes with the left action:
\[
  [\widehat H,U(a)]=0,
  \qquad
  U(a)^{-1}e^{-T\widehat H}U(a)=e^{-T\widehat H}.
\]
Therefore simultaneous left multiplication gives
\begin{equation}
  \begin{aligned}
  K_q(a\mathcal G_f,a\mathcal G_i;T)
  &=
  \langle q|
  U(\mathcal G_f)^{-1}U(a)^{-1}
  e^{-T\widehat H}
  U(a)U(\mathcal G_i)
  |q\rangle  =
  K_q(\mathcal G_f,\mathcal G_i;T).
  \end{aligned}
  \label{eq:left-invariance-propagator}
\end{equation}
{On the other hand, right multiplication is instead read through the transformation of the reference states.}
For a multiplet of reference states the propagator is matrix-valued,
\[
  K_{ij}(\mathcal G_f,\mathcal G_i;T)
  =
  \langle \mathcal G_f,i|e^{-T\widehat H}|\mathcal G_i,j\rangle ,
\]
and right multiplication inserts the corresponding $\rho$ matrices at the endpoints,
\[
  K_{ij}(\mathcal G_f h_f,\mathcal G_i h_i;T)
  =
  \rho(h_f^{-1})_i{}^k\,
  K_{kl}(\mathcal G_f,\mathcal G_i;T)\,
  \rho(h_i)^l{}_{j}.
\]
For the one-dimensional $N_+$ character this becomes
\begin{equation}
  \begin{aligned}
  K_q(\mathcal G_f e^{\alpha_f\mathsf L_+},
  \mathcal G_i e^{\alpha_i\mathsf L_+};T)
  &=
  \langle q|
  U(e^{\alpha_f\mathsf L_+})^{-1}
  U(\mathcal G_f)^{-1}
  e^{-T\widehat H}
  U(\mathcal G_i)
  U(e^{\alpha_i\mathsf L_+})
  |q\rangle       \\
  &=
  e^{-q\alpha_f}\,
  K_q(\mathcal G_f,\mathcal G_i;T)\,
  e^{q\alpha_i}.
  \end{aligned}
  \label{eq:right-covariance-propagator-state}
\end{equation}
{This is the Hilbert-space form of the endpoint character derived in \eqref{eq:right-nplus-propagator-equivariance}.}

\paragraph{Inner product and reduced measure.}
{It remains to explain which measure should be used for these equivariant wavefunctions, so that we can define inner product and Hilbert space.}

{We claim that the physical inner product must divide by the right $N_+$ gauge volume.}
{This is most transparent in the path-integral derivation.}
After the Gaussian integral over the normalized momentum, the boundary path integral is over group trajectories and the Lagrange multiplier:
\begin{equation}
  \mathcal D\mu_{\rm naive}
  =
  \prod_u
  d\mathcal G(u)\,
  d\lambda(u),
  \qquad
  d\mathcal G=e^\rho dx\,d\rho\,d\gamma ,
  \label{eq:unreduced-pg-path-integral-measure}
\end{equation}
{Here $d\mathcal G$ is the Haar measure on $SL(2,\mathbb R)$ in the Gauss patch.}
{This is not yet the physical measure: it still contains the local right $N_+$ redundancy.}
That action is invariant under the local right transformation
\[
  \mathcal G(u)\to \mathcal G(u)e^{\alpha(u)\mathsf L_+},
  \qquad
  \gamma(u)\to\gamma(u)+\alpha(u),
  \qquad
  \lambda(u)\to\lambda(u)+\dot\alpha(u) .
\]
Thus $\gamma$ and $\lambda$ should not be counted as two independent physical functions: the quotient removes one arbitrary function $\alpha(u)$.
The group current changes nontrivially under the same right shift:
\[
  \mathcal G^{-1}\partial_u\mathcal G
  \longrightarrow
  e^{-\alpha\mathsf L_+}
  \left(\mathcal G^{-1}\partial_u\mathcal G\right)
  e^{\alpha\mathsf L_+}
  +\dot\alpha\,\mathsf L_+ .
\]
{This is the redundancy that must be removed from the path integral.}\footnote{For a boundary interval with specified endpoints, the pure gauge transformations are the shifts with $\alpha$ vanishing at the endpoints.
The endpoint values themselves are kept as external transformations on the endpoint states and give the characters in \eqref{eq:right-nplus-propagator-equivariance}.
On a closed boundary, all time-dependent shifts are divided out.}
The quotient form of the history measure is therefore
\begin{equation}
  \mathcal D\mu_{\rm phys}
  =
  \frac{
    \prod_u d\mathcal G(u)\,d\lambda(u)
  }{
    \operatorname{Vol}(\mathcal H_+)
  },
  \qquad
  \mathcal H_+=\{e^{\alpha(u)\mathsf L_+}\}.
  \label{eq:path-integral-quotient-measure}
\end{equation}

{Equivalently, one may choose a gauge slice for this quotient.}
Let us consider a more general gauge for the current $\mathscr J(u)=\mathcal G^{-1}\dot{\mathcal G}=A\mathsf L_-+s\mathsf L_0+t\mathsf L_+$, the right transformation gives $A\to A$ and $s\to s-2A\alpha$.
On the charge surface $A=q$, the condition $s(u)=0$ is a good gauge slice in the sense that whatever other gauge can always be brought back to this gauge. 
Indeed, in Gauss coordinates $s=\dot\rho-2\gamma e^\rho\dot x=\dot\rho-2A\gamma$, with $A=e^\rho\dot x$, so $s=0$ fixes $\gamma=\dot\rho/(2A)$, or $\gamma=\dot\rho/(2q)$ after imposing $A=q$.\footnote{We can also think about this from the constraint point of view. $s=0$ is not a second momentum-map constraint; it is the gauge-fixed representation of the quotient by $\mathcal H_+$. We are dealing with the first class constraint, and usually a first class constraint removes two degrees of freedom in the phase space. The $A=q$ condition freeze $A$, and the $s=0$ gauge choice freezes the other.}
\begin{equation}
  \frac{
    \mathcal D\mathcal G\,\mathcal D\lambda
  }{
    \operatorname{Vol}(\mathcal H_+)
  }
  =
  \mathcal D\mathcal G\,\mathcal D\lambda\,
  \delta[s]\,\Delta_{\rm FP},
  \qquad
  \Delta_{\rm FP}
  =
  \det\left(
    \frac{\delta s^\alpha}{\delta\alpha}
  \right)
  =
  \det(-2A).
  \label{eq:gauge-fixed-quotient-measure}
\end{equation}
{On the support of the charge constraint $A=q$, this determinant is the constant $\det(-2q)$, which is why it is invisible in the final bosonic Schwarzian action.}
{The two common orders of integration are equivalent: gauge fixing supplies $\delta[s]\Delta_{\rm FP}$, while the $\lambda$-integral supplies $\delta[A-q]$, since $\mathsf L_+$ is null and $\lambda$ appears linearly after $\mathcal P$ is eliminated.}
In either order one reaches
\begin{equation}
  \int
  \mathcal D\mathcal G\,
  \delta[A-q]\,
  \delta[s]\,
  \Delta_{\rm FP}\,
  e^{-I_{\rm pG}[\mathcal G]} .
  \label{eq:strict-current-gauge-fixed-path-integral}
\end{equation}
Thus $\delta[A-q]$ comes from the Lagrange multiplier, whereas $\delta[s]\Delta_{\rm FP}$ is the gauge-fixed form of the quotient by $\mathcal H_+$.

{For the Hilbert-space inner product, one can directly obtain it from from the reduced path integral  measure at a time slice.}
\begin{equation}
  d\mu_{\rm phys}
  =
  \frac{d\mathcal G}{d\gamma}
  =
  e^\rho\,dx\,d\rho .
  \label{eq:reduced-haar-measure}
\end{equation}
The corresponding position-space resolution of the identity is
\begin{equation}
  \mathbf 1_q
  =
  \int dx\,d\rho\,e^\rho\,
  |x,\rho;q\rangle\langle x,\rho;q| ,
  \label{eq:reduced-position-identity}
\end{equation}
and the physical inner product for reduced wavefunctions $F(x,\rho)$ is
\begin{equation}
  \langle F_1,F_2\rangle_{\rm phys}
  =
  \int dx\,d\rho\,e^\rho\,
  \overline{F_1(x,\rho)}F_2(x,\rho).
  \label{eq:reduced-inner-product}
\end{equation}
In unreduced notation the same statement can be written as
\[
  \langle \Psi_1,\Psi_2\rangle_{\rm phys}
  \sim
  \frac{1}{\operatorname{Vol}(N_+)}
  \int_G d\mathcal G\,
  \overline{\Psi_1(\mathcal G)}\Psi_2(\mathcal G).
\]
{This completes the $\mathcal N=0$ discussion of the particle-on-a-group construction.}

\subsection{Matter Green's functions and bilocals}
\label{subsec:n0-matter-greens-wilson-bilocal}
{We now add a matter field to the same $\mathcal N=0$ BF background.}
{A bulk Green's function is the two-point function for this matter field, so it must solve the matter equation of motion in each argument.}
In the BF language a matter field of dimension $\Delta$ transforms in a representation $\mathcal R_\Delta$, and the linear equation of motion is
\[
  D_{\mathcal A}\Phi_\Delta
  =
  d\Phi_\Delta+\mathcal R_\Delta(\mathcal A)\Phi_\Delta
  =
  0 .
\]
{The natural group-language candidate for the Green's function is therefore the Wilson line \cite{Blommaert:2018oro,Fitzpatrick:2016mtp,Fan:2021wsb},}
\begin{equation}
  \mathcal U_\Delta(i,j)
  =
  \operatorname{P}\exp\!\left[
    -\int_j^i \mathcal R_\Delta(\mathcal A)
  \right].
  \label{eq:n0-matter-wilson-line}
\end{equation}
On a BF saddle the connection is flat, so locally $\mathcal A=\mathcal G^{-1}d\mathcal G$.
Then the Wilson line can be written only in terms of the endpoint group elements $\mathcal G_i,\mathcal G_j$ as\footnote{A short derivation is as follows.
Let $y(\tau)$ be a path from $j$ to $i$, with $\mathcal A_\tau=\mathcal A_\mu\dot y^\mu$.
The Wilson line $\mathcal U_\Delta(\tau,j)$ solves $\left(\frac{d}{d\tau}+\mathcal R_\Delta(\mathcal A_\tau)\right)\mathcal U_\Delta(\tau,j)=0$, with $\mathcal U_\Delta(j,j)=\mathbf 1$.
For $\mathcal A=\mathcal G^{-1}d\mathcal G$, one has $\mathcal A_\tau=\mathcal G(\tau)^{-1}\dot{\mathcal G}(\tau)$.
The endpoint expression obeys $\frac{d}{d\tau}\mathcal R_\Delta(\mathcal G(\tau)^{-1}\mathcal G_j)=-\mathcal R_\Delta(\mathcal A_\tau)\mathcal R_\Delta(\mathcal G(\tau)^{-1}\mathcal G_j)$, with $\mathcal R_\Delta(\mathcal G_j^{-1}\mathcal G_j)=\mathbf 1$.
It therefore obeys the same first-order equation and initial condition as the path-ordered exponential, so the two are equal.}
\begin{equation}
  \mathcal U_\Delta(i,j)
  =
  \mathcal R_\Delta(\mathcal G_i^{-1}\mathcal G_j).
  \label{eq:n0-flat-wilson-line-endpoints}
\end{equation}
{This endpoint formula has exactly the covariance expected of a matter Green's function.}
Under a local $SL(2,\mathbb R)$ gauge transformation $h$, with $\mathcal G\to \mathcal G h$, it transforms as
\[
  \mathcal U_\Delta(i,j)
  \longrightarrow
  \mathcal R_\Delta(h_i)^{-1}
  \mathcal U_\Delta(i,j)
  \mathcal R_\Delta(h_j).
\]
A matter state at endpoint $k$ transforms as
\[
  |\Phi_{\Delta,k}\rangle
  \longrightarrow
  \mathcal R_\Delta(h_k)^{-1}|\Phi_{\Delta,k}\rangle .
\]
{Therefore $\mathcal U_\Delta(i,j)|\Phi_{\Delta,j}\rangle$ transforms as a matter state at endpoint $i$.}
{Thus the Wilson line and the matter Green's function are naturally identified in the BF description.}

{We now specialize to the particle on a group after the reduction and extract the bilocal.}
To extract a scalar from the Wilson line, choose endpoint ``polarizations'' $\langle v_i|$ and $|v_j\rangle$, which are vectors in the Hilbert space carrying the $\mathcal R_\Delta$ representations.
In a natural basis, these states label which matter component is inserted at the endpoints.
For the scalar primary we use $|v_0\rangle_\Delta$ and ${}_\Delta\langle v_0|$; here $\Delta$ labels the representation and $0$ the primary level, while the descendants $|v_n\rangle_\Delta$ span the same $\mathcal R_\Delta$ representation space.\footnote{This is why the AdS/CFT dictionary is usually stated for primary operators: the primary labels the independent bulk field.
In $\mathrm{AdS}_2/\mathrm{CFT}_1$, the normalizable scalar modes have $\omega_n=\Delta+n$, and $|v_n\rangle_\Delta$ is the representation-space version of the $SL(2,\mathbb R)$ descendant $L_{-1}^n\mathcal O|0\rangle$ under the state--operator map.}
The scalar primary polarization obeys the $\mathsf L_+$ annihilation condition, with the dual condition at the bra endpoint, so the $e^{\gamma\mathsf L_+}$ factors in the full Gauss representative act trivially in this bilocal.
The primary bilocal is
\begin{equation}
  \begin{aligned}
  \mathcal W_\Delta(i,j)
  &=
  {}_\Delta\langle v_0|
  \mathcal R_\Delta(\mathcal G_i^{-1}\mathcal G_j)
  |v_0\rangle_\Delta                                      \\
  &=
  {}_\Delta\langle v_0|
  \mathcal R_\Delta\!\left(
    e^{-\rho_i\mathsf L_0}
    e^{(x_j-x_i)\mathsf L_-}
    e^{\rho_j\mathsf L_0}
  \right)
  |v_0\rangle_\Delta .
  \end{aligned}
  \label{eq:n0-scalar-wilson-line-explicit}
\end{equation}
The primary assumption in this step is essential.  In the full Gauss
representative, $\mathcal G_i^{-1}\mathcal G_j$ contains the endpoint
factors associated with the right $\mathsf L_+$ coordinate.  For the scalar
primary polarization these factors drop out because $|v_0\rangle_\Delta$ is
annihilated by $\mathsf L_+$, with the corresponding dual annihilation
condition at the bra endpoint.  This is why the reduced Wilson-line matrix
element collapses to the single invariant power in
\eqref{eq:n0-schwarzian-bilocal-before-source}.  For a descendant polarization
this simplification is no longer true.  If
\[
  |v_n\rangle_\Delta\propto(\mathsf L_-)^n|v_0\rangle_\Delta,
\]
then $\mathsf L_+|v_n\rangle_\Delta\neq0$ for $n>0$, and the same
endpoint factors must be kept in the Wilson-line matrix element
\begin{equation}
  \mathcal W_{\Delta;n_i,n_j}(i,j)
  =
  {}_{\Delta}\langle v_{n_i}|
  \mathcal R_\Delta(\mathcal G_i^{-1}\mathcal G_j)
  |v_{n_j}\rangle_{\Delta}.
  \label{eq:n0-descendant-wilson-line}
\end{equation}
After imposing the JT source constraint, the positive Gauss coordinate is
expressed in terms of the boundary curve; in the bosonic convention used here
$\gamma=-\ddot x/(2q\dot x)$.  Descendant bilocals therefore contain
additional $\dot x,\ddot x,\ldots$ dependence, equivalently the
conformal-covariant derivatives of the primary bilocal.  Thus the Wilson-line
construction applies to all components of the representation, but only the
primary matrix element gives the simple bilocal
$(\dot x_i\dot x_j/(x_i-x_j)^2)^\Delta$.\footnote{In the supersymmetric
discussion the analogous statement is stronger.  A superconformal primary must
be annihilated not only by $\mathsf L_+$ but also by the positive fermionic
generators $\mathsf F_{+,p}$ and $\bar{\mathsf F}_{+}^{p}$.  Some component
fields can be ordinary $SL(2)$ primaries without being superconformal
primaries.  Their reduced bilocals are obtained by acting with the physical,
compensated worldline supercharges described in
Section~\ref{subsec:physical-worldline-supercharges}, not by acting with
bare bulk fermionic generators.}

{Concretely, the bilocal is a boundary-anchored restriction of this Green's function.}
\begin{equation}
  \mathcal O_\Delta(i,j)
  =
  \left[
    \frac{\dot x_i\dot x_j}{(x_i-x_j)^2}
  \right]^\Delta .
  \label{eq:n0-schwarzian-bilocal-before-source}
\end{equation}
The JT source constraint in \eqref{eq:fg-schwarzian-relation} gives $\dot x_i=q e^{-\rho_i}$ and $\dot x_j=q e^{-\rho_j}$.
From here, not surprisingly, we can write the Wilson-line operators as a function of invariants
\begin{equation}
  \mathcal O_\Delta(i,j)
  =
  q^{2\Delta}\mathcal W_\Delta(i,j)
  =
  \left(
    qL_{ij}^{(0)}
  \right)^{2\Delta},
  \qquad
  L_{ij}^{(0)}
  =
  \frac{e^{-\frac12(\rho_i+\rho_j)}}{x_j-x_i}.
  \label{eq:n0-bilocal-from-wilson-line}
\end{equation}
{The factor of $q$ is therefore not part of the bare matter Wilson line.}
{It enters only after the boundary source condition converts the endpoint frame factors into the Schwarzian Jacobians $\dot x_i$ and $\dot x_j$.}

\section{Composition kernel and IBP identities}
\label{app:I3-derivation-normalization}
\label{app:IBP-four-point-kernels}

This appendix records the integrand-level manipulations behind the composition kernel.
We use the same invariant lengths as in Section~\ref{sec:correlators-length-kernel}, but set the fixed charge to one throughout this appendix.
Thus all Bessel arguments are $2L$.

\paragraph{IBP general discussion.}
For a composition block $i\to j\to k$, define
\begin{equation}
  S_{ijk}
  =
  \frac{L_{ij}L_{jk}}{L_{ik}}
  +
  \frac{L_{jk}L_{ik}}{L_{ij}}
  +
  \frac{L_{ik}L_{ij}}{L_{jk}},
  \qquad
  E_{ijk}=e^{-S_{ijk}} .
  \label{eq:app-I3-exponential-factor}
\end{equation}
The basic IBP move uses the Bessel equation
\begin{equation}
  \mathcal L_{\nu,L}
  :=
  L\partial_L(L\partial_L)
  -
  \left(4L^2+\nu^2\right),
  \qquad
  \mathcal L_{\nu,L}K_\nu(2L)=0 .
  \label{eq:app-IBP-bessel-operator}
\end{equation}
With respect to the measure $dL$, the adjoint operator is
\begin{equation}
  \mathcal L_{\nu,L}^{\dagger}F
  :=
  \partial_L\!\left(L\partial_L(LF)\right)
  -
  \left(4L^2+\nu^2\right)F .
  \label{eq:app-IBP-bessel-adjoint}
\end{equation}
The Green identity is
\begin{equation}
  K_\nu(2L)\mathcal L_{\nu,L}^{\dagger}F
  -
  F\mathcal L_{\nu,L}K_\nu(2L)
  =
  \partial_L\mathcal B_L(K_\nu,F),
  \label{eq:app-IBP-green-identity}
\end{equation}
where
\begin{equation}
  \mathcal B_L(u,F)
  =
  L^2\left(u\partial_LF-(\partial_Lu)F\right)+LuF .
  \label{eq:app-IBP-boundary-concomitant}
\end{equation}
We also use
\begin{equation}
  \frac12\partial_LK_0(2L)
  =
  -K_1(2L),
  \qquad
  \frac12\partial_L\!\left[2LK_1(2L)\right]
  =
  -2LK_0(2L).
  \label{eq:app-IBP-basic-derivatives}
\end{equation}
These identities convert mixed $K_0K_1$ terms into total derivatives plus $K_0K_0$ or $K_1K_1$ terms.

\paragraph{\texorpdfstring{$\mathcal N=2$}{N=2} first-order IBP and its OTOC.}
The $\mathcal N=2$ kernel in \eqref{eq:section4-N2-I3} has a useful first-order form.
For the block $1\to2\to3$, let $E_{123}$ be the exponential in \eqref{eq:app-I3-exponential-factor} and define
\begin{equation}
  \mathcal O_{ab}:=2L_{ab}-L_{ab}\partial_{L_{ab}} .
  \label{eq:app-N2-first-order-operator}
\end{equation}
Then
\begin{equation}
  I_3^{\mathcal N=2}(L_{12},L_{23},L_{13})
  =
  \left[
    2\left(
      \mathcal O_{12}
      +
      \mathcal O_{23}
      +
      \mathcal O_{13}
    \right)-1
  \right]E_{123}.
  \label{eq:app-N2-I3-first-order}
\end{equation}
Indeed, acting on the exponential gives
\begin{equation}
  \left[
    2\left(
      \mathcal O_{12}
      +
      \mathcal O_{23}
      +
      \mathcal O_{13}
    \right)-1
  \right]E_{123}
  =
  \left[
    -1
    +
    \frac{2}{L_{12}L_{23}L_{13}}
    (L_{12}L_{23}+L_{23}L_{13}+L_{13}L_{12})^2
  \right]E_{123}.
  \label{eq:app-N2-I3-first-order-check}
\end{equation}

As a check, apply this identity to the $\mathcal N=2$ composition integral.
The zero-energy radial factor obeys
\begin{equation}
  \left(
    \partial_L+2-\frac{1}{2L}
  \right)\Psi_0^{\mathcal N=2}(L)=0,
  \qquad
  \Psi_0^{\mathcal N=2}(L)=\frac{2L}{\sqrt{\pi}}K_{1/2}(2L).
  \label{eq:app-N2-wavefunction-first-order}
\end{equation}
Equivalently, with respect to the measure $dL/L$, the adjoint of $\mathcal O_{ab}$ acts on the external wavefunction as
\begin{equation}
  \mathcal O_{ab}^{\dagger}\Psi_0^{\mathcal N=2}(L_{ab})
  =
  \frac12\Psi_0^{\mathcal N=2}(L_{ab}) .
  \label{eq:app-N2-wavefunction-adjoint}
\end{equation}
Therefore the two adjacent derivatives in \eqref{eq:app-N2-I3-first-order} can be moved onto the two external wavefunctions:
\begin{equation}
  \begin{aligned}
  &\int_0^\infty\frac{dL_{12}}{L_{12}}
  \frac{dL_{23}}{L_{23}}\,
  \Psi_0^{\mathcal N=2}(L_{12})
  \Psi_0^{\mathcal N=2}(L_{23})
  I_3^{\mathcal N=2}(L_{12},L_{23},L_{13})
  \\
  &\qquad =
  \left(1+2\mathcal O_{13}\right)
  \int_0^\infty\frac{dL_{12}}{L_{12}}
  \frac{dL_{23}}{L_{23}}\,
  \Psi_0^{\mathcal N=2}(L_{12})
  \Psi_0^{\mathcal N=2}(L_{23})
  E_{123}.
  \end{aligned}
  \label{eq:app-N2-composition-check}
\end{equation}
After some manipulation the remaining scalar integral is identical to Appendix I of \cite{Lin:2022zxd}.

We now use the same first-order identity for the four-point function.
For the two composition blocks in \eqref{eq:section4-N2-pre-IBP-two-I3}, set
\[
  F_\Delta=
  \left(
    \frac{L_{12}L_{23}L_{34}L_{41}}
         {L_{12}L_{34}+L_{23}L_{41}}
  \right)^{2\Delta}.
\]
\stepcounter{equation}
Then the product of kernels before doing the $L_{13}$-integral is
\begin{align}
  &F_\Delta\,
  I_3^{\mathcal N=2}(L_{12},L_{23},L_{13})\,
  I_3^{\mathcal N=2}(L_{34},L_{41},L_{13})
  \nonumber\\
  &\qquad =
  F_\Delta\,
  \left[
    2\left(
      \mathcal O_{12}+\mathcal O_{23}+\mathcal O_{13}
    \right)-1
  \right]E_{123}
  \left[
    2\left(
      \mathcal O_{34}+\mathcal O_{41}+\mathcal O_{13}
    \right)-1
  \right]E_{341},
  \label{eq:app-N2-fourpoint-operator-form}
\end{align}
This is the form used for the IBP reduction of \eqref{eq:section4-N2-pre-IBP-two-I3}.
When $\mathcal O_{12},\mathcal O_{23},\mathcal O_{34},\mathcal O_{41}$ are moved by parts, the derivatives act on $F_\Delta$.
The basic derivatives are
\[
  \begin{aligned}
  L_{12}\partial_{L_{12}}\log F_\Delta
  =
  L_{34}\partial_{L_{34}}\log F_\Delta
  =
  2\Delta\,\frac{L_{23}L_{41}}
  {L_{12}L_{34}+L_{23}L_{41}},
  \\
  L_{23}\partial_{L_{23}}\log F_\Delta
  =
  L_{41}\partial_{L_{41}}\log F_\Delta
  =
  2\Delta\,\frac{L_{12}L_{34}}
  {L_{12}L_{34}+L_{23}L_{41}} .
  \end{aligned}
\]
\stepcounter{equation}
The diagonal operator $\mathcal O_{13}$ acts only on the two exponentials, since $F_\Delta$ is independent of $L_{13}$.
After the $L_{13}$-integral is performed, the three pieces in \eqref{eq:app-N2-fourpoint-operator-form} give the five terms in \eqref{eq:section4-N2-T1}--\eqref{eq:section4-N2-T5}: the adjacent-adjacent action gives the $\Delta(2\Delta+1)$ terms $T_1,T_2,T_3$, the mixed adjacent-diagonal action gives $T_4$, and the diagonal-diagonal action gives $T_5$.
In this way the IBP reduction turns the two $I_3^{\mathcal N=2}$ kernels into the four-length expression \eqref{eq:section4-N2-IBP-integrand}.

\paragraph{\texorpdfstring{$\mathcal N=4$}{N=4} IBP structure.}
For the $\mathcal N=4$ identities we use the same endpoint notation as in the main text.
For one composition block $i\to j\to k$, we use the exponential $E_{ijk}$ defined in \eqref{eq:app-I3-exponential-factor}.
The two component matrices $\mathbb I_{ijk}^{(0)}$ and $\mathbb I_{ijk}^{(1)}$ in \eqref{eq:section4-I3-matrix-zero}--\eqref{eq:section4-I3-matrix-one} give two scalar densities after they are placed between the $K_0/K_1$ Bessel function structures.
The direct expressions are long, but the IBP form is short.
The source factors are
\begin{equation}
\begin{aligned}
  R_{ijk}^{(0)}
  &=
  -\frac{L_{ik}(L_{ij}^2+L_{jk}^2)}
  {L_{ij}L_{jk}}\,E_{ijk},
  &
  Q_{ij}^{(0)}
  &=
  \left(
  \frac14+\frac{L_{ij}L_{ik}}{2L_{jk}}
  \right)E_{ijk},
  &
  Q_{jk}^{(0)}
  &=
  \left(
  \frac14+\frac{L_{jk}L_{ik}}{2L_{ij}}
  \right)E_{ijk},
  \\
  R_{ijk}^{(1)}
  &=
  0,
  &
  Q_{ij}^{(1)}
  &=
  \frac{L_{ij}L_{ik}}{2L_{jk}}E_{ijk},
  &
  Q_{jk}^{(1)}
  &=
  \frac{L_{jk}L_{ik}}{2L_{ij}}E_{ijk}.
\end{aligned}
  \label{eq:app-IBP-Q-functions}
\end{equation}
Let $I_{ijk}^{(0)}$ and $I_{ijk}^{(1)}$ denote the two resulting densities, with the Bessel  function factors included and with respect to the ordinary measure $dL_{ij}\,dL_{jk}$.
The pointwise IBP identities are
\begin{align}
  I_{ijk}^{(0)}
  &=
  \left[
    R_{ijk}^{(0)}
    +
    \mathcal L_{0,L_{ij}}^{\dagger}Q_{ij}^{(0)}
    +
    \mathcal L_{0,L_{jk}}^{\dagger}Q_{jk}^{(0)}
  \right]
  K_0(2L_{ij})K_0(2L_{jk})
  \nonumber\\
  &\quad
  +\frac12\partial_{L_{ij}}J_{ij}^{(0)}
  +\frac12\partial_{L_{jk}}J_{jk}^{(0)},
  \label{eq:app-IBP-I0-pointwise}
  \\
  I_{ijk}^{(1)}
  &=
  -
  \left[
    \mathcal L_{1,L_{ij}}^{\dagger}Q_{ij}^{(1)}
    +
    \mathcal L_{1,L_{jk}}^{\dagger}Q_{jk}^{(1)}
  \right]
  K_1(2L_{ij})K_1(2L_{jk})
  \nonumber\\
  &\quad
  +\frac12\partial_{L_{ij}}J_{ij}^{(1)}
  +\frac12\partial_{L_{jk}}J_{jk}^{(1)}.
  \label{eq:app-IBP-I1-pointwise}
\end{align}
Here the currents $J$ are explicit total-derivative terms.
Their detailed form is not needed for the composition check; what matters is that they must be kept until the boundary behavior is understood.
The important distinction is already visible in \eqref{eq:app-IBP-I0-pointwise}--\eqref{eq:app-IBP-I1-pointwise}: the $K_0$ channel has a genuine bulk remainder $R_{ijk}^{(0)}$, while the $K_1$ channel is adjoint-exact in the bulk.

\paragraph{\texorpdfstring{$\mathcal N=4$}{N=4} composition kernel check.}
We now apply the preceding identities to the composition law.
The result is the pair of radial projections
\begingroup
\small
\setlength{\arraycolsep}{1pt}
\begin{align}
  &\int_0^\infty \frac{dL_{ij}}{L_{ij}}\frac{dL_{jk}}{L_{jk}}\,
  \frac{E_{ijk}}{2L_{ik}}\!
  \begin{pmatrix}(2L_{ij})^2K_0(2L_{ij})&(2L_{ij})^2K_1(2L_{ij})\end{pmatrix}
  \mathbb I^{(0)}_{ijk}
  \begin{pmatrix}(2L_{jk})^2K_0(2L_{jk})\\ (2L_{jk})^2K_1(2L_{jk})\end{pmatrix}
  =
  -2L_{ik}K_0(2L_{ik}),
  \label{eq:app-I3-zero-action}
  \\
  &\int_0^\infty \frac{dL_{ij}}{L_{ij}}\frac{dL_{jk}}{L_{jk}}\,
  \frac{E_{ijk}}{2L_{ik}}\!
  \begin{pmatrix}(2L_{ij})^2K_0(2L_{ij})&(2L_{ij})^2K_1(2L_{ij})\end{pmatrix}
  \mathbb I^{(1)}_{ijk}
  \begin{pmatrix}(2L_{jk})^2K_0(2L_{jk})\\ (2L_{jk})^2K_1(2L_{jk})\end{pmatrix}
  =
  2L_{ik}K_1(2L_{ik}).
  \label{eq:app-I3-one-action}
\end{align}
\endgroup

Let us indicate how the IBP identities give these two lines.
For the $K_0$ component, the adjoint Bessel operators in \eqref{eq:app-IBP-I0-pointwise} can be moved onto $K_0(2L_{ij})$ and $K_0(2L_{jk})$.
The bulk part of the Bessel operator action vanishes by the Bessel equation, and the regulated boundary terms vanish; the possible corner $L_{ij},L_{jk}\sim\epsilon$ does not leave a finite contribution because $K_0(z)\sim-\log z$.
Thus only the simple bulk remainder $R_{ijk}^{(0)}$ term survives.
\begin{equation}
  \int_0^\infty dL_{ij}\,dL_{jk}\,I_{ijk}^{(0)}
  =
  -\int_0^\infty dL_{ij}\,dL_{jk}\,
  \frac{L_{ik}(L_{ij}^2+L_{jk}^2)}{L_{ij}L_{jk}}\,
  E_{ijk}K_0(2L_{ij})K_0(2L_{jk})
  =
  -2L_{ik}K_0(2L_{ik}) .
  \label{eq:app-I3-K0-IBP-check}
\end{equation}
For the $K_1$ component, the bulk is adjoint-exact and there is no analogue of $R_{ijk}^{(0)}$.
The whole contribution comes from the regulated corner.
More explicitly, regulate the quadrant by
\[
  D_\epsilon=\{(L_{ij},L_{jk}):L_{ij}\geq\epsilon,\ L_{jk}\geq\epsilon\}.
\]
The boundary terms at infinity vanish exponentially, and the only finite contribution comes from the two small-length edges near the corner $L_{ij},L_{jk}\sim\epsilon$.
The boundary one-form which appears here is not a new input.
It is the boundary term obtained from the two Green identities in \eqref{eq:app-IBP-I1-pointwise}, together with the explicit total derivatives in the same equation.
Here $\mathcal B_L(u,F)=L^2(u\partial_LF-(\partial_Lu)F)+LuF$.
\begingroup
\small
\begin{equation}
  dI_{1,\partial}
  =
  \left[-K_1(2L_{jk})\mathcal B_{L_{ij}}\!\left(K_1(2L_{ij}),Q_{ij}^{(1)}\right)+\frac12J_{ij}^{(1)}\right]dL_{jk}
  -
  \left[-K_1(2L_{ij})\mathcal B_{L_{jk}}\!\left(K_1(2L_{jk}),Q_{jk}^{(1)}\right)+\frac12J_{jk}^{(1)}\right]dL_{ij}.
  \label{eq:app-I1-boundary-one-form}
\end{equation}
\endgroup
Thus $dI_{1,\partial}$ is just the boundary form of the adjoint-exact $K_1$ density.
Parametrizing the small corner by $L_{ij}=\epsilon,\ L_{jk}=\epsilon t$, and adding the exchanged edge, gives
\begin{equation}
  \begin{aligned}
  \int_0^\infty dL_{ij}\,dL_{jk}\,I_{ijk}^{(1)}
  &=
  \lim_{\epsilon\to0}
  \int_{\partial D_\epsilon} dI_{1,\partial}
  =
  \int_1^\infty
  L_{ik}\left(1+\frac1{t^2}\right)
  e^{-L_{ik}(t+1/t)}dt
  \\
  &=
  L_{ik}\int_0^\infty
  t^{-2}e^{-L_{ik}(t+1/t)}dt
  =
  2L_{ik}K_1(2L_{ik}) .
  \end{aligned}
  \label{eq:app-IBP-corner-cK1}
\end{equation}

\paragraph{\texorpdfstring{$\mathcal N=4$}{N=4} IBP for correlation integrands.}
For correlation functions, the same block is multiplied by the remaining length dependence before the $L_{ij},L_{jk}$ integrals are done.
Let $T(L_{ij},L_{jk})$ denote this extra factor, with $L_{ik}$ held fixed while the IBP is performed on the two lengths $L_{ij}$ and $L_{jk}$.
Define
\begin{equation}
  B_\alpha^{ijk}[T]
  :=
  \int_0^\infty dL_{ij}\,dL_{jk}\,
  I_{ijk}^{(\alpha)}\,T(L_{ij},L_{jk}).
  \label{eq:app-IBP-block-functional}
\end{equation}
Substituting \eqref{eq:app-IBP-I0-pointwise}--\eqref{eq:app-IBP-I1-pointwise} gives
\begin{align}
  B_\alpha^{ijk}[T]
  &=
  \int dL_{ij}\,dL_{jk}\,
  \widehat I_{ijk}^{(\alpha)}\,T
  -
  \int dL_{ij}\,dL_{jk}\,
  \left(
    \frac12J_{ij}^{(\alpha)}\partial_{L_{ij}}T
    +
    \frac12J_{jk}^{(\alpha)}\partial_{L_{jk}}T
  \right)
  \nonumber\\
  &\quad
  +
  B_{\alpha,\partial}^{ijk}[T],
  \label{eq:app-IBP-spectator-rule}
\end{align}
where $\widehat I_{ijk}^{(\alpha)}$ denotes the non-current part of \eqref{eq:app-IBP-I0-pointwise} or \eqref{eq:app-IBP-I1-pointwise}.
For $T=1$ the current terms integrate to boundary contributions.
For the three-point function in \eqref{eq:section4-N4-integrated-three-point}, the relevant choice is the monomial
\[
  T_3(L_{ij},L_{jk};L_{ik})
  =
  L_{ij}^{\alpha_{ij}}L_{jk}^{\alpha_{jk}}L_{ik}^{\alpha_{ik}},
  \qquad
  \partial_{L_{ij}}T_3=\frac{\alpha_{ij}}{L_{ij}}T_3,
  \qquad
  \partial_{L_{jk}}T_3=\frac{\alpha_{jk}}{L_{jk}}T_3 .
\]
Thus the IBP reduction is still an integrand-level statement.
It rewrites the $I_3^{\mathcal N=4}$ numerator as a finite sum of terms with shifted powers of $L_{ij}$, $L_{jk}$, and $L_{ik}$, multiplying the same exponential $E_{ijk}$ and the same $K_0/K_1$ wavefunctions.
For positive powers $\alpha_{ab}$ the small-length boundary terms are suppressed, so the three-point calculation is reduced to ordinary Bessel function length integrals of this finite list of shifted monomials.
In this case the final bulk representative is still short:
\begin{align}
  B_0^{ijk}[T_3]
  &=
  \frac14
  \int dL_{ij}\,dL_{jk}\,
  E_{ijk}K_0(2L_{ij})K_0(2L_{jk})\,\mathcal M_0,
  \nonumber\\
  B_1^{ijk}[T_3]
  &=
  \frac14
  \int dL_{ij}\,dL_{jk}\,
  E_{ijk}K_1(2L_{ij})K_1(2L_{jk})\,\mathcal M_1,
  \label{eq:app-N4-threepoint-IBP-representative}
\end{align}
where
\begin{align}
  \mathcal M_0
  &=
  L_{ij}^{\alpha_{ij}-1}L_{jk}^{\alpha_{jk}-1}L_{ik}^{\alpha_{ik}-1}
  \Big[
  L_{ij}^2L_{jk}^2
  (\alpha_{ij}+\alpha_{ij}^2+\alpha_{jk}-\alpha_{ij}\alpha_{jk}+\alpha_{jk}^2)
  \nonumber\\
  &\qquad
  -L_{ij}L_{jk}L_{ik}(2+\alpha_{ij}+\alpha_{jk})
  (\alpha_{ij}+\alpha_{ij}^2+\alpha_{jk}+\alpha_{jk}^2)
  \nonumber\\
  &\qquad
  -L_{ik}^2(2+\alpha_{ij}+\alpha_{jk})
  \left(L_{ij}^2(1+\alpha_{jk})+(1+\alpha_{ij})L_{jk}^2\right)
  \Big],
  \nonumber\\
  \mathcal M_1
  &=
  L_{ij}^{\alpha_{ij}-1}L_{jk}^{\alpha_{jk}-1}L_{ik}^{\alpha_{ik}-1}
  \Big[
  -L_{ij}^2L_{jk}^2(\alpha_{ij}^2-\alpha_{ij}\alpha_{jk}+\alpha_{jk}^2)
  \nonumber\\
  &\qquad
  +L_{ij}L_{jk}L_{ik}(2+\alpha_{ij}+\alpha_{jk})
  (\alpha_{ij}^2+\alpha_{jk}^2)
  \nonumber\\
  &\qquad
  +L_{ik}^2(2+\alpha_{ij}+\alpha_{jk})
  \left(\alpha_{jk}L_{ij}^2+\alpha_{ij}L_{jk}^2\right)
  \Big].
  \label{eq:app-N4-threepoint-IBP-polynomials}
\end{align}
This is the form used before doing the length integrals which lead to \eqref{eq:section4-N4-three-point-answer}.

For the scalar OTOC the remaining factor is instead the product of the second composition block and the scalar bilocals.
In that case the derivatives in \eqref{eq:app-IBP-spectator-rule} act both on the second kernel and on $\mathcal O_\Delta(1,3)\mathcal O_\Delta(2,4)$, producing the longer finite sum of four-point integrands used in Section~\ref{sec:correlators-length-kernel}.

\section{\texorpdfstring{$\mathcal N=0$}{N=0} OTOC from length integrals}
\label{app:N0-I3-gamma-Wilson}

This appendix gives an alternative derivation of the bosonic JT gravity OTOC.
The starting point is the two-$I_3$ length-integral representation set up in Section~\ref{sec:correlators-length-kernel}; in the $\mathcal N=0$ limit it reduces to the scalar $I_3I_3$ integral below.
We then perform the length integrals in two steps: first rewriting the OTOC kernel as a two-variable Gamma integral, and then reducing that Gamma integral by Barnes identities.
The result is the standard Wilson-function expression for the JT crossing kernel, equivalently the $SL(2,\mathbb R)$ $6j$ symbol.
This provides a self-contained check that the length-integral representation reproduces the known bosonic JT OTOC kernel.
All Barnes contours below are vertical contours chosen in a nonempty convergence strip.
For an integrand containing $\Gamma(A+z)\Gamma(B-z)$, this means that the poles $z=-A-n$ are on the left of the contour and the poles $z=B+n$ are on the right; the final expressions are understood by meromorphic continuation.

\paragraph{The length integral.}

In the bosonic normalization used in this appendix, $\Psi_s(L)=\frac{2}{\pi}K_{2is}(2q L)$ and $\rho(s)=2s\sinh(2\pi s)$, so that
\[
  \int_0^\infty\frac{dL}{L}\,
  \Psi_s(L)\Psi_{s'}(L)
  =
  \frac{\delta(s-s')}{\rho(s)} .
\]
This fixes the bosonic JT density convention used below.
The composition-normalized scalar three-length kernel is
\begin{equation}
  I_3(L_a,L_b,L_c)
  =
  \frac{1}{\pi}
  \exp\!\left[
  -q\left(
    \frac{L_aL_b}{L_c}
    +\frac{L_bL_c}{L_a}
    +\frac{L_cL_a}{L_b}
  \right)\right],
  \label{eq:app-N0-scalar-I3}
\end{equation}
Equivalently,
\begin{equation}
\begin{aligned}
  I_3(L_a,L_b,L_c)
  &=
  \int_0^\infty ds\,\rho(s)\,
  \Psi_s(L_a)\Psi_s(L_b)\Psi_s(L_c)
  \\
  &=
  \frac{4}{\pi^2}
  \int_{-\infty}^{+\infty}
  \frac{ds}{\Gamma(2is)\Gamma(-2is)}
  K_{2is}(2q L_a)
  K_{2is}(2q L_b)
  K_{2is}(2q L_c).
\end{aligned}
  \label{eq:app-N0-scalar-I3-KL}
\end{equation}
with the same normalization convention.
Gluing two copies of \eqref{eq:app-N0-scalar-I3} gives
\begin{equation}
  \begin{aligned}
  I_4(L_1,L_2,L_3,L_4)
  &=
  \int_0^\infty\frac{dL}{L}\,
  I_3(L_1,L_2,L)I_3(L_3,L_4,L)
  \\
  &=
  \frac{2}{\pi^2}K_0\!\left(
  2q
  \sqrt{
    \frac{
    (L_1L_2+L_3L_4)
    (L_1L_3+L_2L_4)
    (L_1L_4+L_2L_3)}
    {L_1L_2L_3L_4}}
  \right).
  \end{aligned}
  \label{eq:app-N0-I3I3-to-I4}
\end{equation}
The same kernel has the full-line Kontorovich--Lebedev representation
\begin{equation}
  I_4(L_1,L_2,L_3,L_4)
  =
  \frac{8}{\pi^3}
  \int_{-\infty}^{+\infty}
  \frac{ds}{\Gamma(2is)\Gamma(-2is)}
  \prod_{j=1}^{4}K_{2is}(2q L_j).
  \label{eq:app-N0-I4-KL}
\end{equation}

For equal external matter dimension $\Delta$, the stripped bosonic length-space OTOC block can therefore be written as
\begin{equation}
  \mathcal A^{(0)}_\Delta
  =
  \int
  \prod_{j=1}^{4}\frac{dL_j}{L_j}\,
  \Psi_{s_j}(L_j)\,
  \left(
    \frac{L_1L_2L_3L_4}{L_1L_3+L_2L_4}
  \right)^{2\Delta}
  I_4(L_1,L_2,L_3,L_4).
  \label{eq:app-N0-length-OTOC}
\end{equation}
The rest of the appendix evaluates this universal kernel.

\paragraph{$\Gamma$ integral and Barnes reduction.}
To turn \eqref{eq:app-N0-length-OTOC} into a Gamma integral, use \eqref{eq:app-N0-I4-KL} together with the identity
\begin{equation}
  \frac{1}{(X+Y)^{2\Delta}}
  =
  \frac{1}{(XY)^\Delta}
  \int_{-\infty}^{+\infty}\frac{d\kappa}{2\pi}
  \left(\frac{X}{Y}\right)^{i\kappa}
  \frac{\Gamma(\Delta+i\kappa)\Gamma(\Delta-i\kappa)}
       {\Gamma(2\Delta)} .
  \label{eq:app-N0-beta-identity}
\end{equation}
for $X,Y>0$ and $\Re\Delta>0$, with $X=L_1L_3$ and $Y=L_2L_4$.
After substituting \eqref{eq:app-N0-beta-identity} and \eqref{eq:app-N0-I4-KL}, the four length integrals factorize.
Using the shorthand for paired Gamma factors, the needed Mellin--Bessel transform is
\begin{equation}
  \int_0^\infty\frac{dL}{L}\,
  L^{\lambda}K_{2is}(2q L)K_{2is'}(2q L)
  =
  \frac{q^{-\lambda}}{8\,\Gamma(\lambda)}
  \Gamma\!\left(\frac{\lambda}{2}\pm is\pm is'\right).
  \label{eq:app-N0-KK-transform}
\end{equation}
The odd legs $1,3$ carry $\lambda=\Delta+i\kappa$, while the even legs $2,4$ carry $\lambda=\Delta-i\kappa$.
The four factors of $q^{-\lambda}$ give the explicit factor $q^{-4\Delta}$ below.
Thus \eqref{eq:app-N0-length-OTOC} becomes
\begin{equation}
  \mathcal A^{(0)}_\Delta
  =
  q^{-4\Delta}
  \left(\frac{2}{\pi}\right)^4
  \frac{1}{2^8\pi^2\,\Gamma(2\Delta)}
  \mathcal G_\Delta(s_1,s_2,s_3,s_4),
  \label{eq:app-N0-A-from-G}
\end{equation}
where the two-variable Gamma integral is
\begin{equation}
  \begin{aligned}
  \mathcal G_\Delta
  &=
  \int_{\mathbb R}\frac{d\kappa}{2\pi}
  \int_{\mathbb R}\frac{ds}{2\pi}\,
  \frac{
  \displaystyle
  \prod_{\ell=1,3}
  \Gamma\!\left(\frac{\Delta+i\kappa}{2}\pm is_\ell\pm is\right)
  \prod_{m=2,4}
  \Gamma\!\left(\frac{\Delta-i\kappa}{2}\pm is_m\pm is\right)
  }{
  \Gamma(\Delta+i\kappa)\Gamma(\Delta-i\kappa)
  \Gamma(2is)\Gamma(-2is)
  } .
  \end{aligned}
  \label{eq:app-N0-two-variable-gamma}
\end{equation}
This is the promised Gamma integral.
Starting from here, we evaluate \eqref{eq:app-N0-two-variable-gamma} in several non-trivial steps, which leave a single Mellin--Barnes integral.

Set
\begin{equation}
  z_1=i\left(\frac{\kappa}{2}+s\right),
  \qquad
  z_2=i\left(\frac{\kappa}{2}-s\right).
  \label{eq:app-N0-z-variables}
\end{equation}
Then $i\kappa=z_1+z_2$ and $2is=z_1-z_2$, and
\begin{equation}
  \mathcal G_\Delta
  =
  \int\frac{dz_1}{2\pi i}\frac{dz_2}{2\pi i}\,
  \frac{F(z_1)F(z_2)}
  {\Gamma(z_1-z_2)\Gamma(z_2-z_1)
   \Gamma(\Delta+z_1+z_2)\Gamma(\Delta-z_1-z_2)},
  \label{eq:app-N0-G-barnes}
\end{equation}
where
\begin{equation}
  F(z)
  =
  \Gamma\!\left(\frac{\Delta}{2}+z\pm is_1\right)
  \Gamma\!\left(\frac{\Delta}{2}+z\pm is_3\right)
  \Gamma\!\left(\frac{\Delta}{2}-z\pm is_2\right)
  \Gamma\!\left(\frac{\Delta}{2}-z\pm is_4\right).
  \label{eq:app-N0-Fz}
\end{equation}

The motivation for the next two steps is to first disentangle the $z_i$'s from the external $s_i$'s, and then recombine some of the $s_i$'s into the factors multiplying the final one-fold integral.
The two diagonal denominators in \eqref{eq:app-N0-G-barnes} are opened by Barnes' first lemma:
\begin{align}
  \frac{
  \prod_{j=1}^{2}\Gamma\!\left(\frac{\Delta}{2}+z_j\pm is_1\right)
  }{
  \Gamma(\Delta+z_1+z_2)
  }
  &=
  \int_C\frac{du}{2\pi i}\,
  \Gamma(z_1+u)\Gamma(z_2+u)
  \Gamma\!\left(\frac{\Delta}{2}-u\pm is_1\right),
  \label{eq:app-N0-B1-open-u}
  \\
  \frac{
  \prod_{j=1}^{2}\Gamma\!\left(\frac{\Delta}{2}-z_j\pm is_2\right)
  }{
  \Gamma(\Delta-z_1-z_2)
  }
  &=
  \int_C\frac{dv}{2\pi i}\,
  \Gamma(v-z_1)\Gamma(v-z_2)
  \Gamma\!\left(\frac{\Delta}{2}-v\pm is_2\right).
  \label{eq:app-N0-B1-open-v}
\end{align}
After \eqref{eq:app-N0-B1-open-u} and \eqref{eq:app-N0-B1-open-v}, the $z_1,z_2$ integrals have exactly Gustafson's $A_2$ form, with
\begin{equation}
  \boldsymbol p=\left(u,\frac{\Delta}{2}+is_3,\frac{\Delta}{2}-is_3\right),
  \qquad
  \boldsymbol q=\left(v,\frac{\Delta}{2}+is_4,\frac{\Delta}{2}-is_4\right).
  \label{eq:app-N0-A2-parameters}
\end{equation}
The $A_2$ integral gives
\begin{equation}
  \begin{aligned}
  \mathcal G_\Delta
  &=
  2\,\Gamma(\Delta\pm is_3\pm is_4)
  \int_C\frac{du}{2\pi i}
  \int_C\frac{dv}{2\pi i}\,
  \Gamma\!\left(\frac{\Delta}{2}-u\pm is_1\right)
  \Gamma\!\left(\frac{\Delta}{2}-v\pm is_2\right)
  \\
  &\hspace{2cm}\times
  \frac{
  \Gamma(u+v)
  \Gamma\!\left(u+\frac{\Delta}{2}\pm is_4\right)
  \Gamma\!\left(v+\frac{\Delta}{2}\pm is_3\right)
  }{
  \Gamma(u+v+2\Delta)
  } .
  \end{aligned}
  \label{eq:app-N0-after-A2}
\end{equation}
The last nontrivial step is Barnes' second lemma on the $v$-contour.
After applying it, one obtains
\begin{equation}
  \mathcal G_\Delta
  =
  \frac{
  2\,\Gamma(\Delta\pm is_3\pm is_4)
  \Gamma(\Delta\pm is_3\pm is_2)
  }{
  \Gamma(2\Delta)
  }
  \,
  \int_C\frac{du}{2\pi i}\,
  \frac{
  \Gamma\!\left(\frac{\Delta}{2}-u\pm is_1\right)
  \Gamma\!\left(u+\frac{\Delta}{2}\pm is_2\right)
  \Gamma\!\left(u+\frac{\Delta}{2}\pm is_4\right)
  }{
  \Gamma\!\left(u+\frac{3\Delta}{2}+is_3\right)
  \Gamma\!\left(u+\frac{3\Delta}{2}-is_3\right)
  } .
  \label{eq:app-N0-G-to-onefold}
\end{equation}
A convenient contour has $-\frac12\operatorname{Re}\Delta<\operatorname{Re}u<\frac12\operatorname{Re}\Delta$, so that the poles of $\Gamma(u+\frac{\Delta}{2}\pm is_2)\Gamma(u+\frac{\Delta}{2}\pm is_4)$ lie on the left and the poles of $\Gamma(\frac{\Delta}{2}-u\pm is_1)$ lie on the right.

\paragraph{Residues and the Wilson-function answer.}

Closing the $u$-contour to the right gives two pole families, from $u=\frac{\Delta}{2}+i\epsilon s_1+n$ with $\epsilon=\pm1$.
For nonresonant spectral labels this gives
\begin{equation}
  \begin{aligned}
  &\int_C\frac{du}{2\pi i}\,
  \frac{
  \Gamma\!\left(\frac{\Delta}{2}-u\pm is_1\right)
  \Gamma\!\left(u+\frac{\Delta}{2}\pm is_2\right)
  \Gamma\!\left(u+\frac{\Delta}{2}\pm is_4\right)
  }{
  \Gamma\!\left(u+\frac{3\Delta}{2}+is_3\right)
  \Gamma\!\left(u+\frac{3\Delta}{2}-is_3\right)
  }
  \\
  &\qquad =
  \sum_{\epsilon=\pm1}
  \frac{\Gamma(-2i\epsilon s_1)}{
  \Gamma(2\Delta+i\epsilon s_1+is_3)
  \Gamma(2\Delta+i\epsilon s_1-is_3)
  }
  \Gamma(\Delta+i\epsilon s_1\pm is_2)
  \Gamma(\Delta+i\epsilon s_1\pm is_4)
  \\
  &\qquad\quad\times
  {}_4F_3\!\left(
  \begin{matrix}
  \Delta+i\epsilon s_1+is_2,\,
  \Delta+i\epsilon s_1-is_2,\,
  \Delta+i\epsilon s_1+is_4,\,
  \Delta+i\epsilon s_1-is_4
  \\
  2\Delta+i\epsilon s_1+is_3,\,
  2\Delta+i\epsilon s_1-is_3,\,
  1+2i\epsilon s_1
  \end{matrix}
  ;1
  \right).
  \end{aligned}
  \label{eq:app-N0-onefold-residue-4F3}
\end{equation}
The two ${}_4F_3(1)$'s are balanced.

Equation \eqref{eq:app-N0-onefold-residue-4F3} is already the Wilson-function answer in a nonstandard $s_3$-channel.
To make the connection to the usual JT OTOC kernel explicit, let $L[\cdots]$ denote Bailey's normalized $L$-function, equivalently the standard Wilson-function package written as a sum of two balanced ${}_4F_3(1)$'s.
Then
\begin{equation}
  \begin{aligned}
  &\int_C\frac{du}{2\pi i}\,
  \frac{
  \Gamma\!\left(\frac{\Delta}{2}-u\pm is_1\right)
  \Gamma\!\left(u+\frac{\Delta}{2}\pm is_2\right)
  \Gamma\!\left(u+\frac{\Delta}{2}\pm is_4\right)
  }{
  \Gamma\!\left(u+\frac{3\Delta}{2}+is_3\right)
  \Gamma\!\left(u+\frac{3\Delta}{2}-is_3\right)
  }
  \\
  &\qquad =
  \pi\,
  \Gamma(\Delta\pm is_1\pm is_2)
  \Gamma(\Delta\pm is_1\pm is_4)
  \\
  &\times
  L\!\left[
  \begin{array}{@{}c@{}}
  \Delta+i(s_1+s_4),\,
  \Delta+i(s_1-s_4),\,
  \Delta+i(s_1+s_2),\,
  \Delta+i(s_1-s_2)
  \\
  1+2is_1,\,
  2\Delta+i(s_1-s_3),\,
  2\Delta+i(s_1+s_3)
  \end{array}
  \right].
  \end{aligned}
  \label{eq:app-N0-onefold-as-L}
\end{equation}
For comparison, the standard Schwarzian/JT crossing kernel is usually written in the JT channel.
A one-fold Mellin--Barnes form is obtained in the Schwarzian limit of the Ponsot--Teschner kernel in \cite[App.~F]{Mertens:2017mtv}; the same $SL(2,\mathbb R)$ $6j$ is the crossing coefficient for boundary-anchored JT Wilson lines in \cite{Blommaert:2018oro}.
With equal external dimensions and with the momenta relabeled to the present notation, the standard JT answer is
\begin{equation}
  \begin{aligned}
  \mathcal A^{\rm JT}_{\Delta}
  &=
  \frac{
  \Gamma(\Delta+is_1\pm is_4)
  \Gamma(\Delta-is_3\pm is_2)
  \Gamma(\Delta-is_1\pm is_2)
  \Gamma(\Delta+is_3\pm is_4)
  }{
  \Gamma(2\Delta)^2
  }
  \\
  &\qquad\times
  \int_{C_u}\frac{du}{2\pi i}\,
  \frac{
  \Gamma(u\pm is_4)\,
  \Gamma(u+i(s_1+s_3)\pm is_2)\,
  \Gamma(\Delta-is_1-u)\,
  \Gamma(\Delta-is_3-u)
  }{
  \Gamma(u+\Delta+is_1)\,
  \Gamma(u+\Delta+is_3)
  } .
  \end{aligned}
  \label{eq:app-N0-JT-full}
\end{equation}
The Barnes integral in \eqref{eq:app-N0-JT-full} gives another Bailey $L$-function,
\begin{equation}
  \begin{aligned}
  L_{\rm JT}(s_1,s_2,s_3,s_4;\Delta)
  &=
  L\!\left[
  \begin{array}{@{}c@{}}
  \Delta-is_1+is_4,\,
  \Delta-is_1-is_4,\,
  \Delta+is_3+is_2,\,
  \Delta+is_3-is_2
  \\
  1+i(s_3-s_1),\,
  2\Delta,\,
  2\Delta+i(s_3-s_1)
  \end{array}
  \right].
  \end{aligned}
  \label{eq:app-N0-L-JT}
\end{equation}
It is related to the $L[\cdots]$ in \eqref{eq:app-N0-onefold-as-L} by standard $L$-function transformations.
Combining \eqref{eq:app-N0-A-from-G}, \eqref{eq:app-N0-G-to-onefold}, and this Bailey relation, the stripped length-space answer is
\begin{equation}
  \mathcal A^{(0)}_\Delta
  =
  q^{-4\Delta}
  \left(\frac{2}{\pi}\right)^4
  \frac{1}{2^7\pi^2}\,
  \mathcal A^{\rm JT}_{\Delta}
  =
  \frac{q^{-4\Delta}}{8\pi^6}\,
  \mathcal A^{\rm JT}_{\Delta}.
  \label{eq:app-N0-final-Wilson-summary}
\end{equation}
Thus, up to the explicit overall factor in \eqref{eq:app-N0-final-Wilson-summary}, the length integral gives the usual bosonic JT $6j$-symbol in the normalization displayed in \eqref{eq:app-N0-JT-full}.

\paragraph{Unequal dimensions.}
The old $I_3$ calculation also records the extension before imposing equal dimensions.
One keeps the two matter factors separately in the diagonal gluing as $L^{2\Delta_1}(L')^{2\Delta_2}$.
Using $LL'=L_1L_2L_3L_4/(L_1L_3+L_2L_4)$, this is
\[
  \left(
    \frac{L_1L_2L_3L_4}{L_1L_3+L_2L_4}
  \right)^{2\Delta_2}
  L^{2(\Delta_1-\Delta_2)} .
\]
Thus the equal-dimension case is special: the last factor is absent, the two $I_3$'s collapse to $I_4$, and one obtains the length integral evaluated above.
The Gamma-integral calculation applies the same Mellin/Bessel steps to this shifted diagonal gluing: the extra factor is expanded into six Mellin--Barnes blocks, and each block is reduced by the same Mellin--Bessel transform, Barnes' first lemma, the $A_2$ integral, and Barnes' second lemma to a one-fold Mellin--Barnes expression.
The result is the unequal-dimension JT $6j$ kernel
\begin{equation}
  \begin{aligned}
  \mathcal A^{\rm JT}_{\Delta_1,\Delta_2}
  &=
  \frac{
  \Gamma(\Delta_1+is_1\pm is_4)
  \Gamma(\Delta_1-is_3\pm is_2)
  }{\Gamma(2\Delta_1)}
  \frac{
  \Gamma(\Delta_2-is_1\pm is_2)
  \Gamma(\Delta_2+is_3\pm is_4)
  }{\Gamma(2\Delta_2)}
  \\
  &\qquad\times
  \int_{C_u}\frac{du}{2\pi i}\,
  \frac{
  \Gamma(u\pm is_4)\,
  \Gamma(u+i(s_1+s_3)\pm is_2)\,
  \Gamma(\Delta_1-is_1-u)\,
  \Gamma(\Delta_2-is_3-u)
  }{
  \Gamma(u+\Delta_1+is_1)\,
  \Gamma(u+\Delta_2+is_3)
  } .
  \end{aligned}
  \label{eq:app-N0-JT-unequal-delta}
\end{equation}
Setting $\Delta_1=\Delta_2=\Delta$ reduces this expression to \eqref{eq:app-N0-JT-full}, which is the case evaluated explicitly above.

\bibliographystyle{ourbst}
\IfFileExists{../references.bib}{%
  \bibliography{../references}%
}{%
  \bibliography{references}%

\providecommand{\href}[2]{#2}\begingroup\raggedright\begin{thebibliography}{10}

\bibitem{Jackiw:1984je}
R.~Jackiw, ``{Lower Dimensional Gravity},''
\href{http://dx.doi.org/10.1016/0550-3213(85)90448-1}{{\em Nucl. Phys.}
  {\bfseries B252} (1985) 343--356}.

\bibitem{Teitelboim:1983ux}
C.~Teitelboim, ``{Gravitation and Hamiltonian Structure in Two Space-Time
  Dimensions},''
\href{http://dx.doi.org/10.1016/0370-2693(83)90012-6}{{\em Phys. Lett.}
  {\bfseries 126B} (1983) 41--45}.

\bibitem{Almheiri:2014cka}
A.~Almheiri and J.~Polchinski, ``{Models of AdS$_{2}$ backreaction and
  holography},'' \href{http://dx.doi.org/10.1007/JHEP11(2015)014}{{\em JHEP}
  {\bfseries 11} (2015) 014},
\href{http://arxiv.org/abs/1402.6334}{{\ttfamily arXiv:1402.6334 [hep-th]}}.

\bibitem{Jensen:2016pah}
K.~Jensen, ``{Chaos in AdS$_2$ Holography},''
  \href{http://dx.doi.org/10.1103/PhysRevLett.117.111601}{{\em Phys. Rev.
  Lett.} {\bfseries 117} no.~11, (2016) 111601},
\href{http://arxiv.org/abs/1605.06098}{{\ttfamily arXiv:1605.06098 [hep-th]}}.

\bibitem{Maldacena:2016upp}
J.~Maldacena, D.~Stanford, and Z.~Yang, ``{Conformal symmetry and its breaking
  in two dimensional Nearly Anti-de-Sitter space},''
  \href{http://dx.doi.org/10.1093/ptep/ptw124}{{\em PTEP} {\bfseries 2016}
  no.~12, (2016) 12C104}, \href{http://arxiv.org/abs/1606.01857}{{\ttfamily
  arXiv:1606.01857 [hep-th]}}.

\bibitem{Engelsoy:2016xyb}
J.~Engelsoy, T.~G. Mertens, and H.~Verlinde, ``{An investigation of AdS$_{2}$
  backreaction and holography},''
  \href{http://dx.doi.org/10.1007/JHEP07(2016)139}{{\em JHEP} {\bfseries 07}
  (2016) 139},
\href{http://arxiv.org/abs/1606.03438}{{\ttfamily arXiv:1606.03438 [hep-th]}}.

\bibitem{Heydeman:2020hhw}
M.~Heydeman, L.~V. Iliesiu, G.~J. Turiaci, and W.~Zhao, ``{The statistical
  mechanics of near-BPS black holes},''
  \href{http://dx.doi.org/10.1088/1751-8121/ac3be9}{{\em J. Phys. A} {\bfseries
  55} no.~1, (2022) 014004}, \href{http://arxiv.org/abs/2011.01953}{{\ttfamily
  arXiv:2011.01953 [hep-th]}}.

\bibitem{Boruch:2022tno}
J.~Boruch, M.~T. Heydeman, L.~V. Iliesiu, and G.~J. Turiaci, ``{BPS and
  near-BPS black holes in AdS$_{5}$ and their spectrum in $ \mathcal{N} $ = 4
  SYM},'' \href{http://dx.doi.org/10.1007/JHEP07(2025)220}{{\em JHEP}
  {\bfseries 07} (2025) 220}, \href{http://arxiv.org/abs/2203.01331}{{\ttfamily
  arXiv:2203.01331 [hep-th]}}.

\bibitem{Heydeman:2025vcc}
M.~Heydeman, X.~Shi, and G.~J. Turiaci, ``{Can black holes preserve
  $\mathscr{N}>4$ supersymmetry?},''
  \href{http://dx.doi.org/10.1007/JHEP01(2026)054}{{\em JHEP} {\bfseries 01}
  (2026) 054}, \href{http://arxiv.org/abs/2504.20146}{{\ttfamily
  arXiv:2504.20146 [hep-th]}}.

\bibitem{Iliesiu:2020qvm}
L.~V. Iliesiu and G.~J. Turiaci, ``{The statistical mechanics of near-extremal
  black holes},'' \href{http://arxiv.org/abs/2003.02860}{{\ttfamily
  arXiv:2003.02860 [hep-th]}}.

\bibitem{Moitra:2019bub}
U.~Moitra, S.~K. Sake, S.~P. Trivedi, and V.~Vishal, ``{Jackiw-Teitelboim
  Gravity and Rotating Black Holes},''
\href{http://arxiv.org/abs/1905.10378}{{\ttfamily arXiv:1905.10378 [hep-th]}}.

\bibitem{Kapec:2023ruw}
D.~Kapec, A.~Sheta, A.~Strominger, and C.~Toldo, ``{Logarithmic Corrections to
  Kerr Thermodynamics},''
  \href{http://dx.doi.org/10.1103/PhysRevLett.133.021601}{{\em Phys. Rev.
  Lett.} {\bfseries 133} no.~2, (2024) 021601},
  \href{http://arxiv.org/abs/2310.00848}{{\ttfamily arXiv:2310.00848
  [hep-th]}}.

\bibitem{Rakic:2023vhv}
I.~Rakic, M.~Rangamani, and G.~J. Turiaci, ``{Thermodynamics of the
  near-extremal Kerr spacetime},''
  \href{http://dx.doi.org/10.1007/JHEP06(2024)011}{{\em JHEP} {\bfseries 06}
  (2024) 011}, \href{http://arxiv.org/abs/2310.04532}{{\ttfamily
  arXiv:2310.04532 [hep-th]}}.

\bibitem{Maulik:2024dwq}
S.~Maulik, L.~A. Pando~Zayas, A.~Ray, and J.~Zhang, ``{Universality in
  logarithmic temperature corrections to near-extremal rotating black hole
  thermodynamics in various dimensions},''
  \href{http://dx.doi.org/10.1007/JHEP06(2024)034}{{\em JHEP} {\bfseries 06}
  (2024) 034}, \href{http://arxiv.org/abs/2401.16507}{{\ttfamily
  arXiv:2401.16507 [hep-th]}}.

\bibitem{Kapec:2024zdj}
D.~Kapec, Y.~T.~A. Law, and C.~Toldo, ``{Quasinormal corrections to
  near-extremal black hole thermodynamics},''
  \href{http://dx.doi.org/10.1007/JHEP06(2025)069}{{\em JHEP} {\bfseries 06}
  (2025) 069}, \href{http://arxiv.org/abs/2409.14928}{{\ttfamily
  arXiv:2409.14928 [hep-th]}}.

\bibitem{Mertens:2019bvy}
T.~G. Mertens, ``{Towards Black Hole Evaporation in Jackiw-Teitelboim
  Gravity},'' \href{http://dx.doi.org/10.1007/JHEP07(2019)097}{{\em JHEP}
  {\bfseries 07} (2019) 097},
\href{http://arxiv.org/abs/1903.10485}{{\ttfamily arXiv:1903.10485 [hep-th]}}.

\bibitem{Blommaert:2020yeo}
A.~Blommaert, T.~G. Mertens, and H.~Verschelde, ``{Unruh detectors and quantum
  chaos in JT gravity},''
\href{http://arxiv.org/abs/2005.13058}{{\ttfamily arXiv:2005.13058 [hep-th]}}.

\bibitem{Bai:2023hpd}
Y.~Bai and M.~Korwar, ``{Near-extremal charged black holes: greybody factors
  and evolution},'' \href{http://dx.doi.org/10.1007/JHEP03(2023)151}{{\em JHEP}
  {\bfseries 03} (2023) 151}, \href{http://arxiv.org/abs/2301.07739}{{\ttfamily
  arXiv:2301.07739 [hep-th]}}.

\bibitem{Brown:2024ajk}
A.~R. Brown, L.~V. Iliesiu, G.~Penington, and M.~Usatyuk, ``{The evaporation of
  charged black holes},'' \href{http://dx.doi.org/10.1007/JHEP01(2026)109}{{\em
  JHEP} {\bfseries 01} (2026) 109},
  \href{http://arxiv.org/abs/2411.03447}{{\ttfamily arXiv:2411.03447
  [hep-th]}}.

\bibitem{Emparan:2023ypa}
R.~Emparan and J.~M. Magan, ``{Tearing down spacetime with quantum
  disentanglement},'' \href{http://dx.doi.org/10.1007/JHEP03(2024)078}{{\em
  JHEP} {\bfseries 03} (2024) 078},
  \href{http://arxiv.org/abs/2312.06803}{{\ttfamily arXiv:2312.06803
  [hep-th]}}.

\bibitem{Kolanowski:2024zrq}
M.~Kolanowski, D.~Marolf, I.~Rakic, M.~Rangamani, and G.~J. Turiaci, ``{Looking
  at extremal black holes from very far away},''
  \href{http://dx.doi.org/10.1007/JHEP04(2025)020}{{\em JHEP} {\bfseries 04}
  (2025) 020}, \href{http://arxiv.org/abs/2409.16248}{{\ttfamily
  arXiv:2409.16248 [hep-th]}}.

\bibitem{Emparan:2025sao}
R.~Emparan, ``{Quantum cross-section of near-extremal black holes},''
  \href{http://dx.doi.org/10.1007/JHEP04(2025)122}{{\em JHEP} {\bfseries 04}
  (2025) 122}, \href{http://arxiv.org/abs/2501.17470}{{\ttfamily
  arXiv:2501.17470 [hep-th]}}.

\bibitem{Biggs:2025nzs}
A.~Biggs, ``{Following the state of an evaporating charged black hole into the
  quantum gravity regime},'' \href{http://arxiv.org/abs/2503.02051}{{\ttfamily
  arXiv:2503.02051 [hep-th]}}.

\bibitem{Maulik:2025hax}
S.~Maulik, X.~Meng, and L.~A. Pando~Zayas, ``{Quantum-corrected Hawking
  radiation from near-extremal Kerr-Newman black holes},''
  \href{http://dx.doi.org/10.1007/JHEP02(2026)205}{{\em JHEP} {\bfseries 02}
  (2026) 205}, \href{http://arxiv.org/abs/2501.08252}{{\ttfamily
  arXiv:2501.08252 [hep-th]}}.

\bibitem{Lin:2025wof}
G.~Lin, L.~V. Iliesiu, and M.~Usatyuk, ``{The evaporation of black holes in
  supergravity},'' \href{http://dx.doi.org/10.1007/JHEP08(2025)220}{{\em JHEP}
  {\bfseries 08} (2025) 220}, \href{http://arxiv.org/abs/2504.21077}{{\ttfamily
  arXiv:2504.21077 [hep-th]}}.

\bibitem{Betzios:2025sct}
P.~Betzios, O.~Papadoulaki, and Y.~Zhou, ``{Near-extremal quantum cross-section
  for charged fields and superradiance},''
  \href{http://dx.doi.org/10.1007/JHEP11(2025)114}{{\em JHEP} {\bfseries 11}
  (2025) 114}, \href{http://arxiv.org/abs/2507.13896}{{\ttfamily
  arXiv:2507.13896 [hep-th]}}.

\bibitem{Shenker:2013pqa}
S.~H. Shenker and D.~Stanford, ``{Black holes and the butterfly effect},''
  \href{http://dx.doi.org/10.1007/JHEP03(2014)067}{{\em JHEP} {\bfseries 03}
  (2014) 067}, \href{http://arxiv.org/abs/1306.0622}{{\ttfamily arXiv:1306.0622
  [hep-th]}}.

\bibitem{Roberts:2014isa}
D.~A. Roberts, D.~Stanford, and L.~Susskind, ``{Localized shocks},''
  \href{http://dx.doi.org/10.1007/JHEP03(2015)051}{{\em JHEP} {\bfseries 03}
  (2015) 051}, \href{http://arxiv.org/abs/1409.8180}{{\ttfamily arXiv:1409.8180
  [hep-th]}}.

\bibitem{Roberts:2014ifa}
D.~A. Roberts and D.~Stanford, ``{Two-dimensional conformal field theory and
  the butterfly effect},''
  \href{http://dx.doi.org/10.1103/PhysRevLett.115.131603}{{\em Phys. Rev.
  Lett.} {\bfseries 115} no.~13, (2015) 131603},
\href{http://arxiv.org/abs/1412.5123}{{\ttfamily arXiv:1412.5123 [hep-th]}}.

\bibitem{Hosur:2015ylk}
P.~Hosur, X.-L. Qi, D.~A. Roberts, and B.~Yoshida, ``{Chaos in quantum
  channels},'' \href{http://dx.doi.org/10.1007/JHEP02(2016)004}{{\em JHEP}
  {\bfseries 02} (2016) 004}, \href{http://arxiv.org/abs/1511.04021}{{\ttfamily
  arXiv:1511.04021 [hep-th]}}.

\bibitem{Maldacena:2015waa}
J.~Maldacena, S.~H. Shenker, and D.~Stanford, ``{A bound on chaos},''
  \href{http://dx.doi.org/10.1007/JHEP08(2016)106}{{\em JHEP} {\bfseries 08}
  (2016) 106}, \href{http://arxiv.org/abs/1503.01409}{{\ttfamily
  arXiv:1503.01409 [hep-th]}}.

\bibitem{Maldacena:2016hyu}
J.~Maldacena and D.~Stanford, ``{Remarks on the Sachdev-Ye-Kitaev model},''
  \href{http://dx.doi.org/10.1103/PhysRevD.94.106002}{{\em Phys. Rev.}
  {\bfseries D94} no.~10, (2016) 106002},
\href{http://arxiv.org/abs/1604.07818}{{\ttfamily arXiv:1604.07818 [hep-th]}}.

\bibitem{Xu:2024otoc}
S.~Xu and B.~Swingle, ``{Scrambling Dynamics and Out-of-Time-Ordered
  Correlators in Quantum Many-Body Systems},''
  \href{http://dx.doi.org/10.1103/PRXQuantum.5.010201}{{\em PRX Quantum}
  {\bfseries 5} no.~1, (2024) 010201}.

\bibitem{Stanford:2023npy}
D.~Stanford, S.~Vardhan, and S.~Yao, ``{Scramblon loops},''
  \href{http://dx.doi.org/10.1007/JHEP10(2024)073}{{\em JHEP} {\bfseries 10}
  (2024) 073}, \href{http://arxiv.org/abs/2311.12121}{{\ttfamily
  arXiv:2311.12121 [hep-th]}}.

\bibitem{Kitaev:2017awl}
A.~Kitaev and S.~J. Suh, ``{The soft mode in the Sachdev-Ye-Kitaev model and
  its gravity dual},'' \href{http://dx.doi.org/10.1007/JHEP05(2018)183}{{\em
  JHEP} {\bfseries 05} (2018) 183},
\href{http://arxiv.org/abs/1711.08467}{{\ttfamily arXiv:1711.08467 [hep-th]}}.

\bibitem{Mertens:2017mtv}
T.~G. Mertens, G.~J. Turiaci, and H.~L. Verlinde, ``{Solving the Schwarzian via
  the Conformal Bootstrap},''
  \href{http://dx.doi.org/10.1007/JHEP08(2017)136}{{\em JHEP} {\bfseries 08}
  (2017) 136}, \href{http://arxiv.org/abs/1705.08408}{{\ttfamily
  arXiv:1705.08408 [hep-th]}}.

\bibitem{Blommaert:2018oro}
A.~Blommaert, T.~G. Mertens, and H.~Verschelde, ``{The Schwarzian Theory - A
  Wilson Line Perspective},''
  \href{http://dx.doi.org/10.1007/JHEP12(2018)022}{{\em JHEP} {\bfseries 12}
  (2018) 022}, \href{http://arxiv.org/abs/1806.07765}{{\ttfamily
  arXiv:1806.07765 [hep-th]}}.

\bibitem{Mertens:2022irh}
T.~G. Mertens and G.~J. Turiaci, ``{Solvable models of quantum black holes: a
  review on Jackiw{\textendash}Teitelboim gravity},''
  \href{http://dx.doi.org/10.1007/s41114-023-00046-1}{{\em Living Rev. Rel.}
  {\bfseries 26} no.~1, (2023) 4},
  \href{http://arxiv.org/abs/2210.10846}{{\ttfamily arXiv:2210.10846
  [hep-th]}}.

\bibitem{Lin:2022rzw}
H.~W. Lin, J.~Maldacena, L.~Rozenberg, and J.~Shan, ``{Holography for people
  with no time},'' \href{http://dx.doi.org/10.21468/SciPostPhys.14.6.150}{{\em
  SciPost Phys.} {\bfseries 14} no.~6, (2023) 150},
  \href{http://arxiv.org/abs/2207.00407}{{\ttfamily arXiv:2207.00407
  [hep-th]}}.

\bibitem{Lin:2022zxd}
H.~W. Lin, J.~Maldacena, L.~Rozenberg, and J.~Shan, ``{Looking at
  supersymmetric black holes for a very long time},''
  \href{http://dx.doi.org/10.21468/SciPostPhys.14.5.128}{{\em SciPost Phys.}
  {\bfseries 14} no.~5, (2023) 128},
  \href{http://arxiv.org/abs/2207.00408}{{\ttfamily arXiv:2207.00408
  [hep-th]}}.

\bibitem{Chen:2024oqv}
Y.~Chen, H.~W. Lin, and S.~H. Shenker, ``{BPS chaos},''
  \href{http://dx.doi.org/10.21468/SciPostPhys.18.2.072}{{\em SciPost Phys.}
  {\bfseries 18} no.~2, (2025) 072},
  \href{http://arxiv.org/abs/2407.19387}{{\ttfamily arXiv:2407.19387
  [hep-th]}}.

\bibitem{Chen:2026vml}
Y.~Chen, S.~Colin-Ellerin, O.~Mamroud, and K.~Papadodimas, ``{Chaos of Berry
  curvature for BPS microstates},''
  \href{http://arxiv.org/abs/2604.23287}{{\ttfamily arXiv:2604.23287
  [hep-th]}}.

\bibitem{Fan:2021wsb}
Y.~Fan and T.~G. Mertens, ``{Supergroup structure of Jackiw-Teitelboim
  supergravity},'' \href{http://dx.doi.org/10.1007/JHEP08(2022)002}{{\em JHEP}
  {\bfseries 08} (2022) 002}, \href{http://arxiv.org/abs/2106.09353}{{\ttfamily
  arXiv:2106.09353 [hep-th]}}.

\bibitem{Belaey:2024dde}
A.~Belaey, F.~Mariani, and T.~G. Mertens, ``{Gravitational wavefunctions in JT
  supergravity},'' \href{http://dx.doi.org/10.1007/JHEP10(2024)037}{{\em JHEP}
  {\bfseries 10} (2024) 037}, \href{http://arxiv.org/abs/2405.09289}{{\ttfamily
  arXiv:2405.09289 [hep-th]}}.

\bibitem{Belaey:2023jtr}
A.~Belaey, F.~Mariani, and T.~G. Mertens, ``{Branes in JT (super)gravity from
  group theory},'' \href{http://dx.doi.org/10.1007/JHEP02(2024)058}{{\em JHEP}
  {\bfseries 02} (2024) 058}, \href{http://arxiv.org/abs/2310.04245}{{\ttfamily
  arXiv:2310.04245 [hep-th]}}.

\bibitem{Yang:2018gdb}
Z.~Yang, ``{The Quantum Gravity Dynamics of Near Extremal Black Holes},''
  \href{http://dx.doi.org/10.1007/JHEP05(2019)205}{{\em JHEP} {\bfseries 05}
  (2019) 205}, \href{http://arxiv.org/abs/1809.08647}{{\ttfamily
  arXiv:1809.08647 [hep-th]}}.

\bibitem{Jafferis:2022wez}
D.~L. Jafferis, D.~K. Kolchmeyer, B.~Mukhametzhanov, and J.~Sonner, ``{JT
  gravity with matter, generalized ETH, and Random Matrices},''
  \href{http://arxiv.org/abs/2209.02131}{{\ttfamily arXiv:2209.02131
  [hep-th]}}.

\bibitem{Jafferis:2022uhu}
D.~L. Jafferis, D.~K. Kolchmeyer, B.~Mukhametzhanov, and J.~Sonner, ``{Matrix
  models for eigenstate thermalization},''
  \href{http://arxiv.org/abs/2209.02130}{{\ttfamily arXiv:2209.02130
  [hep-th]}}.

\bibitem{Penington:2024sum}
G.~Penington and E.~Witten, ``{Algebras and states in super-JT gravity},''
  \href{http://arxiv.org/abs/2412.15549}{{\ttfamily arXiv:2412.15549
  [hep-th]}}.

\bibitem{Stanford:2017thb}
D.~Stanford and E.~Witten, ``{Fermionic Localization of the Schwarzian
  Theory},'' \href{http://dx.doi.org/10.1007/JHEP10(2017)008}{{\em JHEP}
  {\bfseries 10} (2017) 008},
\href{http://arxiv.org/abs/1703.04612}{{\ttfamily arXiv:1703.04612 [hep-th]}}.

\bibitem{Turiaci:2017zwd}
G.~Turiaci and H.~Verlinde, ``{Towards a 2d QFT Analog of the SYK Model},''
  \href{http://dx.doi.org/10.1007/JHEP10(2017)167}{{\em JHEP} {\bfseries 10}
  (2017) 167},
\href{http://arxiv.org/abs/1701.00528}{{\ttfamily arXiv:1701.00528 [hep-th]}}.

\bibitem{Lam:2018pvp}
H.~T. Lam, T.~G. Mertens, G.~J. Turiaci, and H.~Verlinde, ``{Shockwave S-matrix
  from Schwarzian Quantum Mechanics},''
  \href{http://dx.doi.org/10.1007/JHEP11(2018)182}{{\em JHEP} {\bfseries 11}
  (2018) 182},
\href{http://arxiv.org/abs/1804.09834}{{\ttfamily arXiv:1804.09834 [hep-th]}}.

\bibitem{Mertens:2018fds}
T.~G. Mertens, ``{The Schwarzian theory {\textemdash} origins},''
  \href{http://dx.doi.org/10.1007/JHEP05(2018)036}{{\em JHEP} {\bfseries 05}
  (2018) 036}, \href{http://arxiv.org/abs/1801.09605}{{\ttfamily
  arXiv:1801.09605 [hep-th]}}.

\bibitem{Iliesiu:2019xuh}
L.~V. Iliesiu, S.~S. Pufu, H.~Verlinde, and Y.~Wang, ``{An exact quantization
  of Jackiw-Teitelboim gravity},''
\href{http://arxiv.org/abs/1905.02726}{{\ttfamily arXiv:1905.02726 [hep-th]}}.

\bibitem{Cardenas:2018krd}
M.~C{\'a}rdenas, O.~Fuentealba, H.~A. Gonz{\'a}lez, D.~Grumiller,
  C.~Valc{\'a}rcel, and D.~Vassilevich, ``{Boundary theories for dilaton
  supergravity in 2D},'' \href{http://dx.doi.org/10.1007/JHEP11(2018)077}{{\em
  JHEP} {\bfseries 11} (2018) 077},
  \href{http://arxiv.org/abs/1809.07208}{{\ttfamily arXiv:1809.07208
  [hep-th]}}.

\bibitem{Alkalaev:2022qfc}
K.~Alkalaev, E.~Joung, and J.~Yoon, ``{Schwarzian for colored Jackiw-Teitelboim
  gravity},'' \href{http://dx.doi.org/10.1007/JHEP09(2022)160}{{\em JHEP}
  {\bfseries 09} (2022) 160}, \href{http://arxiv.org/abs/2204.09010}{{\ttfamily
  arXiv:2204.09010 [hep-th]}}.

\bibitem{Valach:2019jzv}
F.~Valach and D.~R. Youmans, ``{Schwarzian quantum mechanics as a
  Drinfeld-Sokolov reduction of $BF$ theory},''
  \href{http://dx.doi.org/10.1007/JHEP12(2020)189}{{\em JHEP} {\bfseries 12}
  (2020) 189}, \href{http://arxiv.org/abs/1912.12331}{{\ttfamily
  arXiv:1912.12331 [hep-th]}}.

\bibitem{Fu:2016vas}
W.~Fu, D.~Gaiotto, J.~Maldacena, and S.~Sachdev, ``{Supersymmetric
  Sachdev-Ye-Kitaev models},''
  \href{http://dx.doi.org/10.1103/PhysRevD.95.026009}{{\em Phys. Rev. D}
  {\bfseries 95} no.~2, (2017) 026009},
  \href{http://arxiv.org/abs/1610.08917}{{\ttfamily arXiv:1610.08917
  [hep-th]}}.

\bibitem{Matsuda:1989kp}
S.~Matsuda and T.~Uematsu, ``{Superschwarzian Derivatives in $N=4$ SU(2)
  Extended Superconformal Algebras},''
  \href{http://dx.doi.org/10.1142/S0217732390000937}{{\em Mod. Phys. Lett. A}
  {\bfseries 5} (1990) 841}.

\bibitem{Aoyama:2018lfc}
S.~Aoyama and Y.~Honda, ``{N = 4 super-Schwarzian theory on the coadjoint orbit
  and PSU(1,1|2)},'' \href{http://dx.doi.org/10.1007/JHEP06(2018)070}{{\em
  JHEP} {\bfseries 06} (2018) 070},
  \href{http://arxiv.org/abs/1801.06800}{{\ttfamily arXiv:1801.06800
  [hep-th]}}.

\bibitem{Lee:2019uen}
S.~Lee and S.~Lee, ``{Notes on superconformal representations in two
  dimensions},'' \href{http://dx.doi.org/10.1016/j.nuclphysb.2020.115033}{{\em
  Nucl. Phys. B} {\bfseries 956} (2020) 115033},
  \href{http://arxiv.org/abs/1911.10391}{{\ttfamily arXiv:1911.10391
  [hep-th]}}.

\bibitem{deBoer:2008qe}
J.~de~Boer, K.~Papadodimas, and E.~Verlinde, ``{Black Hole Berry Phase},''
  \href{http://dx.doi.org/10.1103/PhysRevLett.103.131301}{{\em Phys. Rev.
  Lett.} {\bfseries 103} (2009) 131301},
  \href{http://arxiv.org/abs/0809.5062}{{\ttfamily arXiv:0809.5062 [hep-th]}}.

\bibitem{Baggio:2017aww}
M.~Baggio, V.~Niarchos, and K.~Papadodimas, ``{Aspects of Berry phase in
  QFT},'' \href{http://dx.doi.org/10.1007/JHEP04(2017)062}{{\em JHEP}
  {\bfseries 04} (2017) 062}, \href{http://arxiv.org/abs/1701.05587}{{\ttfamily
  arXiv:1701.05587 [hep-th]}}.

\bibitem{deBoer:2008ss}
J.~de~Boer, J.~Manschot, K.~Papadodimas, and E.~Verlinde, ``{The Chiral ring of
  AdS(3)/CFT(2) and the attractor mechanism},''
  \href{http://dx.doi.org/10.1088/1126-6708/2009/03/030}{{\em JHEP} {\bfseries
  03} (2009) 030}, \href{http://arxiv.org/abs/0809.0507}{{\ttfamily
  arXiv:0809.0507 [hep-th]}}.

\bibitem{Baggio:2012rr}
M.~Baggio, J.~de~Boer, and K.~Papadodimas, ``{A non-renormalization theorem for
  chiral primary 3-point functions},''
  \href{http://dx.doi.org/10.1007/JHEP07(2012)137}{{\em JHEP} {\bfseries 07}
  (2012) 137}, \href{http://arxiv.org/abs/1203.1036}{{\ttfamily arXiv:1203.1036
  [hep-th]}}.

\bibitem{Strominger:1996sh}
A.~Strominger and C.~Vafa, ``{Microscopic origin of the Bekenstein-Hawking
  entropy},'' \href{http://dx.doi.org/10.1016/0370-2693(96)00345-0}{{\em Phys.
  Lett. B} {\bfseries 379} (1996) 99--104},
  \href{http://arxiv.org/abs/hep-th/9601029}{{\ttfamily arXiv:hep-th/9601029}}.

\bibitem{LarsenMartinec:1999}
F.~Larsen and E.~J. Martinec, ``{$U(1)$ Charges and Moduli in the D1--D5
  System},'' {\em JHEP} {\bfseries 06} (1999) 019,
  \href{http://arxiv.org/abs/hep-th/9905064}{{\ttfamily arXiv:hep-th/9905064}}.

\bibitem{David:1999ec}
J.~R. David, G.~Mandal, and S.~R. Wadia, ``{D1 / D5 moduli in SCFT and gauge
  theory, and Hawking radiation},''
  \href{http://dx.doi.org/10.1016/S0550-3213(99)00620-3}{{\em Nucl. Phys. B}
  {\bfseries 564} (2000) 103--127},
  \href{http://arxiv.org/abs/hep-th/9907075}{{\ttfamily arXiv:hep-th/9907075}}.

\bibitem{Seiberg:1999xz}
N.~Seiberg and E.~Witten, ``{The D1 / D5 system and singular CFT},''
  \href{http://dx.doi.org/10.1088/1126-6708/1999/04/017}{{\em JHEP} {\bfseries
  04} (1999) 017}, \href{http://arxiv.org/abs/hep-th/9903224}{{\ttfamily
  arXiv:hep-th/9903224}}.

\bibitem{Grant:2008sk}
L.~Grant, P.~A. Grassi, S.~Kim, and S.~Minwalla, ``{Comments on 1/16 BPS
  Quantum States and Classical Configurations},''
  \href{http://dx.doi.org/10.1088/1126-6708/2008/05/049}{{\em JHEP} {\bfseries
  05} (2008) 049}, \href{http://arxiv.org/abs/0803.4183}{{\ttfamily
  arXiv:0803.4183 [hep-th]}}.

\bibitem{Chang:2013fba}
C.-M. Chang and X.~Yin, ``{$1/16$ BPS states in $\mathcal N=4$
  super-Yang--Mills theory},''
  \href{http://dx.doi.org/10.1103/PhysRevD.88.106005}{{\em Phys. Rev. D}
  {\bfseries 88} no.~10, (2013) 106005},
  \href{http://arxiv.org/abs/1305.6314}{{\ttfamily arXiv:1305.6314 [hep-th]}}.

\bibitem{Chang:2022mjp}
C.-M. Chang and Y.-H. Lin, ``{Words to describe a black hole},''
  \href{http://dx.doi.org/10.1007/JHEP02(2023)109}{{\em JHEP} {\bfseries 02}
  (2023) 109}, \href{http://arxiv.org/abs/2209.06728}{{\ttfamily
  arXiv:2209.06728 [hep-th]}}.

\bibitem{Choi:2022caq}
S.~Choi, S.~Kim, E.~Lee, and J.~Park, ``{The shape of non-graviton operators
  for SU(2)},'' \href{http://dx.doi.org/10.1007/JHEP09(2024)029}{{\em JHEP}
  {\bfseries 09} (2024) 029}, \href{http://arxiv.org/abs/2209.12696}{{\ttfamily
  arXiv:2209.12696 [hep-th]}}.

\bibitem{Choi:2023znd}
S.~Choi, S.~Kim, E.~Lee, S.~Lee, and J.~Park, ``{Towards quantum black hole
  microstates},'' \href{http://dx.doi.org/10.1007/JHEP11(2023)175}{{\em JHEP}
  {\bfseries 11} (2023) 175}, \href{http://arxiv.org/abs/2304.10155}{{\ttfamily
  arXiv:2304.10155 [hep-th]}}.

\bibitem{Budzik:2023vtr}
K.~Budzik, H.~Murali, and P.~Vieira, ``{Following Black Hole States},''
  \href{http://arxiv.org/abs/2306.04693}{{\ttfamily arXiv:2306.04693
  [hep-th]}}.

\bibitem{Chang:2023zqk}
C.-M. Chang, L.~Feng, Y.-H. Lin, and Y.-X. Tao, ``{Decoding stringy
  near-supersymmetric black holes},''
  \href{http://dx.doi.org/10.21468/SciPostPhys.16.4.109}{{\em SciPost Phys.}
  {\bfseries 16} no.~4, (2024) 109},
  \href{http://arxiv.org/abs/2306.04673}{{\ttfamily arXiv:2306.04673
  [hep-th]}}.

\bibitem{Choi:2023vdm}
J.~Choi, S.~Choi, S.~Kim, J.~Lee, and S.~Lee, ``{Finite $N$ black hole
  cohomologies},'' \href{http://dx.doi.org/10.1007/JHEP12(2024)029}{{\em JHEP}
  {\bfseries 12} (2024) 029}, \href{http://arxiv.org/abs/2312.16443}{{\ttfamily
  arXiv:2312.16443 [hep-th]}}.

\bibitem{Chang:2024zqi}
C.-M. Chang and Y.-H. Lin, ``{Holographic covering and the fortuity of black
  holes},'' \href{http://arxiv.org/abs/2402.10129}{{\ttfamily arXiv:2402.10129
  [hep-th]}}.

\bibitem{Turiaci:2023jfa}
G.~J. Turiaci and E.~Witten, ``{$ \mathcal{N} $ = 2 JT supergravity and matrix
  models},'' \href{http://dx.doi.org/10.1007/JHEP12(2023)003}{{\em JHEP}
  {\bfseries 12} (2023) 003}, \href{http://arxiv.org/abs/2305.19438}{{\ttfamily
  arXiv:2305.19438 [hep-th]}}.

\bibitem{Iliesiu:2021are}
L.~V. Iliesiu, M.~Kologlu, and G.~J. Turiaci, ``{Supersymmetric indices
  factorize},'' \href{http://dx.doi.org/10.1007/JHEP05(2023)032}{{\em JHEP}
  {\bfseries 05} (2023) 032}, \href{http://arxiv.org/abs/2107.09062}{{\ttfamily
  arXiv:2107.09062 [hep-th]}}.

\bibitem{Chu:1994hm}
M.-f. Chu and P.~Goddard, ``{Quantization of a particle moving on a group
  manifold},'' \href{http://dx.doi.org/10.1016/0370-2693(94)90977-6}{{\em Phys.
  Lett. B} {\bfseries 337} (1994) 285--293},
  \href{http://arxiv.org/abs/hep-th/9407116}{{\ttfamily arXiv:hep-th/9407116}}.

\bibitem{Alekseev:1990mp}
A.~Alekseev and S.~L. Shatashvili, ``{From geometric quantization to conformal
  field theory},'' \href{http://dx.doi.org/10.1007/BF02097053}{{\em Commun.
  Math. Phys.} {\bfseries 128} (1990) 197--212}.

\bibitem{Fitzpatrick:2016mtp}
A.~L. Fitzpatrick, J.~Kaplan, D.~Li, and J.~Wang, ``{Exact Virasoro Blocks from
  Wilson Lines and Background-Independent Operators},''
  \href{http://dx.doi.org/10.1007/JHEP07(2017)092}{{\em JHEP} {\bfseries 07}
  (2017) 092}, \href{http://arxiv.org/abs/1612.06385}{{\ttfamily
  arXiv:1612.06385 [hep-th]}}.

\end{thebibliography}\endgroup
}
\end{document}